\documentclass[12pt,letterpaper]{article}

\usepackage{xcolor}
\definecolor{navy}{rgb}{0,0.1,0.4}
\definecolor{navy-cap}{rgb}{0,0.1,0.5}
\definecolor{lgray}{gray}{0.90}
\definecolor{dkgreen}{rgb}{0,0.6,0}
\definecolor{gray}{rgb}{0.5,0.5,0.5}
\definecolor{mauve}{rgb}{0.58,0,0.82}

\usepackage{sectsty}
\sectionfont{\fontsize{12}{15}\selectfont \color{navy}}
\subsectionfont{\fontsize{12}{15}\selectfont \color{navy}}
\subsubsectionfont{\fontsize{12}{15}\selectfont \color{navy}}

\usepackage[margin = 1 in]{geometry}

\usepackage{titlesec}
\usepackage{tikz}
\titlespacing{\section}{0pt}{0.75em}{0.25em}
\titlespacing{\subsection}{0pt}{0.5em}{0.25em}
\titlespacing{\subsubsection}{0pt}{0.5em}{0.25em}

\usepackage{array,longtable}
\usepackage{xltabular}

\usepackage[font={color=navy-cap, footnotesize}, labelfont={bf}, labelsep=space, justification=justified, skip=0pt]{caption}
\usepackage{subcaption}

\usepackage{booktabs} 
\usepackage{graphicx}
\usepackage{wrapfig}
\usepackage{enumitem}
\usepackage{hyperref}
\usepackage{textcomp}
\usepackage{url}
\usepackage{amssymb}
\usepackage{bm}
\usepackage{amsmath}
\usepackage{dirtytalk}
\usepackage{pdfpages} 
\usepackage{soul} 
\usepackage{fancyhdr}
\usetikzlibrary{patterns}
\usepackage{pgfplots}
\usepackage{floatrow}
\usepackage{setspace}

\usepackage{algorithm2e}

\usepackage[
backend=biber,
style=numeric-comp,
sorting=none,
maxnames = 10
]{biblatex}
\newcommand{\cmt}[1]{} 

\title{\Large Deep Learning for Multiscale Damage Analysis via Physics-Informed Recurrent Neural Networks}
\date{\vspace{-5ex}}
\usepackage{authblk}
\author[1] {\normalsize Shiguang Deng}
\author[2]{Shirin Hosseinmardi}
\author[1]{Diran Apelian}
\author[2]{Ramin Bostanabad\thanks{\noindent Corresponding Author: Raminb@uci.edu\\}}
\affil [1]{ACRC, Materials Science and Engineering, University of California, Irvine}
\affil [2]{Department of Mechanical and Aerospace Engineering, University of California, Irvine}

\begin{document}

\maketitle
\pagenumbering{arabic}

\sloppy 
\noindent \textcolor{navy}{\textbf{Abstract}}

Direct numerical simulation of hierarchical materials via homogenization-based concurrent multiscale models poses critical challenges for 3D large scale engineering applications, as the computation of highly nonlinear and path-dependent material constitutive responses at the fine scale causes prohibitively high computational costs. In this work, we propose a physics-informed data-driven deep learning model as an efficient surrogate to emulate the effective responses of heterogeneous microstructures under irreversible elasto-plastic hardening and softening deformation. Our contribution contains several major innovations. First, we propose a novel training scheme to generate arbitrary loading sequences in the sampling space confined by deformation constraints where the simulation cost of homogenizing microstructural responses per sequence is dramatically reduced via mechanistic reduced-order models. Second, we develop a new sequential learner that incorporates thermodynamics consistent physics constraints by customizing training loss function and data flow architecture. We additionally demonstrate the integration of trained surrogate within the framework of classic multiscale finite element solver. Our numerical experiments indicate that our model shows a significant accuracy improvement over pure data-driven model and a dramatic efficiency boost than reduced-order models. We believe our data-driven model provides a computationally efficient and mechanics consistent alternative for classic constitutive laws, which is beneficial for potential high-throughput simulations that needs material homogenization of irreversible behaviors. 

\section{Introduction} \label{sec:intro}

\begin{figure}[!t]
    \centering
    \includegraphics[width = 1.0\textwidth]{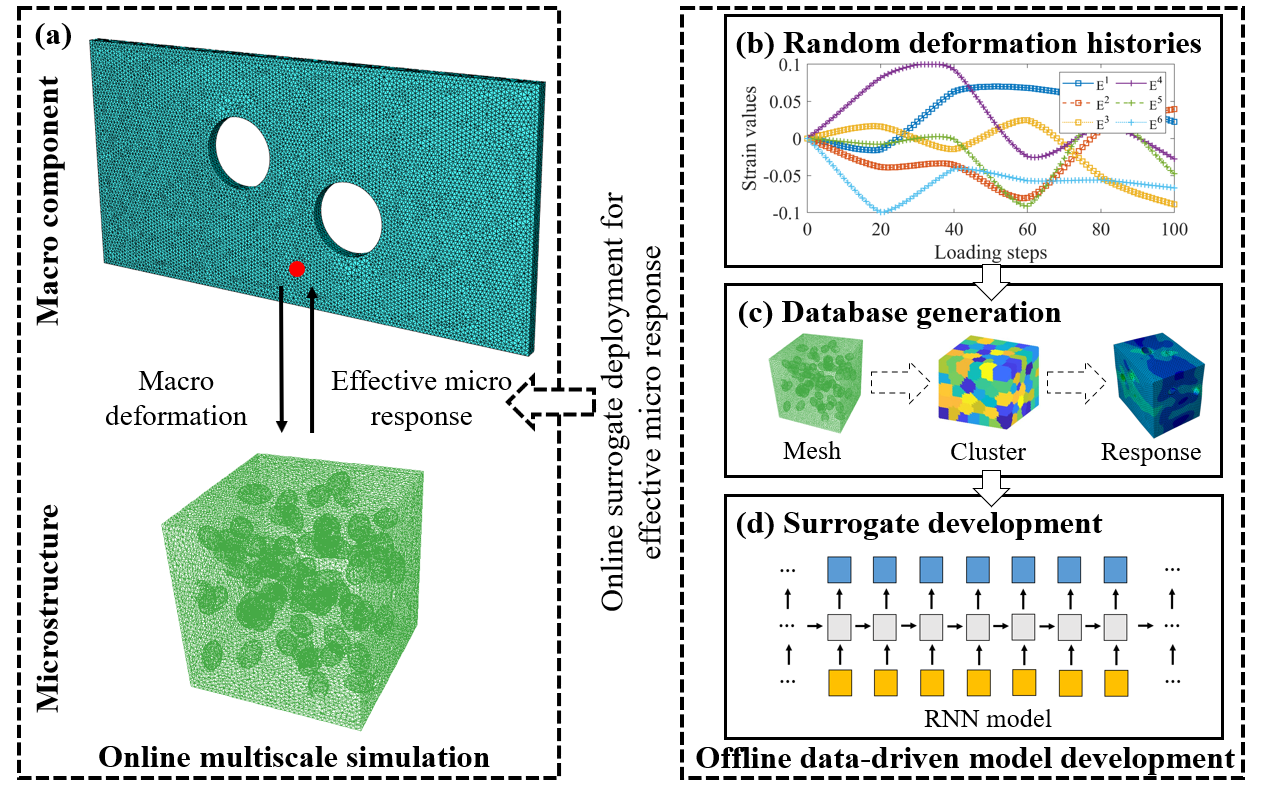}
    \caption{\textbf{Framework of the proposed data-driven material model:} Our physics-informed data-driven model, replacing the microstructural effective responses amid online multiscale simulations, is composed of three major offline components: microstructural representation, response database generation and machine learning model development.
    } \label{fig:framework}
\end{figure}

Heterogeneous materials are increasing used in many engineering applications. Analyzing the behavior of such materials often relies on multiscale simulations such as the FE\textsuperscript{2} method \cite{feyel2000fe2} which is a popular homogenization-based concurrent multiscale model that uses the finite element method (FEM) at two scales. Despite the recent advancements in software/hardware and mechanics theory \cite{kanoute2009multiscale}, the simulation of hierarchical materials via FE\textsuperscript{2} is still prohibitively costly. Consider the multiscale model in Figure \ref{fig:framework}(a) where each integration point (IP) of the macroscale component represents a microstructure with complex local morphology. In this model, the two-scale spatial discretization requires a large memory storage and also results in long runtimes since the solver repeatedly iterates between the scales. These challenges are exacerbated in the presence of microstructural deformations that are path-dependent and involve damage. That is, evaluation of the microstructural responses are the primary computational bottleneck. 
Our goal in this paper is to address such bottlenecks by developing a deep learning (DL) model that surrogates the microstructural analyses in 3D multiscale simulations that involve plasticity and damage. 

Mechanistic reduced-order models (ROMs) are attractive alternatives to expensive methods such as the FE\textsuperscript{2} and direct numerical simulations (DNS). The main idea behind ROMs is to reduce the number of unknown variables (e.g., stresses, strains, or internal variables such as the damage parameters) while striking a balance between accuracy and efficiency. 
For example, the transformed field analysis method \cite{dvorak1992transformation} and its non-uniform variant \cite{roussette2009nonuniform} employ proper orthogonal decomposition to reduce material state variables by expressing arbitrary strain fields as a subspace representation of pre-computed eigenstrains. 
Clustering-based ROMs reduce unknown variables by agglomerating a large number of material points into a few clusters. For instance, the self-consistent clustering \cite{liu2016self} and its variant the virtual clustering analysis \cite{tang2018virtual} assume material points with similar elastic responses behave similarly during inelastic deformations and solve incremental Lippmann-Schwinger equations to approximate the evolution of cluster-wise material responses. Deflated clustering analysis (DCA) \cite{deng2022reduced} utilizes clusters to decompose both macroscale and microscale domains where macro analysis is faithfully accelerated in a deflation space while the effective microstructural responses are approximated in a coarse-graining manner where close-by material points are presumed to share the same behaviors. 
DCA's robustness and efficiency are further improved in a later work \cite{deng2022concurrent} where both spatial and temporal dimensions are adaptively reduced for elasto-plastic deformations with softening.While ROMs dramatically accelerate multiscale simulations, their runtimes are still quite high (especially in the presence of softening). Additionally, ROMs lack solution transferability in that the expensive data of one instance of the model is not reused (e.g., the full strain-stress history obtained for a microstructure corresponding to a macroscopic IP is not reused in another multiscale simulation). 

Artificial neural networks (ANNs) provide feasible solutions for the simulation data transferability. ANNs are essentially consisted by layers of aggregated artificial neurons that transform numerical signals via nonlinear functions before passing onto the subsequent neurons. ANNs have been extensively studied in solid computational mechanics field to approximate material constitutive laws. For instance, various types of ANNs have been developed to surrogate the material constitutive model for visco-plasticity \cite{furukawa1998implicit}, cyclic plasticity \cite{furukawa2004accurate}, interface mechanism \cite{fernandez2020application}, and anisotropic electrical behaviors \cite{lu2017multiscale}. In recent developments, Mianroodi et al. \cite{mianroodi2021teaching} develop a deep neural network (DNN, which is a type of ANN and uses multiple hidden layers between input and output layers for complex representations) to calculate local stress distributions in non-homogeneous microstructures with elasto-plastic behaviors. Haghighata et al. \cite{haghighat2021physics} incorporate the momentum balance and constitutive relations into a DNN model and demonstrate the improved extrapolation capability for single scale elasto-plastic simulations. Peivaste et al. \cite{peivaste2022machine} develop a convolutional neural network (CNN, which is another type of ANN that slides along input features to output translational equivalent features) to surrogate computationally costly phase-field models to simulate material grain evolutions in microstructures. Although most of ANNs are shown efficient in approximating material constitutive laws and able to provide simulation data transferability, they are generally not ideal for surrogating path-dependent irreversible material behaviors, as they lack the fundamental mechanism to capture the impacts of loading histories that are extremely important in such scenarios. 

Data-driven material models are increasingly built on recurrent neural networks (RNNs) to learn the path-dependent constitutive laws for elasto-plastic deformations. RNN is a class of DL that maps a sequence of inputs to a sequence of outputs where deformation history is passed down the model as memory-like variables in the network. 
For example, Mozaffar et al. \cite{mozaffar2019deep} successfully use an RNN to learn plasticity with distortional hardening on 2D fiber composite microstructures. Wang et al. \cite{wang2018multiscale} develop an RNN to link information from different scales via recursive homogenization to capture the multiscale hydro-mechanical coupling effects of heterogeneous media with various pore sizes. The RNN surrogate developed by Wu et al. \cite{wu2020recurrent} is trained on a database whose sampling sequences are generated via a random walking algorithm to simulate the microstructural effective elasto-plastic hardening behaviors under cyclic and non-proportional loading paths. An on-demand sampling strategy is adopted by Ghavamian et al. \cite{ghavamian2019accelerating} that reduces sampling space by running prior macro models to collect the strain-stress sequences for the subsequent RNN's learning process. This strategy reduces sampling efforts and improves prediction accuracy but reduces the generalization power since the trained model can be only applied to the macro component that is used to collect the training sequences. In a recent work \cite{logarzo2021smart}, Logarzo et al. use an RNN to learn the hardening behavior of a 2D composite microstructure under a wide range of deformation histories that are sampled from the space of principal strains. All aforementioned RNN surrogates are black-box or pure data-driven models whose accuracy relies on large training datasets. Building such datasets is very challenging for 3D microstructural analyses that involve softening. While infusing physical laws into the training process can improve the reliance on data and energy consistency, this direction has not been rigorously explored.

Our contribution of this work is to propose a physics-constrained RNN model to surrogate the micro analyses amid online multiscale simulations. Comparing to the aforementioned ROMs and pure data-driven models, our proposed surrogate is computationally efficient, memory light, physics consistent and transferable. Our surrogate is developed in an offline stage, and it consists of three major components as shown in Figure \ref{fig:framework} (b)-(d):
\vspace{-0.2cm}
\begin{itemize}
  \item \textbf{Random deformation histories.} The essence of data transferability of our model comes from the generality of versatile deformation paths our microstructures are deformed by. To maximize the diversity of our sequential dataset, we utilize design of experiment (DoE) method to generate a set of random 3D deformation histories with six independent strain components over a series of loading steps.
  \item \textbf{Database generation.} Development of the surrogate of 3D large scale microstructural responses faces two major challenges. The first challenge comes from the prohibitively high dimensions of the functional space representing sequential data, and the second challenge is the demanding computational costs pertaining to the softening simulation per deformation path in the sampling space. To address such difficulties, we generate a moderate size of response training database by adopting DoE constraints to remove unnecessary sampling sequences, and we deploy mechanistic ROM to faithfully and efficiently compute the microstructural responses where softening-induced solver divergence is addressed by hybrid time integration scheme. 
  \item \textbf{Surrogate development.} To improve the accuracy of our surrogate whose training lacks abundant sequential data, we develop and incorporate thermodynamics consistent physics constraints within our RNN by modifying loss function and data flow architecture. By integrating our trained surrogate in the Newton Raphson algorithm, our model is able to provide highly accurate iterative surrogate estimations towards convergence at each loading step in online multiscale simulations.  
\end{itemize}

The rest of the paper is organized as follows. 
In Section \ref{sec:multiscale}, we briefly review the homogenization-based concurrent multiscale damage analysis along with the numerical techniques that facilitate the convergence of softening simulations. We also demonstrate two thermodynamics-consistent physics constraints that a generic microstructural response should always satisfy under arbitrary deformations. 
In Section \ref{sec:RNN}, we propose our physics-informed data-driven model for the surrogate of microstructural effective elasto-plastic responses that may involve damage and fracture. 
In Section \ref{sec:experiments}, we illustrate the efficiency and accuracy of our data-driven model by comparing its prediction not only to the microstructural effective responses subject to random deformation paths, but also on a number of multiscale structures subject to complex cyclic loading conditions with hardening and softening material behaviors. 
We conclude our paper with some notes on the contributions and future work in Section \ref{sec:conclusions}.
\section{Homogenization-based multiscale damage analysis} \label{sec:multiscale}

Our multiscale damage analysis is based on the the first-order homogenization model which we first review in Section \ref{subsec:multiscaleModel}. Then, we illustrate the numerical instability issue of strain softening models in simulating damage evolution in Section \ref{subsec:strainSoftening}, and we present a hybrid time integration scheme in Section \ref{subsec:hybridIntegration} to address the instability issue. To define the physics constraints that materials must satisfy during plastic deformations, we perform an energy analysis to derive thermodynamic consistency conditions that a generic microstructure fulfills for an arbitrary iso-thermal elasto-plastic deformation in Section \ref{subsec:thermoDynamics}. We apply these conditions in Section \ref{sec:RNN} to reduce the reliance of our data-driven model to data while increasing its prediction accuracy. 

\subsection{Multiscale modeling}\label{subsec:multiscaleModel}

Our multiscale models in this work are based on the first-order computational homogenization method which assumes scale separation between a macro-component and its micro-features. In solving multiscale systems, the solutions at the macroscale and microscale are coupled via the Hill-Mandel condition \cite{otero2018multiscale} that indicates the density of virtual internal work of a macroscale IP equals the volume average of the virtual work in the associated microstructure subject to any kinematically admissible displacement field:
\begin{equation}
    {\textbf{S}_M}:{{\delta}\textbf{E}_M} = \frac{1}{|\Omega_{0m}|}{ \int_{\Omega_{0m}} {\textbf{S}_m:{\delta}\textbf{E}_m} \,d\Omega }
    \label{eqn:ScaleTrans_energy}
\end{equation}
\noindent where $\textbf{S}_M$, $\delta \textbf{E}_M$, $\textbf{S}_m$ and $\delta \textbf{E}_m$ represent the macroscopic and microscopic stress and virtual strain tensors, respectively. The subscripts $M$ and $m$ indicate the macroscale and microscale, respectively. The $:$ operator represents the double dot product contracting a pair of repeated indices. In addition, $\Omega_{0m}$ and $|\Omega_{0m}|$ indicate the reference microstructural domain and its volume, respectively. Following the virtual energy condition, the macroscopic effective stress and virtual strain can be expressed as the volume average of their micro counterparts as: 
\begin{gather}
    {\textbf{S}_M} = \frac{1}{|\Omega_{0m}|}{ \int_{\Omega_{0m}} {\textbf{S}_m} \,d\Omega }; \hspace{0.8cm} {{\delta}\textbf{E}_M} = \frac{1}{|\Omega_{0m}|}{ \int_{\Omega_{0m}} {{\delta}\textbf{E}_m} \,d\Omega }  
    \label{eqn:ScaleTrans_StressStrain}
\end{gather}

The stress and strain fields at both the macro- and micro- scales need to satisfy equilibrium equations at their length scale. For instance, under the infinitesimal deformation assumption, the macro-solutions at an arbitrary macroscopic IP $\textbf{P}$ can be computed by solving the following boundary value problem (BVP):
\begin{subequations}
\begin{align}
    {\textbf{S}_M}(\textbf{P}) \cdot \nabla_0 + \textbf{b}_M &= \textbf{0} &\forall\textbf{P} \in \Omega_{0M} \\
    \textbf{u}_M(\textbf{P}) &= \bar{\textbf{u}}_M &\forall\textbf{P} \in \Gamma^{D}_{0M} \\
    {\textbf{S}_M}(\textbf{P}) \cdot \textbf{n}_M &= \bar{\textbf{t}}_M &\forall\textbf{P} \in \Gamma^{N}_{0M}    
\end{align}
\end{subequations}
\noindent where $\textbf{u}_M$ is the unknown macroscopic displacement in $\Omega_{0M}$ and $\bar{\textbf{u}}_M$ is the prescribed displacement on the Dirichlet boundary $\Gamma^{D}_{0M}$ over the undeformed macroscopic domain $\Omega_{0M}$ with an outward unit vector $\textbf{n}_M$. Also, $\nabla_0$ indicates the gradient operator with respect to the original configuration. $\textbf{b}_M$ and $\bar{\textbf{t}}_M$ represent the body force and prescribed surface traction on the Neumann boundary $\Gamma^{N}_{0M}$, respectively. 

In a similar manner, the strong form of the microscale equilibrium equations can be written as a BVP for the microstructure or representative volume element (RVE) composed of micro IPs \textbf{p} as:
\begin{subequations}
\begin{align}
    {\textbf{S}_m}(\textbf{p}) \cdot \nabla_0 &= \textbf{0} &\forall\textbf{p} \in \Omega_{0m} \\
    {\textbf{S}_m}(\textbf{p}) \cdot \textbf{n}_m &= \bar{\textbf{t}}_m &\forall\textbf{p} \in \Gamma_{0m}
\end{align}
\end{subequations}
\noindent where $\bar{\textbf{t}}_m$ indicates the surface traction per unit area over the reference microstructural boundary $\Gamma_{0m}$ with an outward unit normal vector $\textbf{n}_m$. 

\subsection{Strain softening}\label{subsec:strainSoftening}

In this work, we adopt isotropic continuum damage model to simulate the strain softening in ductile metals whose load-carrying capacity drops due to the degradation of yield stress and stiffness. To simulate the onset of softening, we choose ductile damage initiation criteria which assumes the effective strain $\bar{E}_{d}^{pl}$ at damage initiation as a function of stress and strain states. We presume $\bar{E}_{d}^{pl}$ is a constant and that damage begins when the equivalent plastic strain is equal or greater than the damage initiation criteria, i.e., $\bar{E}^{pl} \geqslant \bar{E}_{d}^{pl}$. Under progressing damage, we formulate the softening response of the ductile metal with an elasto-plastic behavior as:
\begin{gather}
    \textbf{S} = (1-D)\textbf{S}^0;  \hspace{0.8cm}
    \textbf{S}^0 = \mathbb{C}^{el}:\textbf{E}^{el} = \mathbb{C}^{el}:(\textbf{E}-\textbf{E}^{pl})
\label{eqn:damage_stress}
\end{gather}
\noindent where $\textbf{S}$ and $\textbf{S}^0$ are the damaged stress and the reference stress that undergoes the same deformation path but in the absence of damage, respectively. $\mathbb{C}^{el}$ represents the fourth-order elasticity tensor. $\textbf{E}$, $\textbf{E}^{el}$ and $\textbf{E}^{pl}$ are the total strain, elastic strain and plastic strain, respectively. $D$ represents the damage parameter that monotonically increases within $[0.0, 1.0]$. We note that in our context of isotropic continuum damage, $D$ is a scalar and it becomes a tensor in anisotropic damage models. 

A major challenge of using the isotropic continuum damage model in Equation \ref{eqn:damage_stress} is the softening-induced non-positive stiffness matrix that results in solution convergence and negative wave speeds \cite{bazant2010can}. Specifically, the ill-posed problem causes equilibrium equations to lose objectivity with respect to mesh sizes by exhibiting spurious mesh sensitivity. We therefore adopt two different damage models to mitigate the mesh dependency at macroscale and microscale, respectively.

For macroscopic softening, we define the evolution of damage parameter $D_M$ as a function of $\bar{E}^{pl}$, $\bar{E}_{d}^{pl}$, and a user-defined non-negative damage evolution rate parameter $\alpha$ \cite{liu2018microstructural} as: 
\begin{gather}
    D_M = 
    \begin{cases} 
    0;  \hspace{0.8cm} &\bar{E}^{pl} \leq \bar{E}_{d}^{pl} \\
    1-{\frac{\bar{E}_{d}^{pl}}{\bar{E}^{pl}}} exp(-\alpha(\bar{E}^{pl}-\bar{E}_{d}^{pl}));  \hspace{0.8cm} &\bar{E}^{pl} > \bar{E}_{d}^{pl}
    \end{cases}
\label{eqn:macro_damageVar}
\end{gather}
\noindent where no damage occurs when the plastic strain is smaller than or equal ti the predefined plastic damage strain criterion. When $\bar{E}^{pl}$ grows to a value that is larger than $\bar{E}_{d}^{pl}$, material damage initiates, and the value of $D_M$ monotonically increases from $0.0$ to $1.0$ in the irreversible damage process. We constrain the progression of $D_M$ by an integral-type non-local damage model to mitigate the spurious mesh dependency as:
\begin{gather}
    \hat{D}_M(\textbf{P}, \textbf{P}') = \int_{B} {\omega(\|\textbf{P}-\textbf{P}'\|)}D_M(\textbf{P}') \,d\textbf{P}'
\label{eqn:nonlocal_damageFunc}
\end{gather}
\noindent where $\hat{D}_M(\textbf{P}, \textbf{P}')$ is the non-local damage parameter at a macroscopic point $\textbf{P}$ surrounded by close-by points $\textbf{P}'$ in a compact neighborhood $B$. $D_M(\textbf{P}')$ represents the local damage parameter at $\textbf{P}'$, and $\omega$ indicates the non-local weighting function depending on the distance $\|\textbf{P}-\textbf{P}'\|$ between the studied point and its supporting points. In this work, We define $\omega$ by a polynomial bell-shape function as:
\begin{gather}
    {\omega(\|\textbf{P}-\textbf{P}'\|)} = \frac{\omega_\infty (\|\textbf{P}-\textbf{P}'\|)} {\int_{B} {\omega_\infty (\|\textbf{P}-\textbf{P}'\|)} \,d\textbf{P}'}; \hspace{0.8cm}
    {\omega_\infty (\|\textbf{P}-\textbf{P}'\|)} =  \left\langle 1 - \frac{4(\|\textbf{P}-\textbf{P}'\|)^2}{l_d^2} \right\rangle^2
\label{eqn:nonlocal_length}
\end{gather}
\noindent where $\langle \dots \rangle$ is the Macauley bracket defined as $\langle x \rangle = max(0,x)$. $l_d$ denotes the strain localization bandwidth whose value represents the non-local interacting radius, and the support domain $B$ is a sphere with a radius of $l_d/2$ in 3D models. 

To address the lack of objectivity to mesh choices in microstructural damage simulations, re-definition of a microscopic strain localization bandwidth would counteract the physical meaning of its macroscopic counterpart $l_d$. Instead, we convert the microstructural softening constitutive equation from stress-strain relation to the stress-displacement relation to drive the micro-damage evolution after initiation as:  
\begin{gather}
    G_f = \int_{\bar{E}_0^{pl}}^{\bar{E}_f^{pl}} l_e S_y \,d{\bar{E}^{pl}} = \int_{0}^{\bar{u}_f^{pl}} S_y \,d{\bar{u}^{pl}}
\end{gather}
\noindent where $l_e$ indicates the element characteristic length in an arbitrary RVE, and $G_f$ represents the dissipated energy after damage initiation that opens a unit area of crack. The equivalent plastic displacement $\bar{u}^{pl}$ is the fracture work conjugate to the yield stress $S_y$ in the fracture evolution from the damage initiation (with the effective plastic strain $\bar{E}_0^{pl}$ and zero plastic displacement $\bar{u}^{pl}$) to the final failure (with the effective fracture strain $\bar{E}_f^{pl}$ and the fracture displacement $\bar{u}_{f}^{pl}$). We can then define the damage evolution rule based on the amount of released energy in an exponential form of the plastic displacement \cite{ABAQUS2009} as in Equation \ref{eqn:micro_damageVar}. We note that $D_m$ approaches $1.0$ asymptotically with infinitely large $\bar{u}^{pl}$. In practice, we set $D_m$ as $1.0$ when the dissipated energy exceeds $0.99G_f$.
\begin{gather}
    D_m = 1 - exp(-\frac{1}{G_f} \int_{0}^{\bar{u}_f^{pl}} S_y \,d{\bar{u}^{pl}})
\label{eqn:micro_damageVar}
\end{gather}

\subsection{Hybrid temporal integration} \label{subsec:hybridIntegration}  

The non-positive definiteness of the stiffness matrix is the primarily reason for the slow convergence of classic implicit time integration schemes that are used in continuum damage simulations. For illustration, consider the constitutive equation of an isotropic damage model integrated by an implicit backward-Euler integration scheme. Its algorithmic tangent operator at an arbitrary macroscopic IP can be written as:
\begin{gather}
    \mathbb{C}_{n+1}^{alg} = \frac{\partial \textbf{S}_{n+1}}{\partial \textbf{E}_{n+1}} = (1-D_{n+1})\mathbb{C}^{el} - \frac{S_{n+1} - H_{n}\bar{E}_{n+1}^{pl}}{(\bar{E}_{n+1}^{pl})^3} \textbf{S}_{n+1}^{0} \otimes \textbf{S}_{n+1}^{0}
\end{gather}
\noindent where $\mathbb{C}_{n+1}^{alg}$, $\bar{E}_{n+1}^{pl}$, $S_{n+1}$, $\textbf{S}_{n+1}^{0}$ and $H_{n}$ represent the fourth-order algorithmic tangent operator, equivalent plastic strain, equivalent stress, referenced stress tensor, and softening modulus, respectively. The subscripts denote time steps and the symbol $\otimes$ represents the cross product between tensors. Softening causes negative values for $H_{n}$ which can result in the loss of positive definiteness of $\mathbb{C}_{n+1}^{alg}$. A non-positive $\mathbb{C}_{n+1}^{alg}$ leads to ill-conditioned elemental stiffness matrix with near-zero or negative eigenvalues, and further deteriorates the global stiffness matrix in the element assembly process. Such ill-posed matrices dramatically reduce the efficiency of iterative solvers (e.g., Newton-Raphson methods) and often causes job abortion before final convergence.

To fundamentally resolve the convergence issue, we adopt a hybrid time integration scheme \cite{oliver2008implicit,deng2022concurrent} to integrate the governing equations of elasto-plastic and softening equations explicitly-implicitly. 
The basic idea of the hybrid integration is to maintain the positive-definiteness of the system's algebraic tangent operator by separately integrating constitutive equations in two consecutive steps via explicit and implicit schemes. In the first step, we explicitly extrapolate internal material state variables at the time step $n+1$ from the previous step $n$ to compute an explicit stress state $\tilde{\textbf{S}}_{n+1}$ that balances the equilibrium equation between internal and external forces. In the second step, we compute an implicit stress state $\textbf{S}_{n+1}$ based on the current strain state $\textbf{E}_{n+1}$ by the classic backward Euler method. We then use the computed implicit stress $\textbf{S}_{n+1}$ to update the trial stress (in the classic elastic predictor and plastic corrector scheme) and yield functions in the next time step $n+2$.

For the elasto-plastic model, we choose the material state variable as the incremental plastic strain tensor $\triangle\tilde{\textbf{E}}_{n+1}^{pl}$ such that the explicitly extrapolated stress $\tilde{\textbf{S}}_{n+1}$ can be computed as:
\begin{gather}
    \tilde{\textbf{S}}_{n+1}(\triangle\tilde{\textbf{E}}_{n+1}^{pl}) = \tilde{\textbf{S}}_{n+1}^{trial} - \mathbb{C}^{el}:\triangle\tilde{\textbf{E}}_{n+1}^{pl} = \mathbb{C}^{el}:\textbf{E}_{n+1} - \mathbb{C}^{el}:\textbf{E}_{n}^{pl} - \mathbb{C}^{el}:\triangle\tilde{\textbf{E}}_{n+1}^{pl} \\
    \triangle\tilde{\textbf{E}}_{n+1}^{pl} = \frac{\triangle t_{n+1}}{\triangle t_n} \triangle \textbf{E}_n^{pl} \nonumber
\end{gather}
\noindent where $\textbf{E}_n^{pl}$ represents the implicit incremental plastic strain tensor from the previous time step $n$, $\triangle t_n$ and $\triangle t_{n+1}$ indicate the lengths of time steps at two consecutive steps. The algorithmic tangent operator (under loading) is therefore computed as in the following equation, whereas it is equal to the elastic modulus in the unloading scenarios:
\begin{gather}
    \tilde{\mathbb{C}}_{n+1}^{alg} = \frac{\partial{\tilde{\textbf{S}}_{n+1}(\triangle\tilde{\textbf{E}}_{n+1}^{pl})}}{\partial{\textbf{E}_{n+1}}} = \frac{\partial(\mathbb{C}^{el}:\textbf{E}_{n+1} - \mathbb{C}^{el}:\textbf{E}_{n}^{pl} - \mathbb{C}^{el}:\triangle\tilde{\textbf{E}}_{n+1}^{pl})}{\partial{\textbf{E}_{n+1}}} = \mathbb{C}^{el}
\label{eqn:hybridInteg_C1}
\end{gather}

In a similar manner, for isotropic continuum damage models, we choose the explicitly interpolated material state variable in the hybrid integration as the incremental plastic multiplier $\triangle \tilde{\lambda}_{n+1}$, i.e., $\triangle \tilde{\lambda}_{n+1} = (\triangle t_{n+1} / \triangle t_n) \triangle \lambda_n$. We can then write its explicit damaged stress and algorithmic tangent operator under loading as in the following equation, while $\tilde{\mathbb{C}}_{n+1}^{alg}=\mathbb{C}^{el}$ holds for unloading scenarios:
\begin{gather}
    \tilde{\textbf{S}}_{n+1} = (1-\tilde{D}_{n+1}) \textbf{S}_{n+1}^0 = (1-\tilde{D}_{n+1}) \mathbb{C}^{el}:\textbf{E}_{n+1}; \hspace{0.8cm}  \tilde{D}_{n+1} = \tilde{D}_{n+1} (D_n, \triangle \tilde{\lambda}_{n+1}) \\
    \tilde{\mathbb{C}}_{n+1}^{alg} = \frac{\partial{\tilde{\textbf{S}}_{n+1}}}{\partial{\textbf{E}_{n+1}}} = (1-\tilde{D}_{n+1}) \mathbb{C}^{el} 
\label{eqn:hybridInteg_C2}
\end{gather}
\noindent where $\textbf{S}_{n+1}^0$ is the effective stress tensor, and $\tilde{D}_{n+1}$ represents the explicit state of the damage variable which is a function of its previous implicit state $D_n$ and the current explicit incremental plastic multiplier $\triangle \tilde{\lambda}_{n+1}$. We note that in the hybrid integration scheme, the loading tangent operators of the elasto-plastic model in Equation \ref{eqn:hybridInteg_C1} and the damage model in Equation \ref{eqn:hybridInteg_C2} are trivially equal to the elastic modulus $\mathbb{C}^{el}$ and $(1-\tilde{D}_{n+1}) \mathbb{C}^{el}$. Hence, the hybrid integration scheme is not only advantageous in preserving the positive-definiteness of the governing equations, but also letting the global stiffness matrix be assembled only once before online simulations. The global stiffness matrix remains constant for the elasto-plastic regime and only needs partial update on matrix entries associated to the softening IPs by Equation \ref{eqn:hybridInteg_C2}. As softening is often highly localized in small regions, the global stiffness can be incrementally updated during the entire elasto-plastic-hardening-softening process \cite{deng2022concurrent}, saving significant memory footprints with robust convergence performance.

\subsection{Energy analysis}\label{subsec:thermoDynamics}

Assuming a microscopic IP in an RVE is subject to an iso-thermal elasto-plastic deformation, we can compute its total work rate per unit volume $\Dot{W}$ via thermodynamics principles \cite{silhavy2013mechanics} as:
\begin{gather}
    \Dot{W} = \Dot{\psi} + \Phi
\end{gather}
\noindent where $\Dot{\psi}$ represents the rate of Helmholtz free energy and $\Phi$ accounts for the rate of dissipated energy including the dissipation from plasticity, damage, damping, etc. For general elasto-plastic material behaviors, we can decompose the rate of work into elastic and plastic parts:
\begin{gather}
    \Dot{W} = \Dot{W}^{el} + \Dot{W}^{pl}
\end{gather}
\noindent where the elastic work rate $\Dot{W}^{el}$ at an arbitrary microscopic IP is equal to the rate of recoverable elastic free energy or strain energy $\Dot{\psi}^{el}$, while the plastic work rate $\Dot{W}^{pl}$ is equal to the sum of the conditionally recoverable plastic free energy $\Dot{\psi}^{pl}$ and the irrecoverable dissipation rate \cite{yang2018energy}. That is:
\begin{gather}
    \Dot{W}^{el} = \Dot{\psi}^{el} = \textbf{S}_m:\Dot{\textbf{E}}^{el}_m; \hspace{0.8cm} 
    \Dot{W}^{pl} = \Dot{\psi}^{pl} + \Phi = \textbf{S}_m:\Dot{\textbf{E}}^{pl}_m
\end{gather}

We write the total rate of work per unit volume $\Dot{W}$ of the RVE as the multiplication of the micro stress $\textbf{S}_m$ and the rate of microscopic total strain $\Dot{\textbf{E}}_m$:
\begin{gather}
    \Dot{W} = \textbf{S}_m:\Dot{\textbf{E}}^{el}_m + \textbf{S}_m:\Dot{\textbf{E}}^{pl}_m = \textbf{S}_m:\Dot{\textbf{E}}_m
\label{eqn:workRate}
\end{gather}
\noindent where we use the additive decomposition rule for the strain. We compute the total work by integrating the work rate over the time interval and spatial domain $\Omega_{0m}$. Additionally, We compute the total work by invoking the Hill-Mandel energy condition from Equation \ref{eqn:ScaleTrans_energy} and assuming the rates of strains to be within the hyperspace of the virtual strains as:
\begin{equation}
\begin{split}
    W &= \int_{t} \int_{\Omega} \Dot{W} \,d\Omega_{0m} \,dt = \int_{t} \int_{\Omega} \textbf{S}_m:\Dot{\textbf{E}}_m \,d\Omega_{0m} \,dt \\
    &= |\Omega_{0m}| \int_{t} \textbf{S}_M:\Dot{\textbf{E}}_M \,dt = |\Omega_{0m}| \int_{t} \textbf{S}_M \,d\textbf{E}_M
\label{eqn:totalEnergy}
\end{split}
\end{equation}

We write the total work $W$ as a summation of total strain energy $W^{el}$ and total plastic work $W^{pl}$ as:
\begin{gather}
    W = W^{el} + W^{pl}
\end{gather}

Assuming linear elasticity, we can show the total strain energy of the RVE as:
\begin{equation}
\begin{split}
    W^{el} &= \int_{t} \int_{\Omega} \textbf{S}_m:\Dot{\textbf{E}}^{el}_m \,d\Omega_{0m} \,dt = \int_{\Omega} \int_{\textbf{E}^{el}_m} \textbf{S}_m \,d\textbf{E}^{el}_m \,d\Omega_{0m} \\
    &= \frac{1}{2} \int_{\Omega} \textbf{E}^{el}_m:\mathbb{C}^{el}_m:\textbf{E}^{el}_m \,d\Omega_{0m} \geq 0
\label{eqn:energy_elastic}
\end{split}
\end{equation}
\noindent where $\mathbb{C}^{el}_m$ represents the elastic modulus at a micro-point. It is equal to $(1-D_m)\mathbb{C}^{el}$ if damage occurs (with a micro damage parameter $D_m$) and $\mathbb{C}^{el}$ if there is no damage. Since $0 \leq D_m \leq 1$, it is straight forward to show $W^{el} \geq 0$. 

Similarly, we can compute $W^{pl}$ by spatiotemporally integrating $\Dot{W}^{pl}$, and it is equal to the sum of total dissipated energy and total plastic free energy as:
\begin{gather}
    W^{pl} = W^{di} + W^{pf} 
\end{gather}
\noindent where $W^{di}$ can be expressed as the spatiotemporal integration of the non-negative dissipation rate as:
\begin{gather}
    W^{di} = \int_{t} \int_{\Omega} \Phi \,d\Omega_{0m} \,dt \geq 0
\label{eqn:energy_dissipation}
\end{gather}
\noindent where the non-negativity is due to the fact that $\Phi \geq 0$. In addition, we note that the total plastic free energy equals the integrated rate of plastic free energy:
\begin{gather}
    W^{pf} = \int_{t} \int_{\Omega} \Dot{\psi}^{pl} \,d\Omega_{0m} \,dt = \int_{\Omega} \psi^{pl} \,d\Omega_{0m} 
\end{gather}
\noindent where $\psi^{pl}$ stands for the density of the plastic free energy in the RVE and it can be decomposed into isotropic and anisotropic parts \cite{yang2018energy} as:
\begin{gather}
    \psi^{pl} = \psi^{pl}_{iso} + \psi^{pl}_{ani}; \hspace{0.8cm} \psi^{pl}_{ani} = \psi^{pl}_{kin} - \psi^{pl}_{dis}
\end{gather}
\noindent where $\psi^{pl}_{iso}$, $\psi^{pl}_{ani}$, $\psi^{pl}_{kin}$, and $\psi^{pl}_{dis}$ represent the constituents of plastic free energy density from isotropic, anisotropic, kinematic, and distortional deformations, respectively ($\psi^{pl}_{dis}$ is related to the distortional strain hardening with directional distortion of the yield surface but exploring this relation is not in the scope of this work). We can calculate $\psi^{pl}_{iso}$ and $\psi^{pl}_{kin}$ via \cite{feigenbaum2007directional}:
\begin{gather}
    \psi^{pl}_{iso} = \frac{c_1}{2\rho} \bar{k}^2; \hspace{0.8cm} \psi^{pl}_{kin} =  \frac{c_2}{2\rho} \bar{\alpha}_{ij} \bar{\alpha}_{ij}
\end{gather}
\noindent where $\bar{k}$ and $\bar{\alpha}_{ij}$ are, respectively, the thermodynamic conjugates to the size of the yield surface and the deviatoric back stress tensor that represents the center of the yield surface, and $\rho$ is the material density. $c_1$ and $c_2$ are two non-negative material constants depending on the type of material models. We can therefore express the total plastic free energy as:
\begin{gather}
    W^{pf} = \int_{\Omega} (\frac{c_1}{2\rho} \bar{k}^2 + \frac{c_2}{2\rho} \bar{\alpha}_{ij} \bar{\alpha}_{ij}) \,d\Omega_{0m} \geq 0
\label{eqn:energy_plasticFree}
\end{gather}

By plugging Equations \ref{eqn:energy_elastic}, \ref{eqn:energy_dissipation}, and \ref{eqn:energy_plasticFree} into Equation \ref{eqn:totalEnergy}, we show that for an arbitrary macroscale IP associated with an RVE that is subject to general plastic hardening and softening deformations, the effective macroscale stress and strain fields satisfy the following constraint:
\begin{gather}
    \int_{t} \textbf{S}_M \,d\textbf{E}_M \geq 0
\label{eqn:energy_constraint}
\end{gather}

We incorporate this constraint together with the non-decreasing damage variable described in Section \ref{subsec:strainSoftening} as the two physics constraints into the data-driven model in Section \ref{sec:RNN}.

\section{Physics-informed data-driven surrogate} \label{sec:RNN}

We propose a computational framework for building physics-constrained data-driven material models that surrogate the fine-scale homogenization procedures in multiscale damage simulations. 
In Section \ref{subsec:subSec_DBGeneration} we elaborate on the data generation process which builds a set of independent and systematically sampled microstructural deformation-response sequences. This dataset is then used in Section \ref{subsec:RNN_architecture} to train an RNN that serves as the data-driven material model at the microscale. To improve this model's accuracy on unseen deformation paths, we incorporate two physics constraints in Section \ref{subsec:RNN_constraints}. In Section \ref{subsec:RNN_multiscale}, we show the integration procedure of our surrogate in multiscale solvers.     

\subsection{Database generation}\label{subsec:subSec_DBGeneration}
The first step of developing a data-driven surrogate is to generate a database that samples the underlying functional space via sufficient sequential data. Comparing to non-sequential variables, the dimension of the sampling space of temporal (sequential) variables is much larger, as it requires a sequence of time-related data rather than a single data point. To better exploit the sampling space of temporally varying deformation paths, we use design of experiment (DoE) to systematically create a set of random strain paths. 

In the DoE, we assume every strain path starts from relaxing state with zero initial strain without residual stress, and it evolves to final state by a number of loading steps $n_{load}$. To reduce sampling efforts, we assume the strain values at any time step should be no larger than a user-defined threshold $\zeta_1$. In addition, we assume the bulk modulus of our material is fairly large such that the deformation-induced material volume change is within a user-defined limit $\zeta_2$. Accordingly, we can express the two sampling constraints as:
\begin{equation}
    |E^i_n| \leq \zeta_1; \hspace{0.8cm} |E^{vol}_n| \leq \zeta_2
    \label{eqn:DoEConstraints}
\end{equation}
\noindent where $E^i_n$ represents the $i^{th}$ component of the strain vector at the time step $n \in \{1,2,...,n_{load}\}$ where $i \in \{1,2,3,4,5,6\}$ indicates the six components of 3D strains in which $i=\{1,2,3\}$ indicates normal strains and $i=\{4,5,6\}$ represents shear strain components. We note that $E^{vol}_n =  E^{1}_n+E^{2}_n+E^{3}_n$ is the volumetric strain standing for the material volume change after deformation. 

To generate random loading sequences, we assume all deformation paths to have the same number of loading steps $n_{load}$. More precisely, we select $n_{c}$ evenly spaced control points along loading steps, and assign them with strain values drawn from random processes, e.g., Sobol sequence or Latin hypercube sampling. In order to enforce the values of all six strain components at the control points to satisfy the DoE constraints in Equation \ref{eqn:DoEConstraints}, we begin with generating random strain sequences with component values satisfying $|E^i_{c}| \leq \zeta_1$ with $i \in \{1,2,3,4,5,6\}$. We note that we generate all the random strain values jointly by the random process, which generally provides us an optimal sampling solution that maximizes the space filling property of the deformation space with no overlapping, minimum correlation, and even distributed samples. In order to satisfy the constraint of the volume change, we generate the random values for the first two normal strain components and the volumetric strain which satisfy the constraints $|E^{\{1,2\}}_{c}| \leq \zeta_1$ and $|E^{vol}_{c}| \leq \zeta_2$. In this way, we can determine the third normal strain component at the $n_{c}$ control points as:
\begin{equation}
    E^3_{c} = E^{vol}_{c}-E^1_{c}-E^2_{c}
\end{equation}

With random strain values generated at $n_{c}$ control points, we use Gaussian Process (GP) method to interpolate the strain values at all $n_{load}$ loading steps. In specific, we compute a random sequence for each of the six strain components by using a one-dimensional GP that is defined by the mean $m(n)$ and the covariance function $c(n,n')$ as:
\begin{equation}
    E^i_n \sim GP(m(n), c(n,n'))
    \label{eqn:GP}
\end{equation}
\noindent where the parameters $n$ and $n'$ represent two different loading steps. We adopt a simple GP with a zero mean and a covariance function with the Gaussian kernel $r$ defined as:
\begin{equation}
    c(n,n') = \sigma^2 r(n,n'); \hspace{0.8cm} r(n,n') = exp(-w (n-n')^2)
\end{equation}
\noindent where the covariance function $c$ depends on the prior variance $\alpha^2$, the roughness parameter $w$, and the distance between two loading steps $n$ and $n'$. Since the GP-interpolated strain values at $n_{load}$ loading steps are continuous and smooth, we can use them to approximate the deformation histories in real mechanistic simulations.    

\begin{figure}
    \centering
    \begin{subfigure}[b]{0.35\textwidth}
        \centering
        \includegraphics[width=\textwidth]{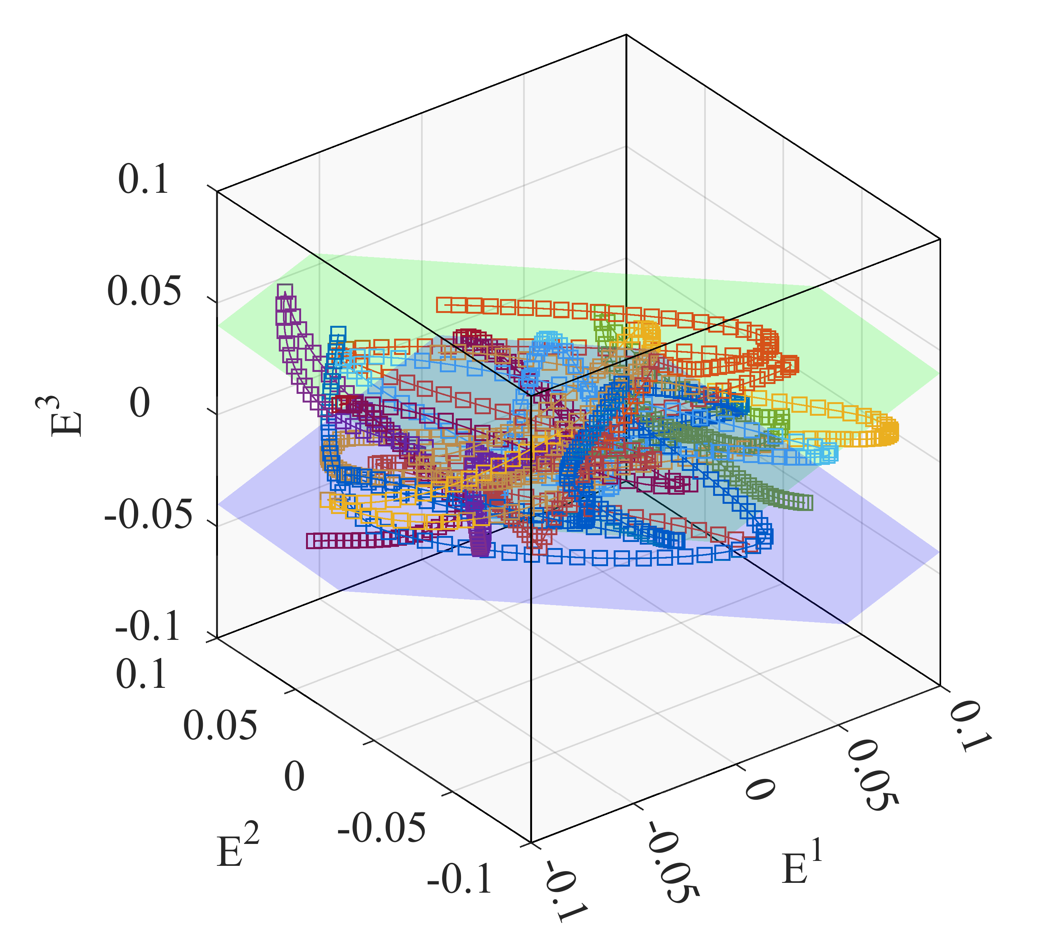}
        \caption{Random normal strains}
    \end{subfigure}
    \begin{subfigure}[b]{0.35\textwidth}
        \centering
        \includegraphics[width=\textwidth]{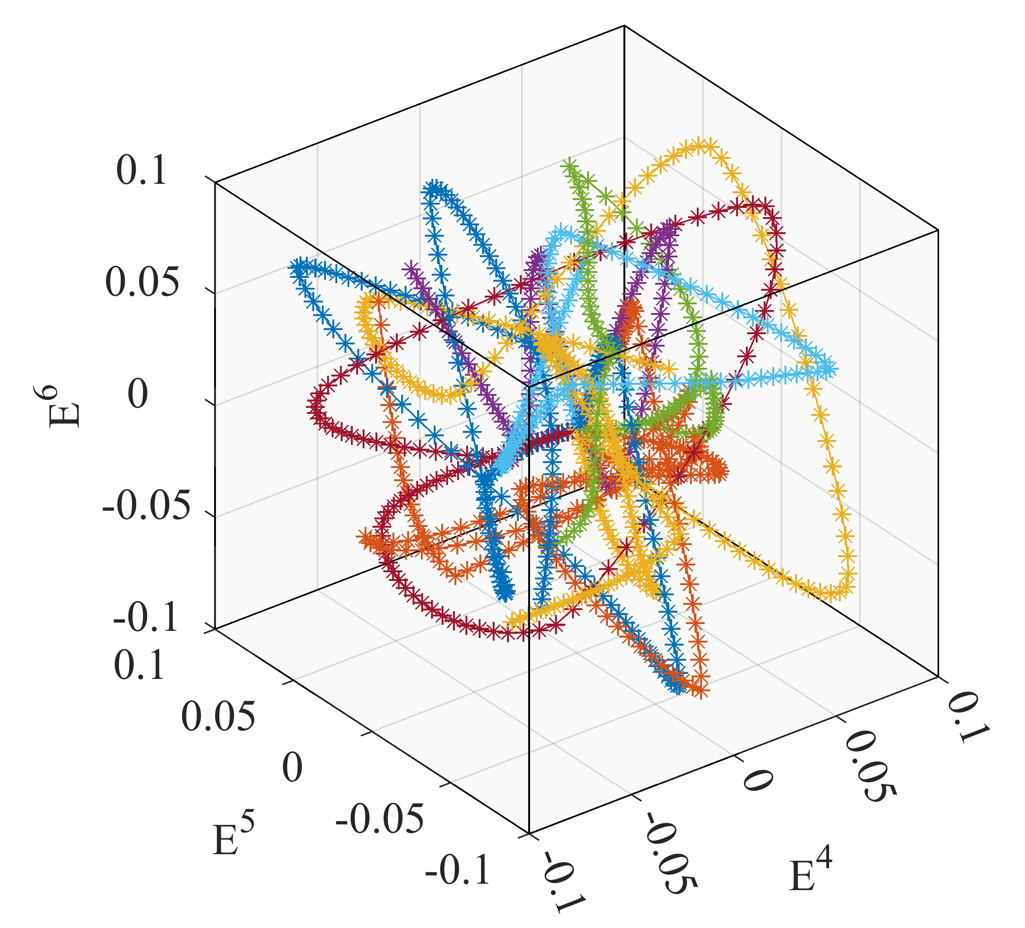}
        \caption{Random shear strains}
    \end{subfigure} 
    \begin{subfigure}[b]{0.35\textwidth}
        \centering
        \includegraphics[width=\textwidth]{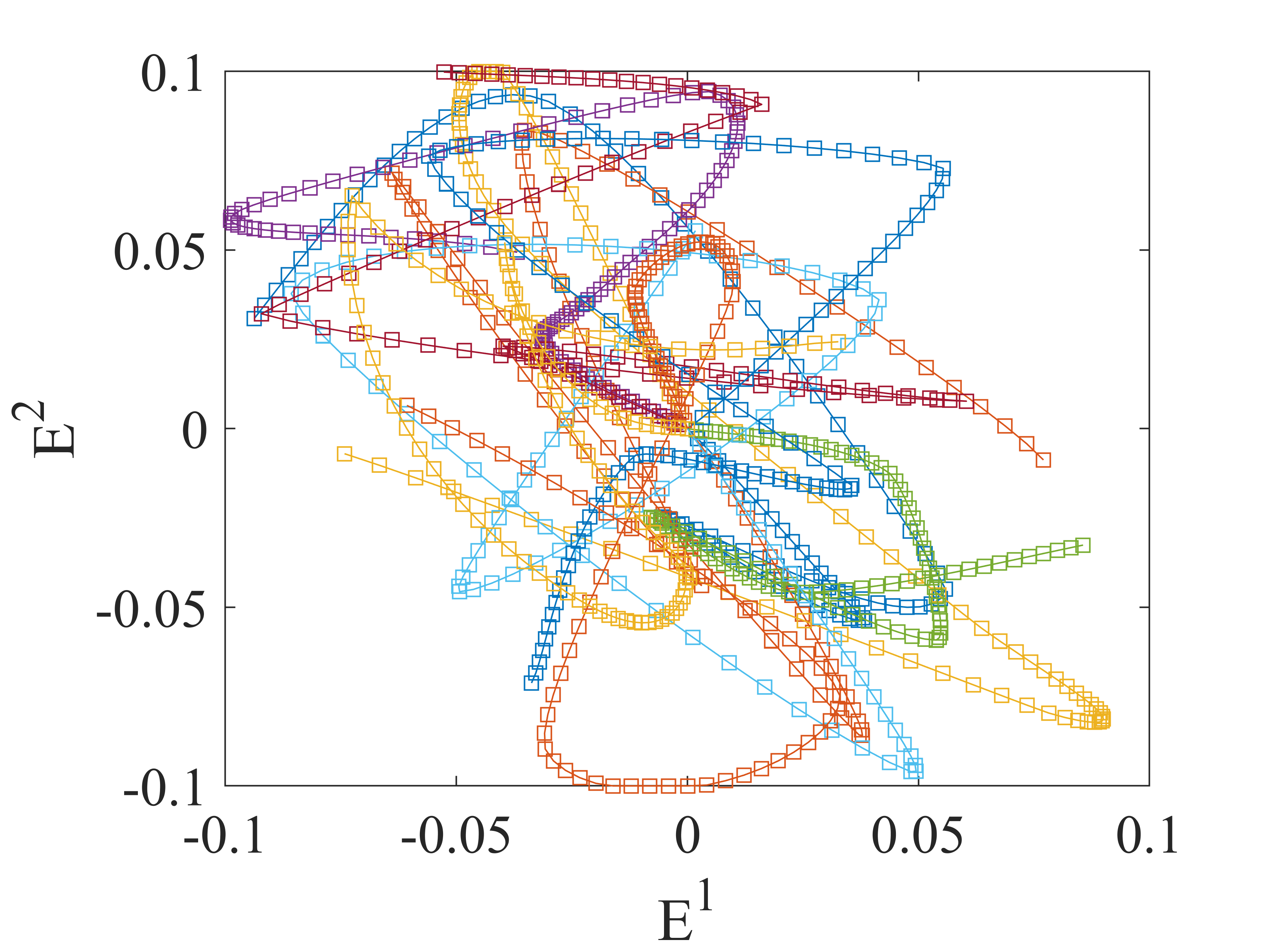}
        \caption{Normal strains $E^1$ and $E^2$}
    \end{subfigure} 
    \begin{subfigure}[b]{0.35\textwidth}
        \centering
        \includegraphics[width=\textwidth]{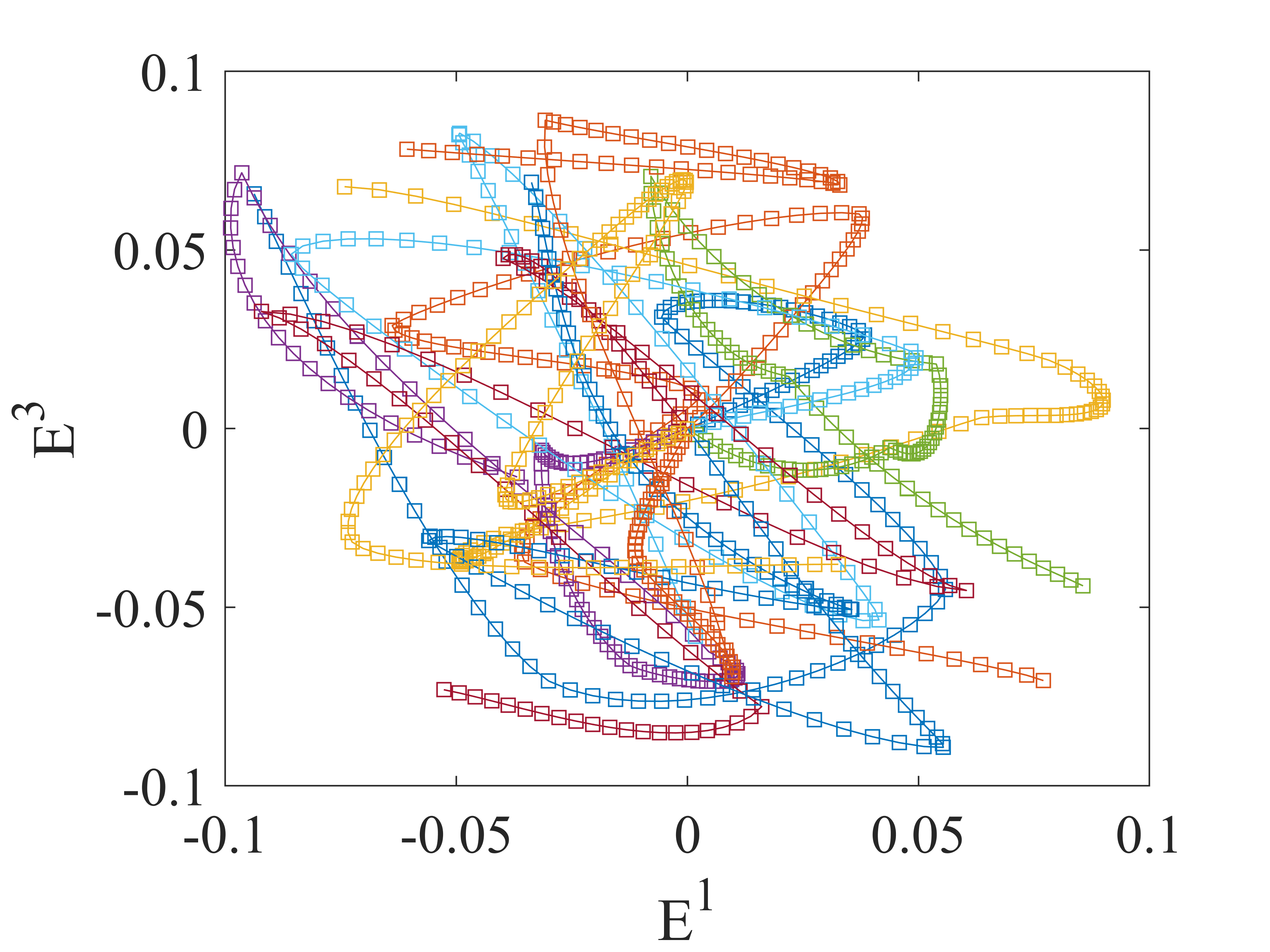}
        \caption{Normal strains $E^1$ and $E^3$}
    \end{subfigure}
    \begin{subfigure}[b]{0.35\textwidth}
        \centering
        \includegraphics[width=\textwidth]{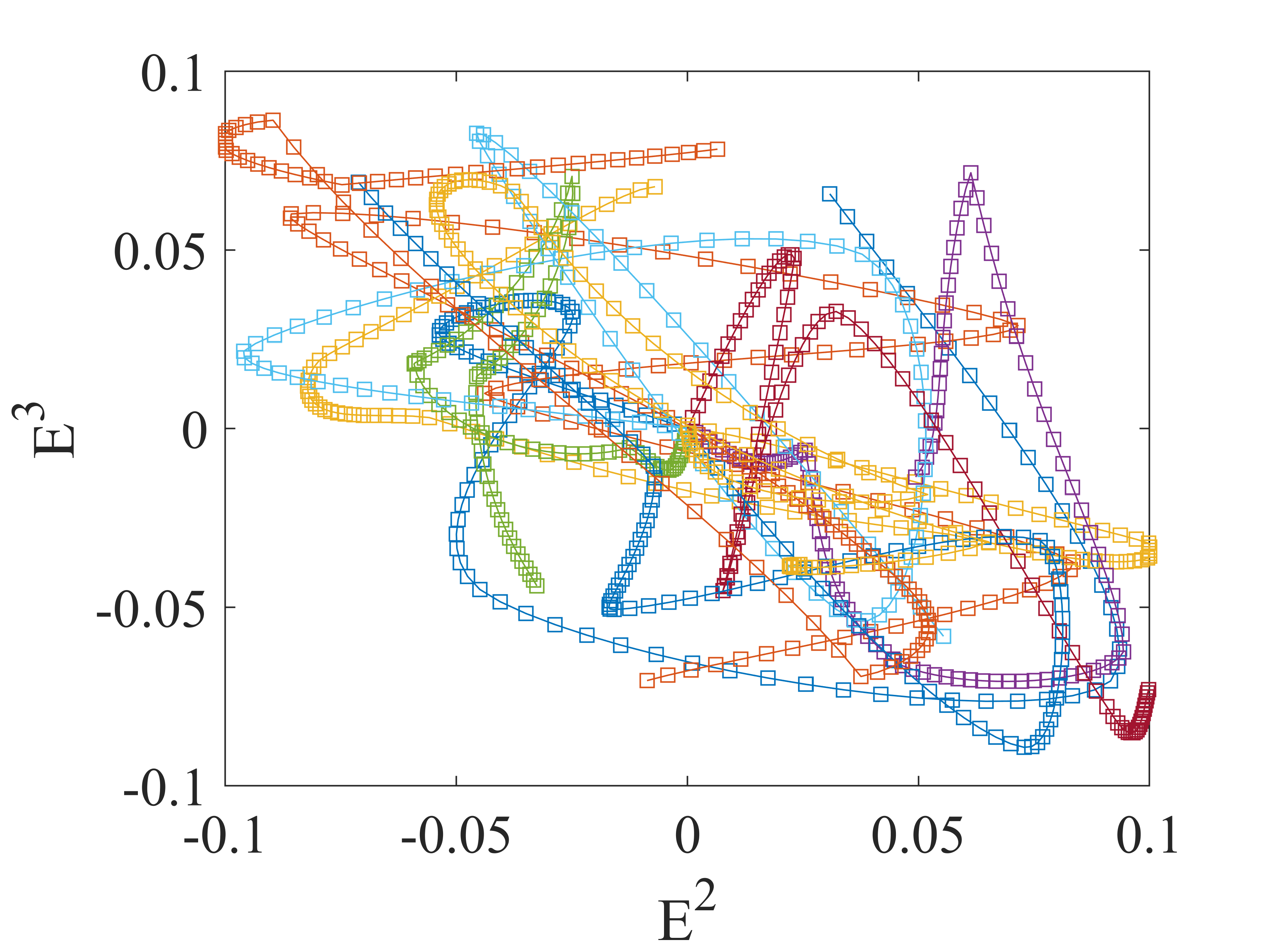}
        \caption{Normal strains $E^2$ and $E^3$}
    \end{subfigure}
    \begin{subfigure}[b]{0.35\textwidth}
        \centering
        \includegraphics[width=\textwidth]{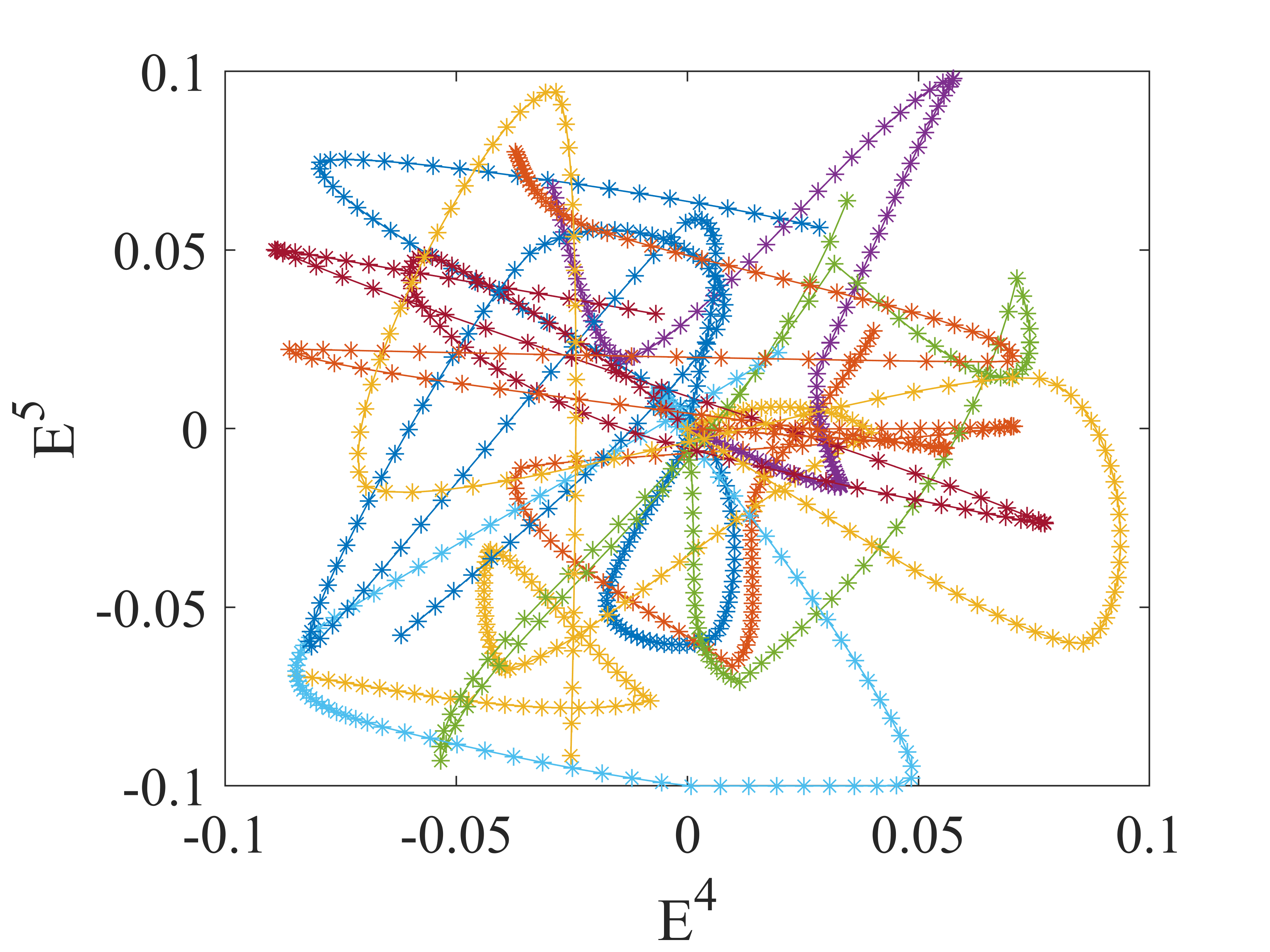}
        \caption{Shear strains $E^4$ and $E^5$}
    \end{subfigure}
    \begin{subfigure}[b]{0.35\textwidth}
        \centering
        \includegraphics[width=\textwidth]{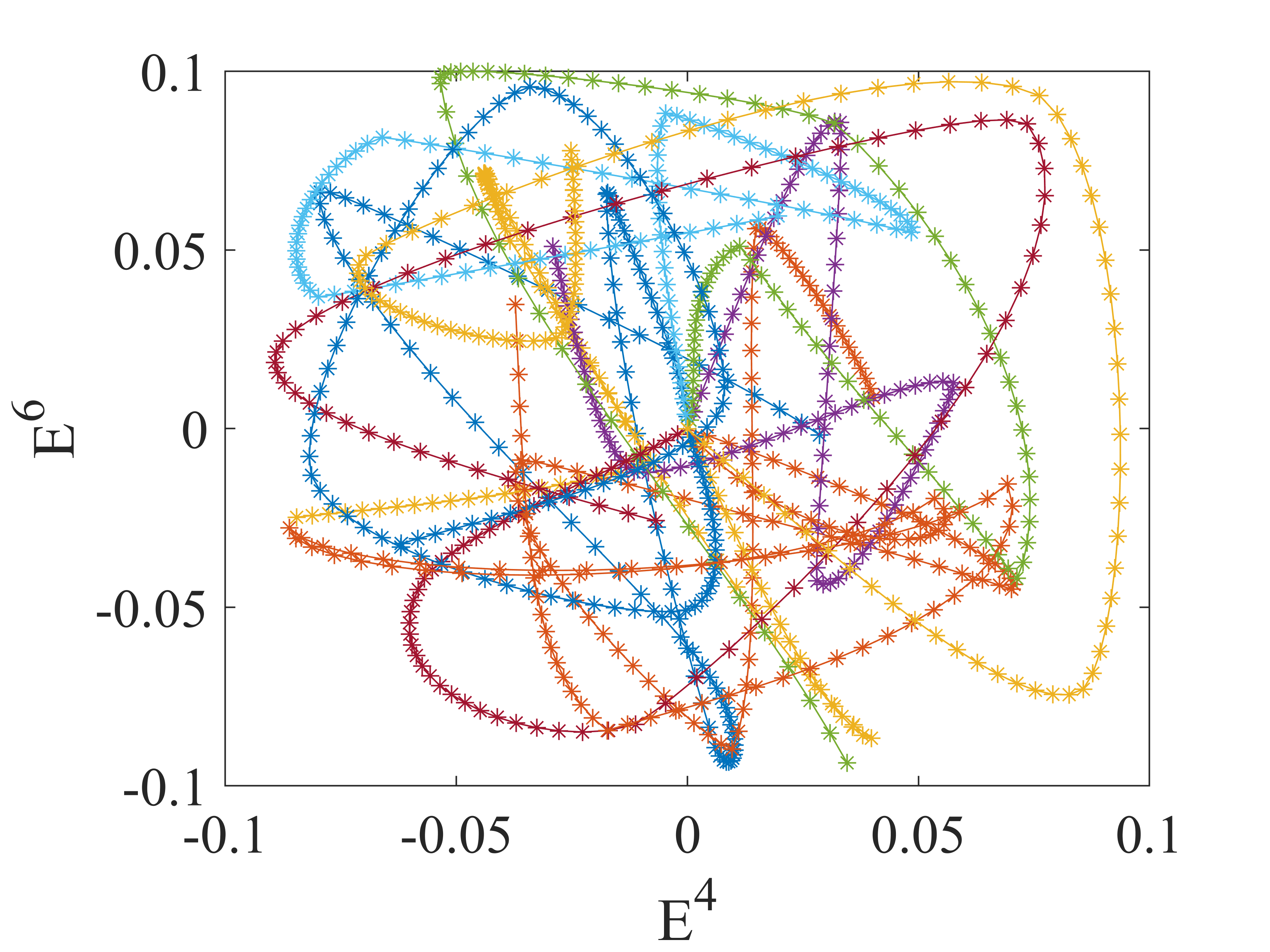}
        \caption{Shear strains $E^4$ and $E^6$}
    \end{subfigure}
    \begin{subfigure}[b]{0.35\textwidth}
        \centering
        \includegraphics[width=\textwidth]{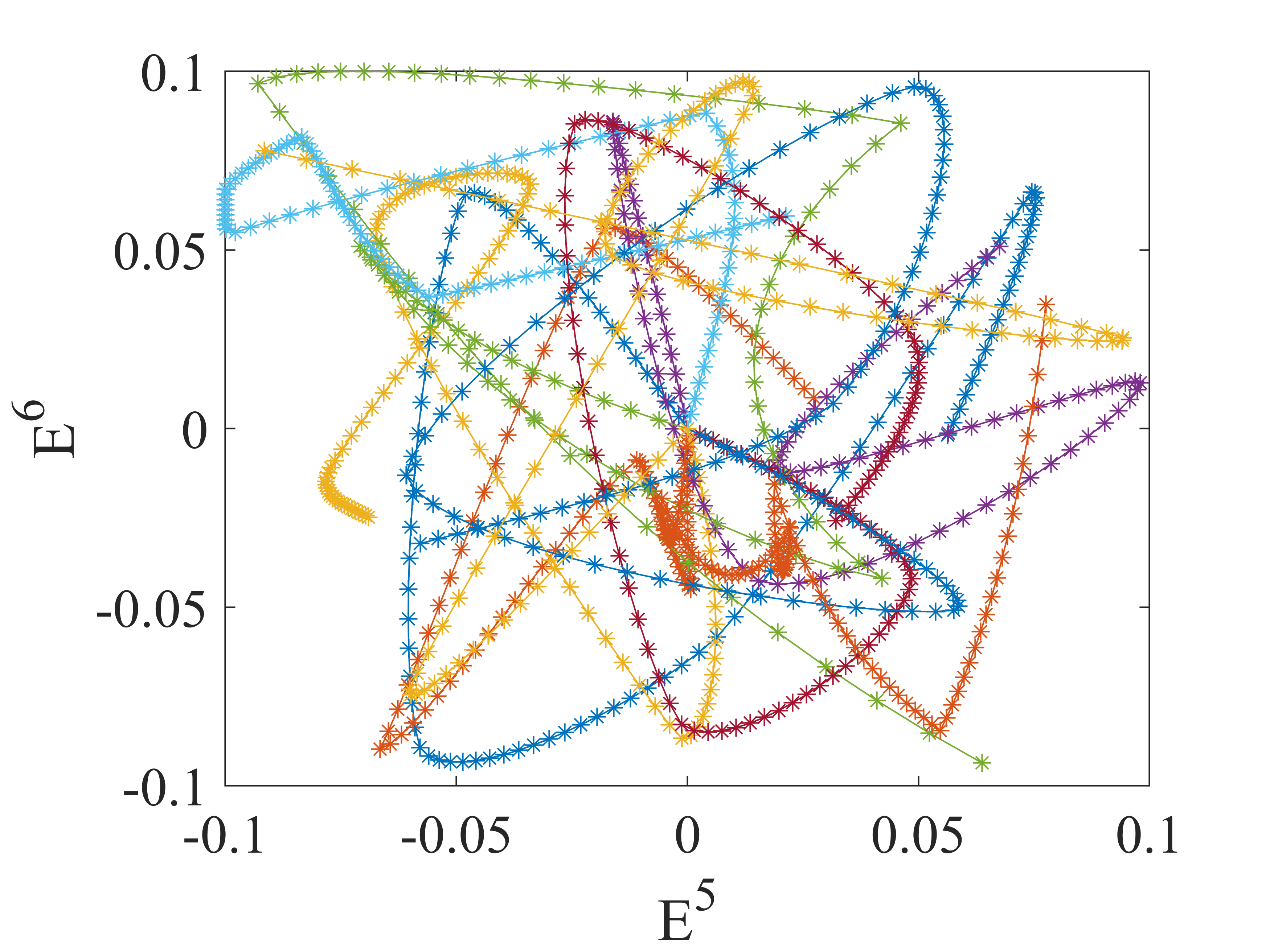}
        \caption{Shear strains $E^5$ and $E^6$}
    \end{subfigure}
    \caption{\textbf{Demonstration of random deformation paths:} We illustrate ten random strain paths that are generated by GP interpolations for (\textbf{a}) normal strain components and (\textbf{b}) shear strain components, and the 2D projections of the random strain components are for (\textbf{c})-(\textbf{e}) normal components and (\textbf{f})-(\textbf{g}) shear components.}
    \label{fig:DoE_StrainPath}
\end{figure}

We create a total of $n_p$ deformation paths by the DoE and the GP interpolations, and each path accounts for the temporal evolution of the six independent strain components. For illustration, we plot ten strain paths in Figure \ref{fig:DoE_StrainPath}. It is evident from the figure that while the random shear strains span the entire hypercube-shaped deformation space defined by the constraint of $\zeta_1$, the normal strain components are additionally confined between the two hyper-planes that represent the volumetric strain constraint of $\zeta_2$. We also plot the 2D projections of the random strain sequences in Figure \ref{fig:DoE_StrainPath} where We can clear see that both the normal and shear components start from the relaxing state without any strain values. In addition, we observe that the highly complex deformation histories consist of multiple loading-unloading-reloading cycles.

After we generate the random strain paths, we need to compute their corresponding effective responses. To compute RVE responses, we use the strain values on the random strain paths to impose the displacement boundary conditions by following the affine boundary condition as:
\begin{equation}
    \textbf{u}_m (\textbf{p}) = \textbf{E}_M \Delta \textbf{p} \hspace{0.8cm} \forall\textbf{p} \in \Gamma_{0m} 
\end{equation}
\noindent where the microstructural displacement boundary condition $\textbf{u}_m$ depends on the macro strain tensors $\textbf{E}_M$ (generated from GP interpolations) and the relative coordinates $\Delta \textbf{p}$ of the nodes on the RVE boundary $\Gamma_{0m}$. From this BVP, we solve the microstructural local stress $\textbf{S}_m$, and compute the effective stress $\textbf{S}_M$ by following Equation \ref{eqn:ScaleTrans_StressStrain}. Additionally, we compute the RVE's effective damage parameter \cite{liu2018microstructural} by: 
\begin{gather}
    D_M = 1 - \frac{\|\textbf{S}_M : \textbf{S}_M^0 \|}{\|\textbf{S}_M^0:\textbf{S}_M^0 \|}
    \label{eqn:effDamageParam}
\end{gather}
\noindent where the homogenized damage parameter $D_M$ indicates the damage status of the RVE. Its value is dependent on the values of the effective stress $\textbf{S}_M$ and the reference stress $\textbf{S}_M^0$ without damage as in Equation \ref{eqn:damage_stress}. 

\subsection{Pure data-driven surrogate}\label{subsec:RNN_architecture}
After we generate the database in Section \ref{subsec:subSec_DBGeneration}, we illustrate how to use the database to train a pure data-driven RNN in this section and discuss the limitations of such pure data-driven surrogates in the end. RNN is a special type of ANN whose working mechanism we illustrate first in the following. ANN consists of a network of artificial neurons to perform weighted sum operations on the network's inputs to compute outputs by activation functions. The most basic type of ANN is the multi-layer perceptron, also known as the feed-forward neural network (FFNN) that only allows information to pass in the forward direction, i.e., from inputs to outputs. In other words, FFNN is a collection of neurons arranged in multiple layers such that each neuron has one-way connections to the neurons of the subsequent layer. If a FFNN is fully connected, every neuron is connected to all neurons of the subsequent layer. 

\begin{figure}
    \centering
    \begin{subfigure}[b]{0.35\textwidth}
        \centering
        \includegraphics[width=\textwidth]{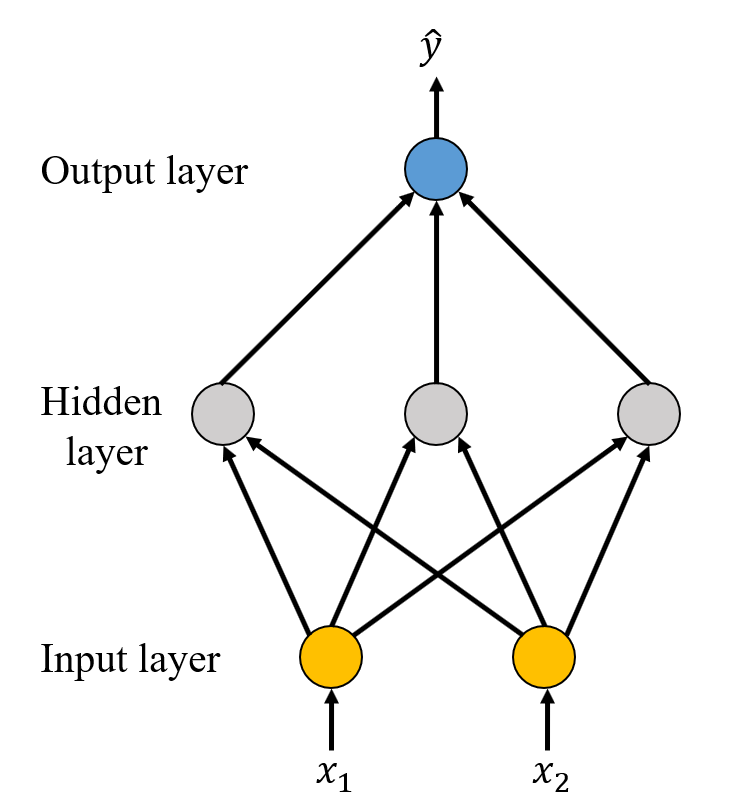}
        \caption{Neural networks}
    \end{subfigure}
    \hspace{1.2cm}
    \begin{subfigure}[b]{0.25\textwidth}
        \centering
        \includegraphics[width=\textwidth]{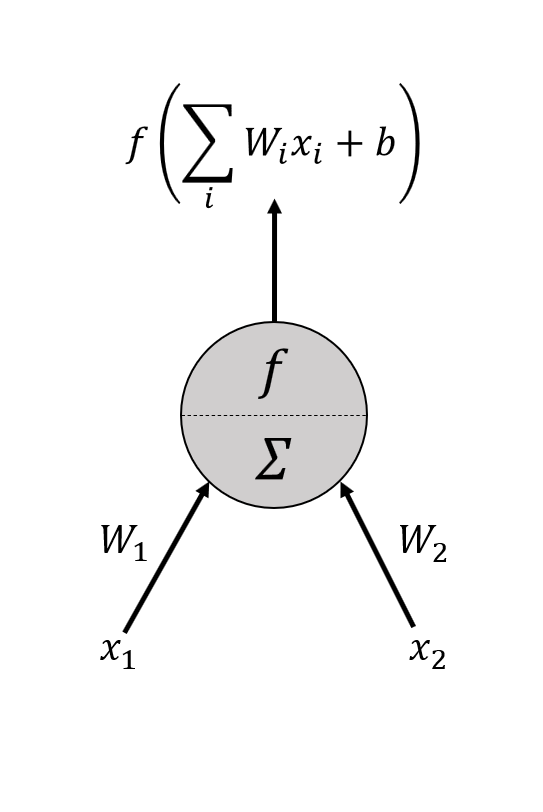}
        \caption{Artificial neuron}
    \end{subfigure}
    \caption{\textbf{Illustration of neural networks:} (\textbf{a}) A shallow FFNN with two input neurons, one hidden layer with three hidden neurons, and single output neuron; and (\textbf{b}) Illustration of the mathematical operations in a hidden neuron.}
    \label{fig:FFNW_neuron}
\end{figure}

A simple fully-connected FFNN with two input neurons, one hidden layer with three neurons, and a single-neuron output layer is shown in Figure \ref{fig:FFNW_neuron} in which each neuron performs a mathematical operation that adds a bias to the weighted sum of its inputs, followed by an activation function \cite{hornik1989multilayer}. While we can choose different types of activation functions based on learning tasks at hand, common choices include hyperbolic tangent, rectified linear unit (ReLU), leaky ReLU, and swish. Mathematically, the neuron in Figure \ref{fig:FFNW_neuron}(b) transforms inputs into outputs by a composition of weighted summation and activation functions as:
\begin{equation}
    \hspace{0.2cm}
    \textbf{x}^{(l-1)}=f \left(\mathbf{W}^{(l)} \mathbf{x}^{(l-1)}+\mathbf{b}^{(l)} \right)
    \label{eqn:ffnn}
\end{equation}
where $f$ is the nonlinear activation function of choice, and $\textbf{x}^{(l-1)}$ and $\textbf{x}^{(l)}$ represent the inputs from the previous layer $l-1$ and the outputs at the current layer $l$, respectively. Additionally, $\textbf{W}^{(l)}$ and $\textbf{b}^{(l)}$ are respectively the weight matrix and the bias vector on the current layer $l$. We note that both weights and biases are known as the network parameters as they are learned in the training process. 

RNN, as a derivation from ANN, was initially developed for ordinal or temporal problems to learn from sequential data \cite{lipton2015critical}. However, there are several major differences between RNN and FFNN. First, the data order of an input sequence is important for RNN such that its output is dependent on both previous and current input values. Second, while FFNN learns training parameters separately for each neuron, RNN shares parameters within each layer of the network, leading to fewer trainable parameters and thus, higher training efficiency. Third, RNN's parameters are typically learned by the algorithms of gradient descent and back propagation through time (BPTT). The BPTT varies from the regular back propagation as it computes the sum of errors from each time step.

\begin{figure}
    \centering
    \begin{subfigure}[b]{0.6\textwidth}
        \centering
        \includegraphics[width=\textwidth]{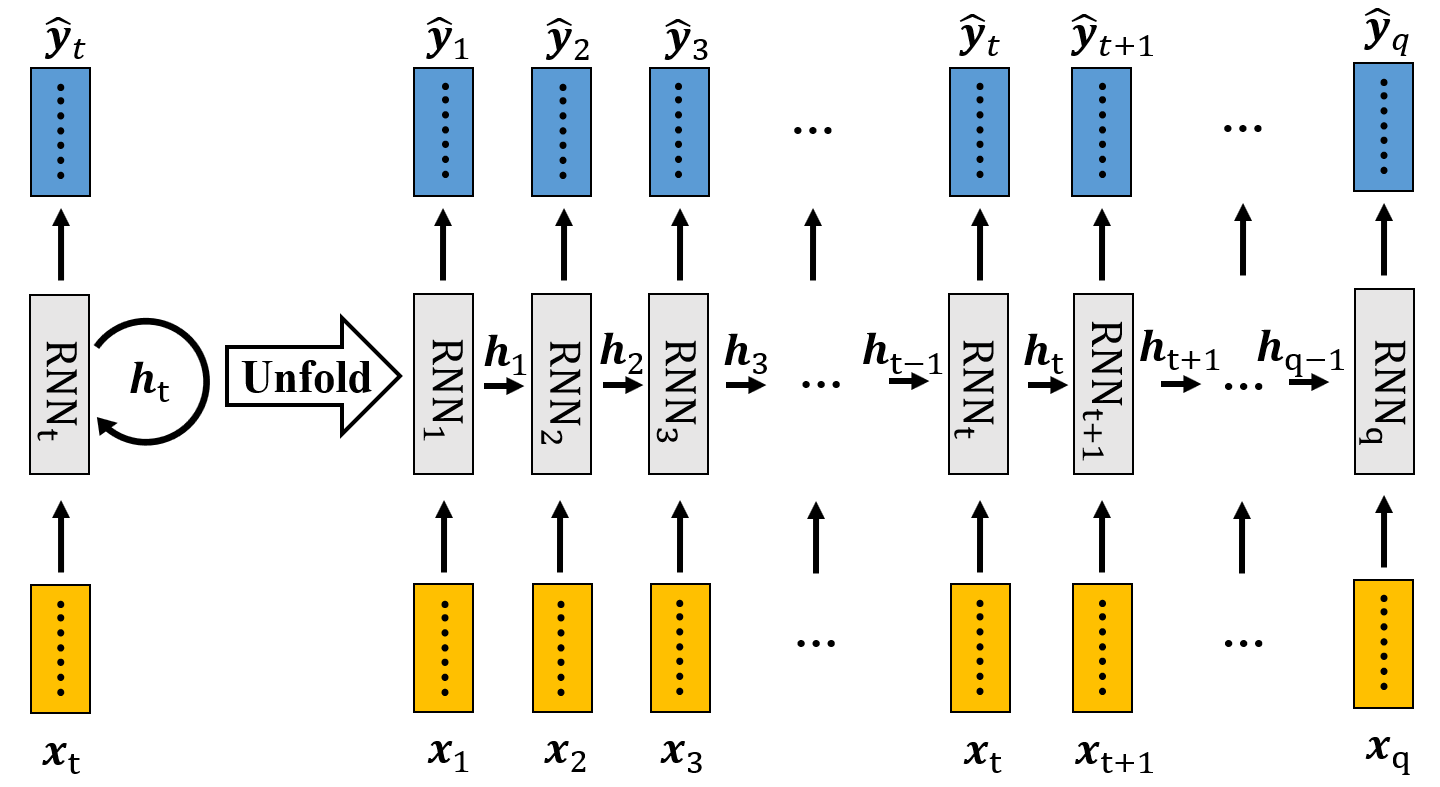}
        \caption{Folded and unfolded RNN}
    \end{subfigure}
    \hfill
    \begin{subfigure}[b]{0.35\textwidth}
        \centering
        \includegraphics[width=\textwidth]{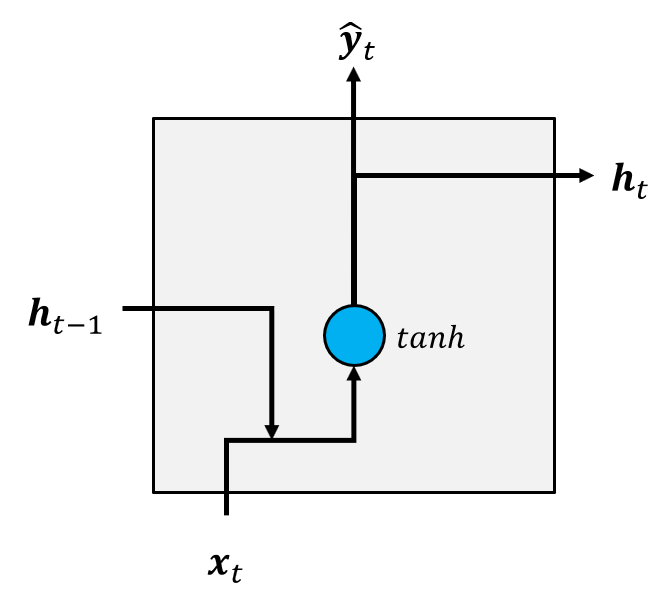}
        \caption{RNN cell}
    \end{subfigure}
    \caption{\textbf{Computational graph of pure data-driven RNN}: \textbf{(a)} Folded and unfolded representations of the RNN architectures that map a sequence of inputs to a sequence of outputs with hidden state variables passing down cells as system memory where $t$ represents an pseudo-time instance in a total of $n_{load}$ time step; and \textbf{(b)} Schematic representation of the internal structure and data flow of one RNN cell at the time instance $t$ where a hyperbolic tangent function maps the weighted current inputs and hidden variables from the previous time step to the current outputs and hidden variables.
    } \label{fig:RNN_compGraph}
\end{figure}

To understand the working mechanism of RNN, let's look at its computational graph in Figure \ref{fig:RNN_compGraph}(a) where a layer of RNN cells relate the input sequence $\textbf{x}_t$ to a series of outputs $\textbf{y}_t$ with $t$ representing an pseudo-time instance (or loading steps) within a total of $n_{load}$ time steps. Specifically, as time propagates, the network unfolds itself such that the temporal-dependent RNN cells sequentially connect one to another to pass down memory-like hidden variables. The mathematical operations in the RNN cell at the time step ${t}$ in Figure \ref{fig:RNN_compGraph}(b) can be expressed as:
\begin{subequations}
\label{eqn:VanillaRNN}
\begin{align}
\mathbf{h}_t &= \operatorname{tanh}(\mathbf{W}_{hh} \mathbf{h}_{t-1}+\mathbf{W}_{xh} \mathbf{x}_t+\mathbf{b}_h) \label{eqn:rnn1} \\
\hat{\mathbf{y}}_t &=\mathbf{W}_{hy} \mathbf{h}_t+\mathbf{b}_y \label{eqn:rnn2}  
\end{align}
\end{subequations}
where hyperbolic tangent is chosen as the activation function, and the hidden state $\mathbf{h}_t$ at the current step is computed from the current input state $\mathbf{x}_t$ and the previous hidden state $\mathbf{h}_{t-1}$. In addition, $\mathbf{W}_{xh}$, $\mathbf{W}_{hh}$, and $\mathbf{W}_{hy}$ are weighting matrices corresponding to input-to-hidden, hidden-to-hidden, and hidden-to-output affine transformations, respectively. $\mathbf{b}_h$ and $\mathbf{b}_y$ are the bias terms associated to the current hidden variables and output estimations.

The major issues that an RNN suffers from when processing long sequential data are the vanishing and exploding gradients \cite{hanin2018neural}. On the one hand, vanishing gradient occurs when the magnitude of gradients continuously decreases as the learned weights eventually become insignificant. On the other hand, the exploding gradient results in extremely large updates to weights during training, and causes unstable learning process. Both issues deteriorate RNN's learning capability and lead to the further development of more advanced sequential learning cells, e.g., long short-term memory (LSTM) \cite{staudemeyer2019understanding} and gated recurrent unit (GRU) \cite{karpathy2015visualizing}. While both LSTM and GRU are considered as the variants of the RNN cell, We adopt GRU due to its high efficiency and we review its detailed working mechanism in Appendix \ref{sec:appendix_B}.  

\subsection{Physics-informed surrogate for material constitutive model} 
\label{subsec:RNN_constraints}

Pure data-driven RNN surrogates often require a large amount of training data due to the large number of network parameters. However, creating a large training dataset in our work requires prohibitively high computational costs, as each of the 3D elasto-plastic simulations is computational expensive. In this section, we propose a physics-constrained RNN architecture that only needs to be trained on a small database, and it outperforms its pure data-driven counterpart. To this end, we explore two types of physics constraints, and incorporate them into our data-driven surrogate by customizing the RNN's loss function and data flow architecture.

\subsubsection{Loss function}

In the training process, we define a loss function to be iteratively minimized to update learning parameters by BPTT. We can define a generic loss function $l_t^0$ for the RNN cell at an arbitrary time instance $t \in \{1,2,\dots,n_{load}\}$ as: 
\begin{equation}
    l_t^0 = \frac{1}{d_{out}} \frac{1}{n_b} \sum_{b=1}^{n_b} \| \textbf{y}_t^b - \hat{\textbf{y}}_t^b \|_{2}
\end{equation}
\noindent where $\textbf{y}_t $ and $\hat{\textbf{y}}_t$ represent the ground truth and predicted values of the outputs including six homogenized stress components $\textbf{S}_t$ and one effective damage parameter $D_t$ at the time instance $t$, i.e., $\textbf{y}_t  = (\textbf{S}_t, D_t)$. We note that $d_{out}$ is the dimension of outputs, and $\| \cdot \|_{2}$ indicates the $l^2$ norm of vectors. In addition, $b$ represents the data index within a training batch of size $n_b$. 

As we have shown in Equation \ref{eqn:energy_constraint} of Section \ref{subsec:thermoDynamics} that the total internal work at an arbitrary macro IP can be computed from its associated RVE's homogenized stress and strain, and its value at any time instance should be always non-negative. We can therefore incorporate this constraint into the generic loss function by using a penalty term as:
\begin{equation}
    \mathcal{L} = \sum_{t=1}^{n_{load}} l_t; \hspace{0.8cm} l_t = l_t^0 + \lambda l_t^1; \hspace{0.8cm} l_t^1 = \frac{1}{n_b} \sum_{b=1}^{n_b} ReLU(- \sum_t (\hat{\textbf{S}}_t^b : \Delta \textbf{E}_t^b) )
    \label{eqn:penalizedLoss}
\end{equation}
\noindent where $\mathcal{L}$ represents the total loss function of our RNN surrogate, $l_t$ is the loss function associated to the RNN cell at the time instance $t$, and $l_t^1$ is the augmented penalty term associated with the internal work in Equation \ref{eqn:energy_constraint}. We approximate the total internal work by the sum of incremental internal work, which is computed by the predicted current stress $\hat{\textbf{S}}_t^b$ and the incremental strain $\Delta \textbf{E}_t^b$ in a training batch, i.e., $\Delta \textbf{E}_t^b = \textbf{E}_t^b - \textbf{E}_{t-1}^b$. 

To penalize any violation of the work constraint, we adopt a non-negative penalty parameter $\lambda$ in Equation \ref{eqn:penalizedLoss}, i.e., $\lambda \geq 0$, and a rectified linear unit (ReLU), i.e., $ReLU(x) = max(x,0) \geq 0$. We use $\lambda$ and ReLU to ensure that a negative internal work increases the value of the total loss function during minimization that results in the penalty on such constraint violation. We note that the value of the penalty parameter $\lambda$ is case-dependent and often needs to be carefully chosen to balance the weights of the generic loss and the penalty term as it affects learning efficiency and could pose challenges to the overall training process.

\subsubsection{RNN architecture}

Our second constraint is based on the fact that our material is not self-healing in the irreversible damage process such that the damage parameter at an arbitrary macro-point is non-decreasing as damage evolves:
\begin{equation}
    \dot{D}_t = \frac{\partial D_t}{\partial t} \geq 0
    \label{eqn:RNNConstraint_damage}
\end{equation}
\noindent where $\dot{D}_t$ is the damage rate, and $D_t$ is the effective macro damage parameter at time step $t$ that can be computed from the RVEs' homogenized damaged and reference stresses in Equation \ref{eqn:effDamageParam}. However, the condition of non-decreasing damage parameter is not necessarily satisfied by the pure data-driven (vanilla) architecture in Figure \ref{fig:RNN_compGraph}(a) and thus, we have to develop a numerical scheme to explicitly enforce the constraint in Equation \ref{eqn:RNNConstraint_damage} to satisfy the irreversible damage process.

\begin{figure} 
    \centering
    \includegraphics[width = 0.7\textwidth]{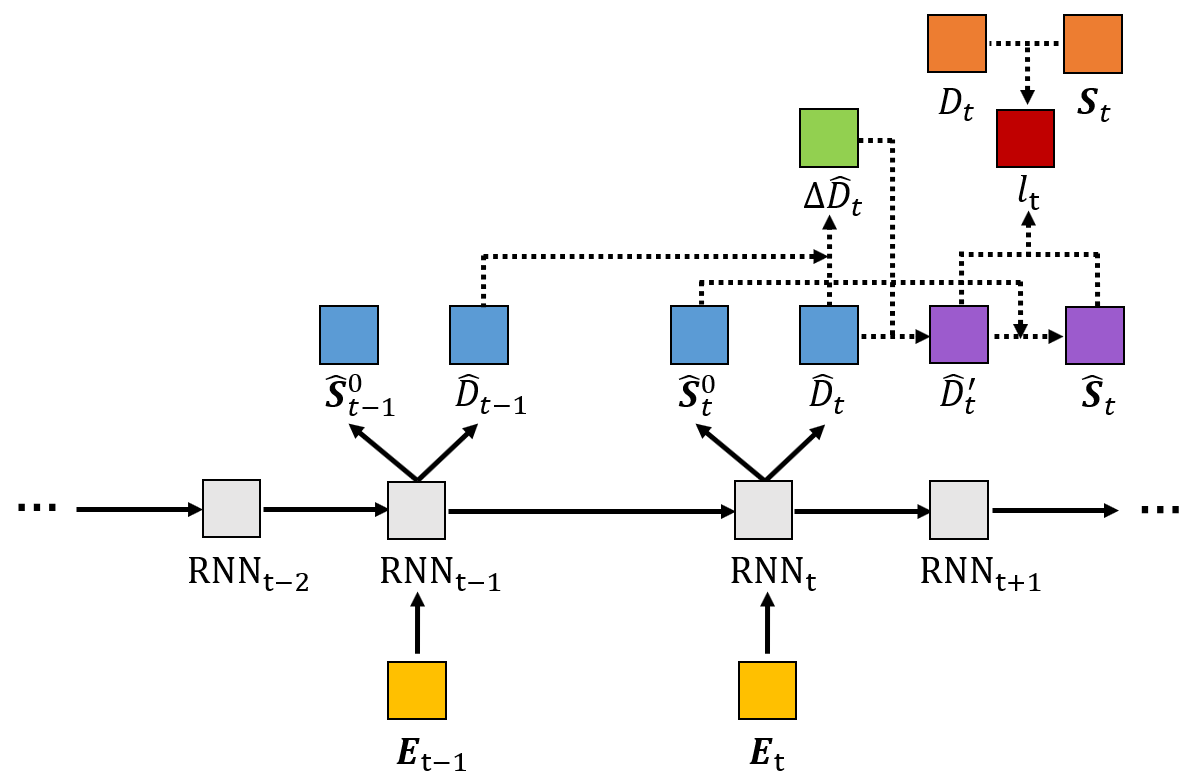}
    \caption{\textbf{Computational graph of physics-constrained RNN architecture:} In this architecture, we illustrate the RNN cells at different time steps by the grey squares where the inputs and outputs of the cells are represented by the yellow and blue squares. The architecture's intermediate variables to compute the loss function (red square) at the time instance $t$ are represented by the green and purple squares, and the ground truth values are represented by the orange squares. In addition, we use solid lines and dash lines to represent the directions of the data flow associated to the RNN cells and the intermediate variables, respectively.
    } \label{fig:GRU_graph}
\end{figure}

To incorporate such damage constraint, we propose a new RNN architecture by introducing several major modifications of the vanilla model as shown in Figure \ref{fig:GRU_graph}. Specifically, we firstly attach two FFNNs to the outputs of RNN cells, and assume the FFNNs' outputs are the effective reference stress in the absence of damage and the damage parameter $\hat{D}_t$ at the time instance $t$. The underlying reason for us to choose the reference stress instead of damaged stress as the FFNNs' outputs is that we intend to let our RNN to learn the stress based on a much simpler elasto-plastic hardening relation given our small size of training dataset. On the contrary, if we select damaged stress as the outputs as in the vanilla model, since the softening simulation involves much more complex behaviors, a representative RNN model requires more learning parameters that needs to be trained on a larger dataset to achieve the same level of prediction accuracy. To enforce the damage irreversibility, we compute the damage increment by comparing its predictions at the current step $\hat{D}_t$ and the previous step $\hat{D}_{t-1}$ and accordingly update the damage parameter by using the following scheme:
\begin{equation}
    \hat{D}'_t = \sum_t \{\hat{D}_t + \triangle\hat{D}_t[sgn(\triangle\hat{D}_t) \times 0.5 - 0.5] \}; \hspace{0.8cm} \triangle\hat{D}_t = \hat{D}_t - \hat{D}_{t-1}
\end{equation}
\noindent where $\hat{D}'_t$ indicates the corrected output of the effective damage parameter at the time step $t$, and $\triangle\hat{D}_t$ represents the incremental difference between two consecutive steps. We apply the sign function $sgn(\cdot)$ to $\triangle\hat{D}_t$ to mark the time instance at which the values of the estimated damage parameter decreases compared to their previous value. We use the sign function to compensate the FFNNs-estimated damage parameter $\hat{D}_t$ by the incremental errors that are accumulated from the initial time instance. It is followed by a normalization function to ensure $\hat{D}'_t$ to stay within the bounds of $[0,1]$. 

With the corrected damage parameter $\hat{D}'_t$ and the estimated reference stress $\hat{\textbf{S}}_t^0$, we can now compute the damaged stress by following the same relation as the continuum damage mechanics in Equation \ref{eqn:damage_stress} as:
\begin{equation}
    \hat{\textbf{S}}_t = (1 - \hat{D}'_t) \hat{\textbf{S}}_t^0
\end{equation}
\noindent where $\hat{\textbf{S}}_t$ corresponds to the corrected stress components. By comparing the corrected damage parameter and the damaged stress values to their ground truths, we are able to compute the loss function from Equation \ref{eqn:penalizedLoss}. 

We emphasize that we enforce the two physics constraints within our RNN architecture by using two different approaches. While we implement the energy constraint in Equation \ref{eqn:penalizedLoss} as a soft constraint by adding an associated penalty term in the loss function, we enforce the damage constraint in Equation \ref{eqn:RNNConstraint_damage} as a hard constraint by imposing architectural modifications and post-processing RNN cells' outputs by using the intermediate variables in the network. Upon comparing the two approaches, We note that although the hard constraint always guarantees constraint's enforcement, it may lead to a stiffer optimization problem \cite{marquez2017imposing} in training. Additionally, its architectural modifications involve significant model development efforts, and such modifications may become infeasible for more complex problems. Therefore, the choice of a hard or soft approach is situational, and it needs more investigations in future study.    

\subsubsection{Teacher forcing}
Teacher forcing \cite{goodfellow2016deep} is an efficient machine learning training technique frequently used in RNNs. It uses ground truth from previous time steps to augment the inputs at the current step in order to force the networks to stay close to the ground truth at each step.

\begin{figure}
    \centering
    \begin{subfigure}[b]{0.8\textwidth}
        \centering
        \includegraphics[width=\textwidth]{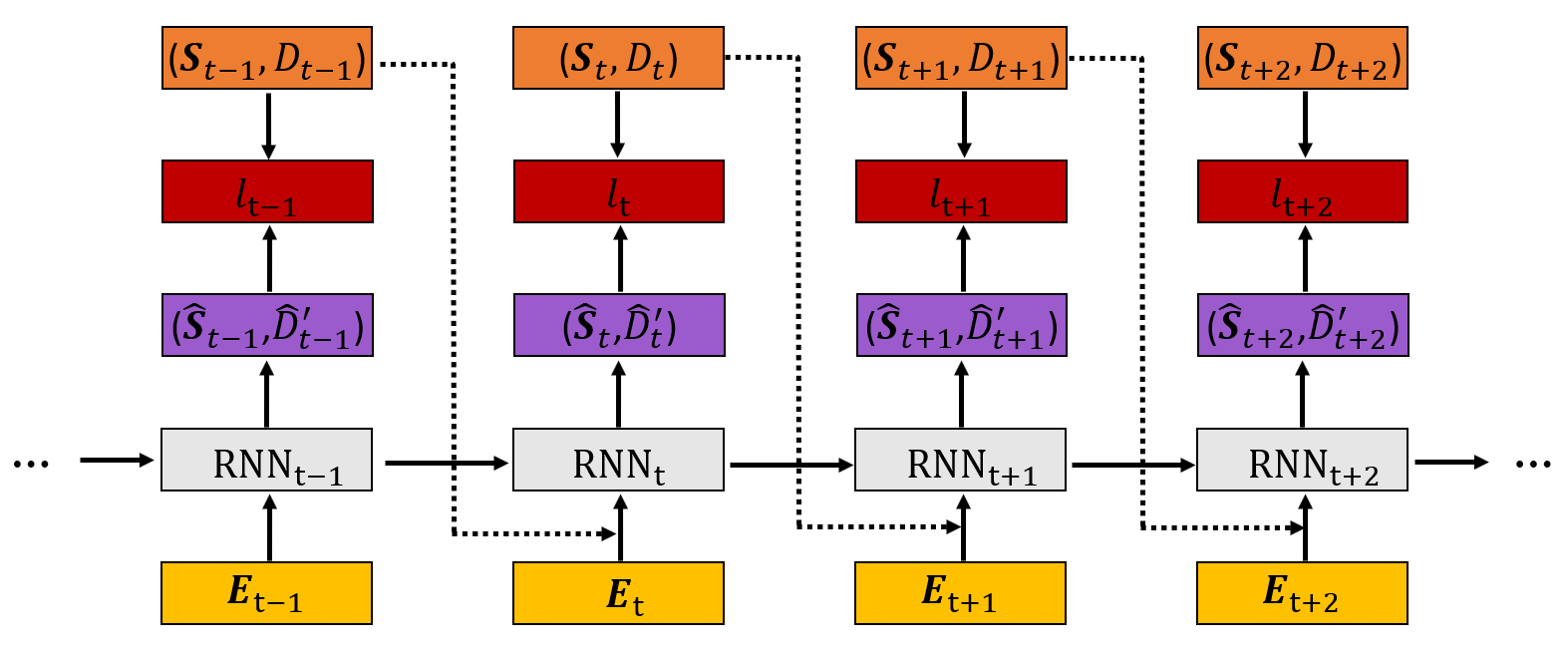}
        \caption{Teacher forcing in training time}
    \end{subfigure} \\
    \vspace*{0.5cm}
    \begin{subfigure}[b]{0.8\textwidth}
        \centering
        \includegraphics[width=\textwidth]{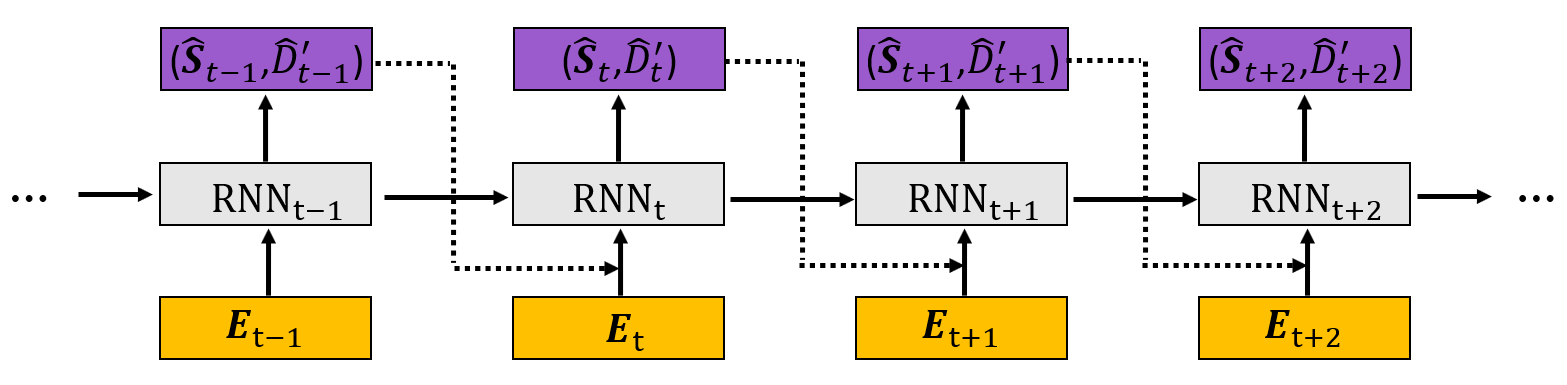}
        \caption{Teacher forcing in testing time}
    \end{subfigure}
    \caption{\textbf{Teacher forcing in our RNN architecture:} (\textbf{a}) In training time, the teacher forcing method with one look back step augments the ground truth from the last time step to the inputs at the current step; and (\textbf{b}) During testing time, the teacher forcing method augments the predictions from the last time step to the inputs at the current step. 
    } \label{fig:teacherforcing_graphs}
\end{figure}

We utilize the teacher forcing differently in the training and testing stages as demonstrated in Figure \ref{fig:teacherforcing_graphs}. In the offline training stage, we use the teacher forcing to augment the previous ground truth (RVE's effective stress and damage variable) to the current input (RVE's macro strain). As the previous ground truth is a part of the input, its associated weight matrices are updated by the BPTT. During the online testing stage, since there is no ground truth available, we feed the previous prediction to the current input by assuming the prediction values are close the ground truth. We note that while the teacher forcing demonstrated in Figure \ref{fig:teacherforcing_graphs} has one look back step that involves feeding the ground truth (training time) or prediction (testing time) from the last step, a generic teacher forcing can have multiple look back steps.

The teacher forcing is advantageous in faster training convergence and higher accuracy by providing a sequence of previous ground truth to the inputs. However, in online computing, we have to feed the previous output back to the current input. When online sequences are dramatically different from those of training, the networks may result in non-negligible discrepancy between ground truths and predictions that leads to errors accumulating along time evolution and causing model divergence. We demonstrate both the advantage and disadvantage of the teacher forcing approach by numerical experiments in Section \ref{sec:experiments}.   

\subsection{Integration of material constitutive surrogate in multiscale solvers}\label{subsec:RNN_multiscale}

In the data-driven multiscale simulations, we need to use the trained RNN as a surrogate to replace the computationally expensive micro analysis models. However, the online deployment of the RNN surrogate within a physical iterative solver poses many difficulties. This is because, during training, RNN has access to the convergent deformation and effective response histories at all $n_{load}$ loading steps. Comparatively, in the online computations, the trained RNN can only access to the convergent strain and response histories in previous steps, as it lacks the convergence information at both the current step and future steps. We address such difficulty by explicitly modifying the data structure of RNN input sequences and implicitly resetting RNN's hidden variables amid iterations. 

We demonstrate the approach to incorporate our trained RNN into a multiscale model by the pseudo-code in Algorithm \ref{alg:RNN_multiscale}. In nonlinear material simulations, Newton Raphson method is a classic numerical approach to iteratively solve for material's path-dependent responses. This method essentially consists of a double-loop structure: while the outer loop accounts for the steps with different loading conditions, the inner loop iterates material responses under a certain loading condition until convergence, i.e., equilibrium between internal and external forces. 

In a typical step of a multiscale simulation, we compute the macro strain at an arbitrary IP from equilibrium equations within the inner loop. By appending the strain at the current iteration to the sequence of previous convergent strains, the length of strain sequence equals the current load step number that is shorter than the required length of RNN input sequence $n_{load}$. We add replicate padding by repeating the value of the current strain multiple times to append to the end of the strain sequence. It not only makes the strain sequence compatible with RNN inputs, but also implicitly enforcing the RNN's hidden variables at the current step to stay constant within the iteration (inner) loop. This is because the values of the current hidden variables are decided by the state of network parameters and the inputs from early time instances, see Equation \ref{eqn:VanillaRNN}. The underlying reason of using the same values of hidden variables at the current step is similar to the classic radial return algorithm in material plastic analysis where material state variables are only updated upon convergence. We also emphasize that the number of the loading steps of the multiscale model should be smaller than or equal to the sequence length of RNN inputs $n_{load}$, as a larger step number would result in data truncation during input data preparation and erroneous RNN inference.
\RestyleAlgo{ruled}
\SetKwComment{Comment}{/* }{ */}
\\
\begin{algorithm}[]
\caption{Integration of RNN in multiscale analysis} 
\label{alg:RNN_multiscale}
$i=1,2,\dots,n_{load}$ \Comment*[r]{Newton step number}
$j=1,2,\dots,n_{iter}$ \Comment*[r]{Newton iteration number}
$\epsilon = 10^{-6}$ \Comment*[r]{Convergence criterion}
\While{$i \leq n_{load}$}{ 
    \While{$j \leq n_{iter}$}{ 
        (1) Read macro strain $\textbf{E}_i^j$ from macro equilibrium equation \\
        (2) Append $\textbf{E}_i^j$ to the convergent strain sequence $\{ \textbf{E}_1^c, \textbf{E}_2^c, \dots, \textbf{E}_{i-1}^c, \textbf{E}_i^j \}$ \\
        (3) Add ($n_{load}$ - $i$) replicate padding of $\textbf{E}_i^j$ to the end of the sequence in (2) \\
        (4) Perform RNN inference on the updated strain sequence \\
        (5) Read RNN's outputs for the effective responses at the step $i$ \\
        (6) Solve macro equilibrium equation \\
        \eIf{$ \| \textbf{f}_{int} - \textbf{f}_{ext} \| \leq \epsilon $} 
        {
            Update convergent strain $\textbf{E}_i^c = \textbf{E}_i^j$ \\
            Continue to the next loading step: $i \leftarrow i+1 $ \\
            Break \Comment*[r]{Iteration convergence}
        }
        {
            Continue to the next iteration: $j \leftarrow j+1 $ \\
        }
    }
}
\end{algorithm}
\section{Numerical experiments} \label{sec:experiments}

In this section, we demonstrate the efficiency and accuracy of the proposed physics-constrained data-driven surrogate for the multiscale damage simulations. This section is organized as follows: we first illustrate the efficacy of our physics-informed RNN in predicting microstructural effective behaviors in Section \ref{subsec:exp_RVE} where we assume micro porosity as the only material defects. We perform the computation of multiscale elasto-plastic hardening and softening simulations in Section \ref{subsec:exp_Lbracket} by integrating our RNN (as a surrogate of microstructural analysis model) with a macro FEM solver. In Section \ref{subsec:exp_DoubleNotch}, we deploy our multiscale surrogate model to perform a mesh convergence study on a component with different spatial discretization levels to simulate macro damage patterns. In the experiments, we record computational costs and perform accuracy analysis to provide an insight of our model's performance.

As for the model implementation, we program our RNN model in Python environment and use the deep learning package Keras. We generate the database of microstructural effective responses on a state-of-the-art high-performance cluster (HPC) by using $60$ CPU cores (AMD EPYC processors) and $360$ GB RAM. We carry out the training procedure of our RNN model on the HPC by using two GPU units (NVIDIA Tesla v100) with $32$ GB RAM. For multiscale simulations, we develop a dedicated program to integrate our RNN model within a multiscale analysis engine which is implemented in Matlab environment. We note that all data-driven multiscale computations in Sections \ref{subsec:exp_Lbracket} and \ref{subsec:exp_DoubleNotch} are conducted on a $64$-bit Windows desktop with four CPU cores (Intel i7-3770) with $16$ GB RAM.

\subsection{Surrogate for microscale damage analysis}\label{subsec:exp_RVE}

\begin{figure}
    \centering
    \includegraphics[width = 0.65\textwidth]{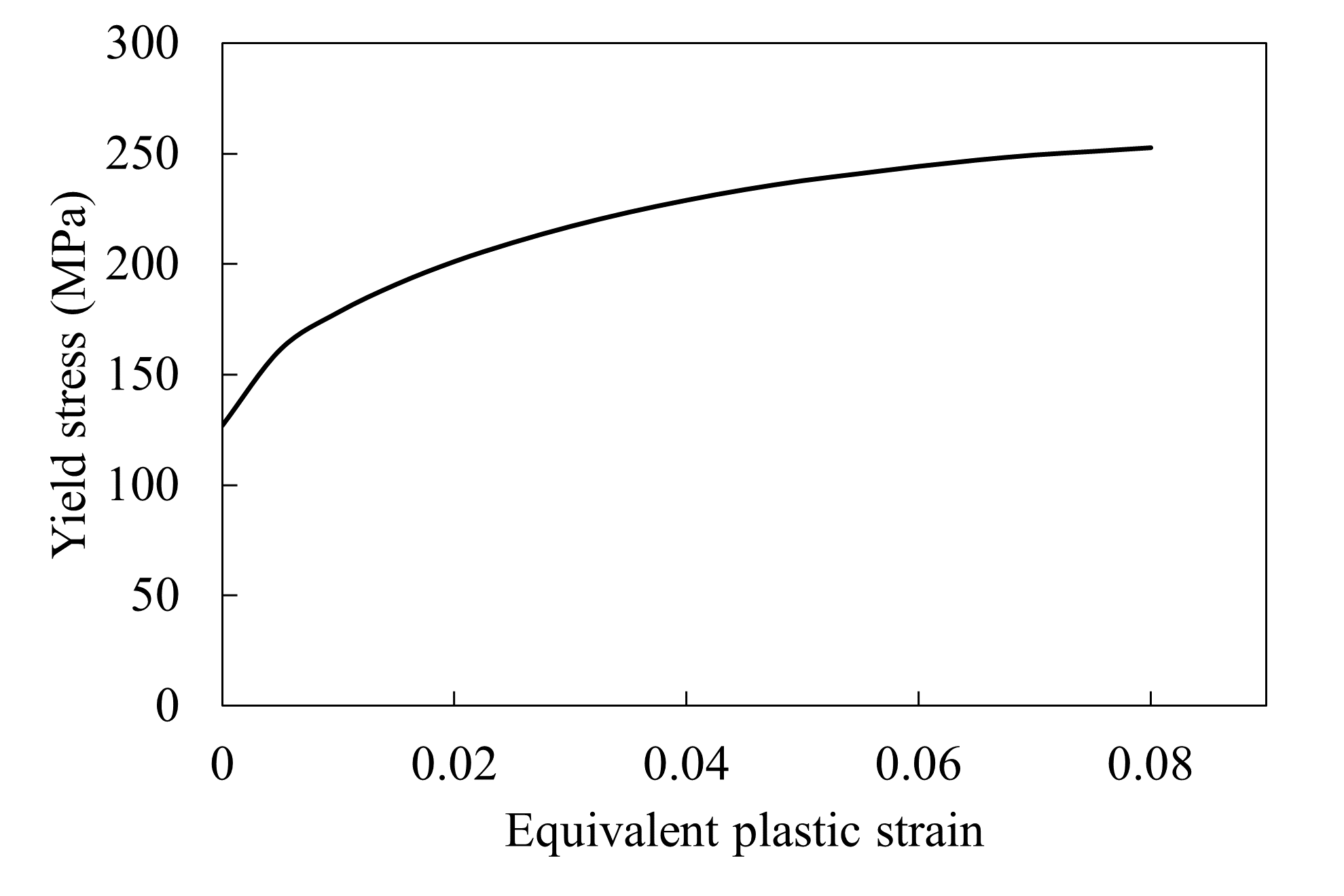}
    \caption{\textbf{Elasto-plastic hardening behavior:} we use a piecewise linear hardening model to define the elasto-plastic behavior of aluminum alloy A356.
    } \label{fig:material_hardening}
\end{figure}

We assume the material as aluminum alloy A356 with elastic modulus of $5.7e4$ MPa and the Poisson's ratio of $0.33$. Its elasto-plastic hardening behavior is assumed as isotropic and follows associated plastic flow rule with the Mises yield surface defined by:

\begin{equation}
    S \leq S_y(\bar{E}^{pl})
    \label{eqn:VM_YieldSurface}
\end{equation}
\noindent where $S$ and $S_y$ are respectively the Mises equivalent stress and yield stress depending on the equivalent plastic strain $\bar{E}^{pl}$. To model strain hardening, we assume the relation between $S_y$ and $\bar{E}^{pl}$ as piecewise-linear as shown by the hardening curve in Figure \ref{fig:material_hardening}. For softening simulations, we employ the damage continuum model as discussed in Section \ref{subsec:strainSoftening} with the fracture strain $E_f$ of $0.067$ and the fracture energy $G_f$ of $1.92e4$ N/m.

\subsubsection{Database generation of RVE effective responses}

\begin{figure}
    \centering
    \includegraphics[width = 0.9\textwidth]{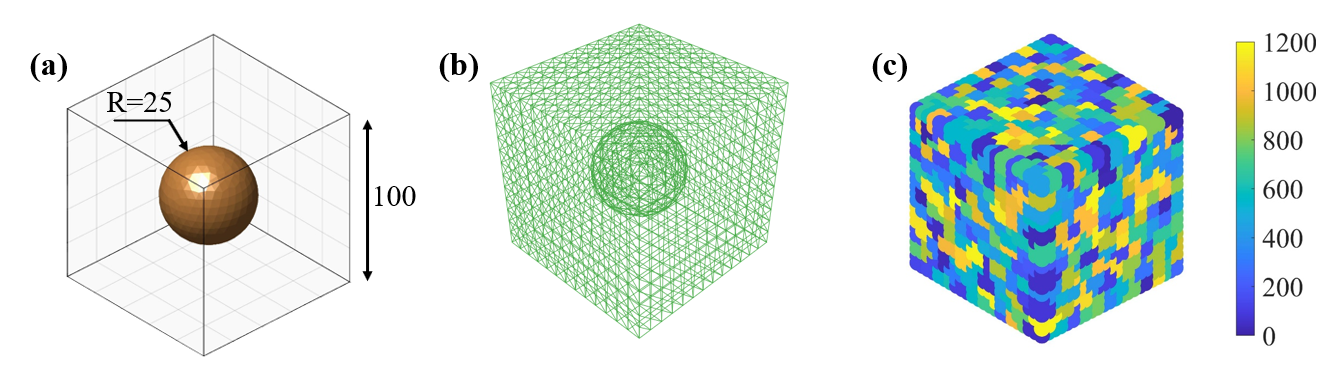}
    \caption{\textbf{The geometry, dimension and mesh of our RVE:} (\textbf{a}) The dimension (unit: $\mu m$) of our RVE that contains a spherical pore at center with a pore volume fraction of $6.25\%$; (\textbf{b}) The RVE is discretized by $15,000$ finite elements; and (\textbf{c}) Our ROM with $1,200$ clusters.
    } \label{fig:RVE_geom_mesh}
\end{figure}

We solve micro BVPs on a simple microstructure whose geometry and mesh are illustrated in Figure \ref{fig:RVE_geom_mesh}(a). The microstructure contains a spherical pore in the center that is surrounded by the material matrix of A356. Even though classic FEM with sufficiently fine mesh, e.g., see Figure \ref{fig:RVE_geom_mesh}(b), can provide high-fidelity solutions to BVPs, it is generally expensive, especially for the response database generation. To improve computational efficiency, we apply our previously developed mechanistic DCA-based ROM \cite{deng2022reduced,deng2022concurrent,deng2022data} to solve the BVPs. Compared to FEM, our ROM strikes a good balance between efficiency and accuracy by agglomerating elements into clusters, e.g., see Figure \ref{fig:RVE_geom_mesh}(c) where material IPs in the same cluster are assumed to share identical elasto-plastic behaviors.  

We note that a mesh independence study is often required in material softening simulations to choose a proper spatial discretization for solution convergence \cite{bazant2010can, deng2022data}, see Section \ref{subsec:strainSoftening}. We conduct the microscale mesh convergence investigation in Appendix \ref{sec:appendix_A} where we systematically compare the softening behaviors between FEM and ROM for the RVE in Figure \ref{fig:RVE_geom_mesh}(a) and it shows that the ROM with $1,200$ clusters is able to provide consistent post-failure behaviors as the FEM, but only requires less than $10\%$ of computational time. It is for this reason that we choose the ROM with $1,200$ clusters for the generation of material softening database. In addition, we consider the ROM as the benchmark when validating our data-driven surrogates in the following experiments. 

 For the database generation, we set the sampling constraint for the strain magnitude as $\zeta_1=10\%$ and the constraint of the volumetric strain as $\zeta_2=4\%$. For the GP interpolation, we set the number of control points with random strain values as $n_c=5$. Our database contains a total of $n_p=30,000$ deformation paths and RVE effective responses where each path includes six strain components, six effective stress components and one effective damage variable at $n_{load}=101$ sequential loading steps. Generating this database costs about ten-day computational time on the HPC by exploiting parallel computing with 60 CPU cores. 

\subsubsection{Impacts of physics constraints}

To demonstrate the impacts of the two physics constraints in Section \ref{subsec:RNN_constraints}, we compare the prediction accuracy of a pure data-driven vanilla model against our RNN model. For this comparison, we randomly choose $200$ deformation-responses sequences as a test set. We further randomly select $6,000$, $12,000$, $18,000$, $24,000$ and $29,800$ sequences from the database to form five different training-validation datasets. For all training-validation datasets, we split them into $80\%$ for training set and $20\%$ for validation set. For example, the dataset of the size of $6,000$ has $4,800$ sequences for training and $1,200$ for validation. We point out the training and validation sets serve different purposes, as the training set is used to iteratively update learning parameters during BPTT, while the validation set is to monitor overall training process by avoiding overfitting or underfitting.  

For the model training, we start with normalizing all data sequences. We use $1,200$ training epochs with a batch size of $n_b=64$, and choose \textit{Adam} as the optimizer with an adaptive learning rate that starts at $10^{-3}$ and reduces by $25\%$ when the validation error is not reduced over $30$ training epochs. We terminate the training process when the training reaches the maximum number of epochs or the loss function is not improved by $10^{-7}$ over $50$ epochs. We use the mean squared error (MSE) to quantify the prediction errors on the testing dataset. We define the MSE as:
\begin{subequations}
\begin{align}
    &\mathrm{MSE} = \frac{1}{n_t n_{load} d_{out}} \sum_{m=1}^{n_t} \sum_{t=1}^{n_{load}} \sum_{i=1}^{d_{out}} \left(y_{i, t}^m - \hat{y}_{i, t}^m\right)^2  \\
    &\mathrm{MSE_S} = \frac{1}{n_t n_{load} d_S} \sum_{m=1}^{n_t} \sum_{t=1}^{n_{load}} \sum_{i=1}^{d_S} \left(S_{i, t}^m - \hat{S}_{i, t}^m\right)^2 \\
    &\mathrm{MSE_D} = \frac{1}{n_t n_{load}} \sum_{m=1}^{n_t} \sum_{t=1}^{n_{load}} \left(D_t^m - \hat{D}_t^m\right)^2
\end{align}
\end{subequations}
\noindent where MSE accounts for the total prediction error including both stress and damage predictions, while $\mathrm{MSE_S}$ and $\mathrm{MSE_D}$ are the prediction error for stress and damage, respectively. We note that $n_t$ and $d_{out}$ are the number of data sequences in the testing set and the dimension of outputs. $y_{i, t}^m$ and $\hat{y}_{i, t}^m$ are the ground truth and prediction for the $i^{th}$ output component at the $t^{th}$ load step for the $m^{th}$ testing sample. In addition, $d_{S}$, $S_{i, t}^m$, $\hat{S}_{i, t}^m$, $D_t^m$ and $\hat{D}_t^m$ are the number of 3D stress components, true stress, predicted stress, true damage and predicted damage variable, respectively, i.e., $y_{i, t}^m=(S_{i, t}^m, D_t^m)$. We note that the values of $n_t$, $d_{out}$ and $d_{S}$ are equal to $200$, $7$ and $6$, respectively. Furthermore, we set the penalty parameter as $\lambda=10^{-6}$ in the customized loss function in Equation \ref{eqn:penalizedLoss} to account for the internal work constraint. 

\begin{figure} 
    \centering
    \includegraphics[width = 0.5\textwidth]{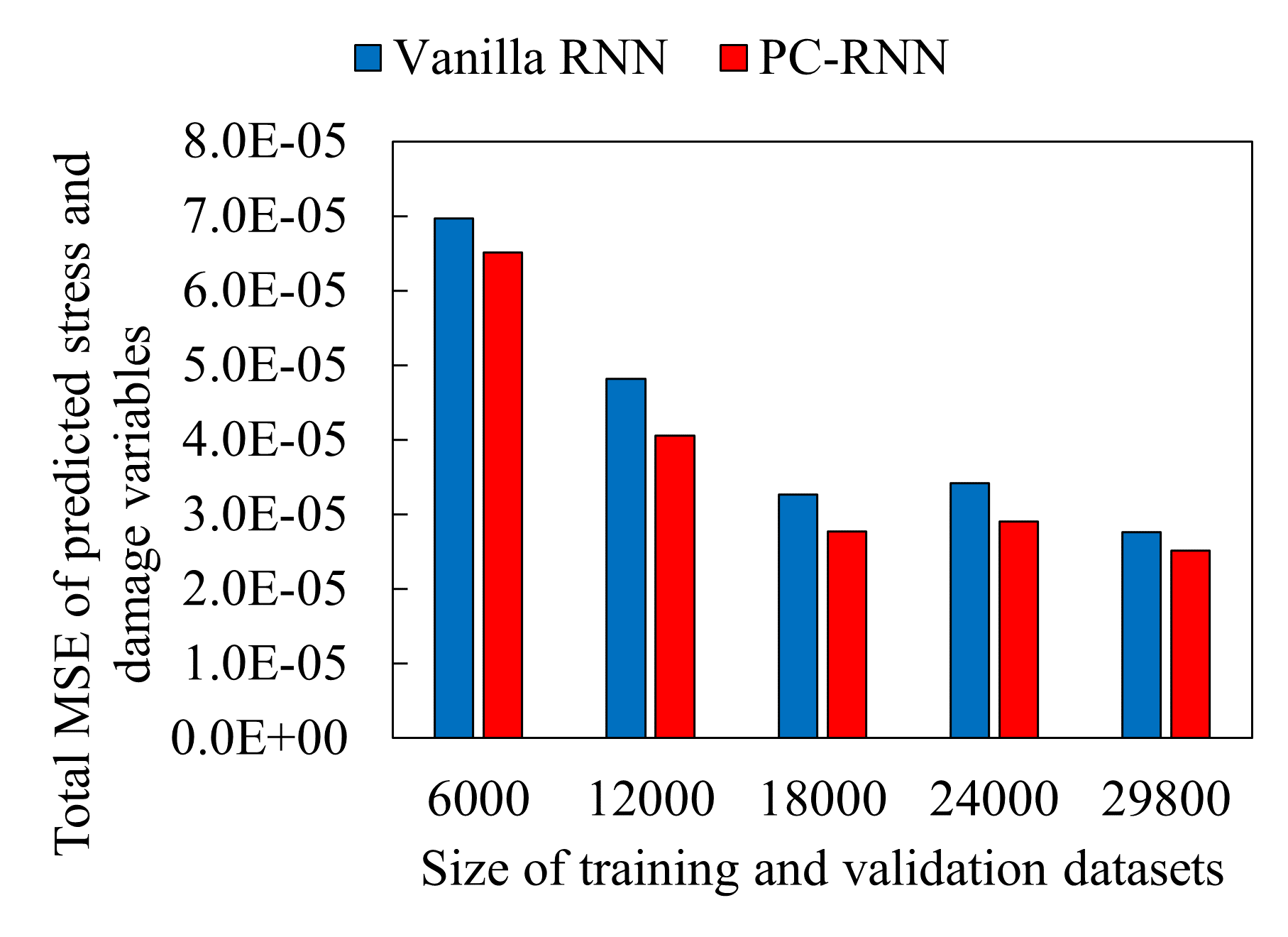}
    \caption{\textbf{Comparison of RNN prediction errors with respect to the sizes of datasets:} Comparison of the total MSE on the predicted stress and damage variables between the pure data-driven vanilla model and the proposed RNN model on the different sizes of training and validation datasets where the errors are computed on the same testing dataset containing $200$ deformation-responses sequences that are unseen amid training process. 
    } \label{fig:GRU_MSE}
\end{figure}

\begin{figure} 
    \centering
    \includegraphics[width = 0.9\textwidth]{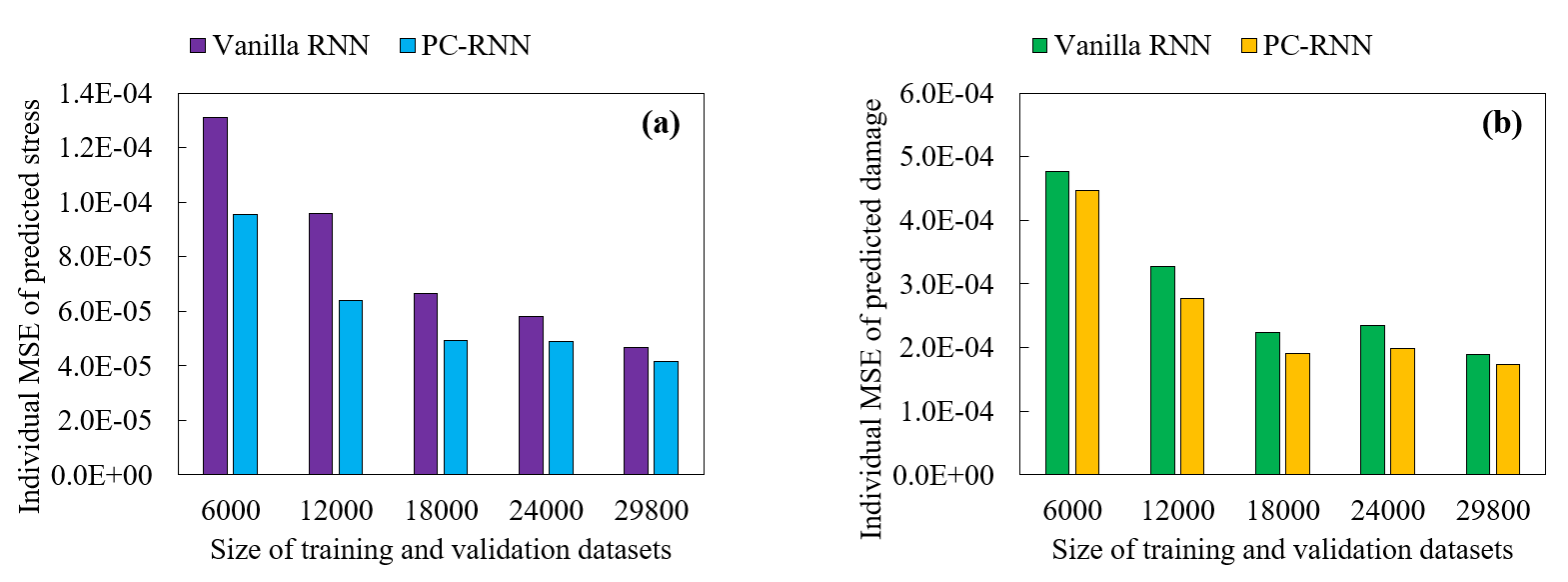}
    \caption{\textbf{Individual MSE on predicted stress and damage:} (\textbf{a}) Comparison of the individual MSE on the predicted stress between the pure data-driven vanilla model and the proposed physics-constrained RNN model on the different sizes of training and validation datasets; and (\textbf{b}) Comparison of individual MSE for the predicted damage variables. 
    } \label{fig:GRU_MSE_stress_damage}
\end{figure}

After models are trained on the five different training-validation datasets, we compare the their prediction errors on the same test set that is unseen amid the entire training process as shown in Figure \ref{fig:GRU_MSE} and Figure \ref{fig:GRU_MSE_stress_damage}. For the overall MSE, we find that as the sizes of training-validation datasets increase from $6,000$ to $29,800$, the MSEs of both the pure data-driven vanilla model and our physics-constrained model (PC-RNN) decrease dramatically from about $7 \times 10^{-5}$ to $3\times 10^{-5}$. It is clear that the overall MSE of our proposed model is always lower than that of the vanilla counterpart. 

For the predictions on the effective stress as shown in Figure \ref{fig:GRU_MSE_stress_damage}(a), the proposed model evidently demonstrates a better accuracy than the vanilla model. Particularly, on the smaller datasets ($6,000$ or $12,000$ sequences) with a limited amount of training data, the MAE of our model is about $60\%$ lower than the vanilla model, which demonstrates the importance of physics constraints in regularizing data-driven models. As the sizes of databases increase, we observe the gap of MSE in stress predictions narrows, as the pure data-driven vanilla model is shown more sensitive to the amount of data. In addition, the MSEs of both models on the effective damage variable seem sensitive to the data sizes. Again, the vanilla model is outperformed by our proposed model in the damage prediction, illustrating the importance of incorporating damage constraint. 

\begin{figure}
    \centering
    \begin{subfigure}[b]{0.85\textwidth}
        \centering
        \includegraphics[width=\textwidth]{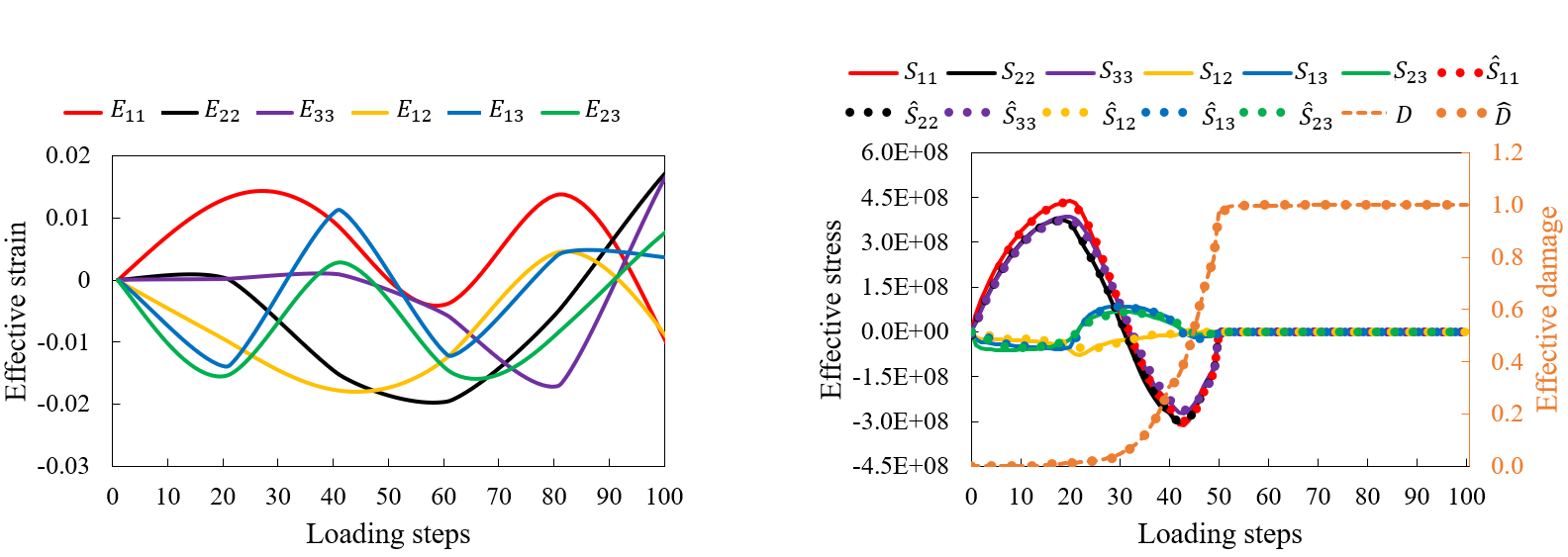}
        \caption{Random strains and effective responses of test sample 1}
    \end{subfigure}
    \begin{subfigure}[b]{0.85\textwidth}
        \centering
        \includegraphics[width=\textwidth]{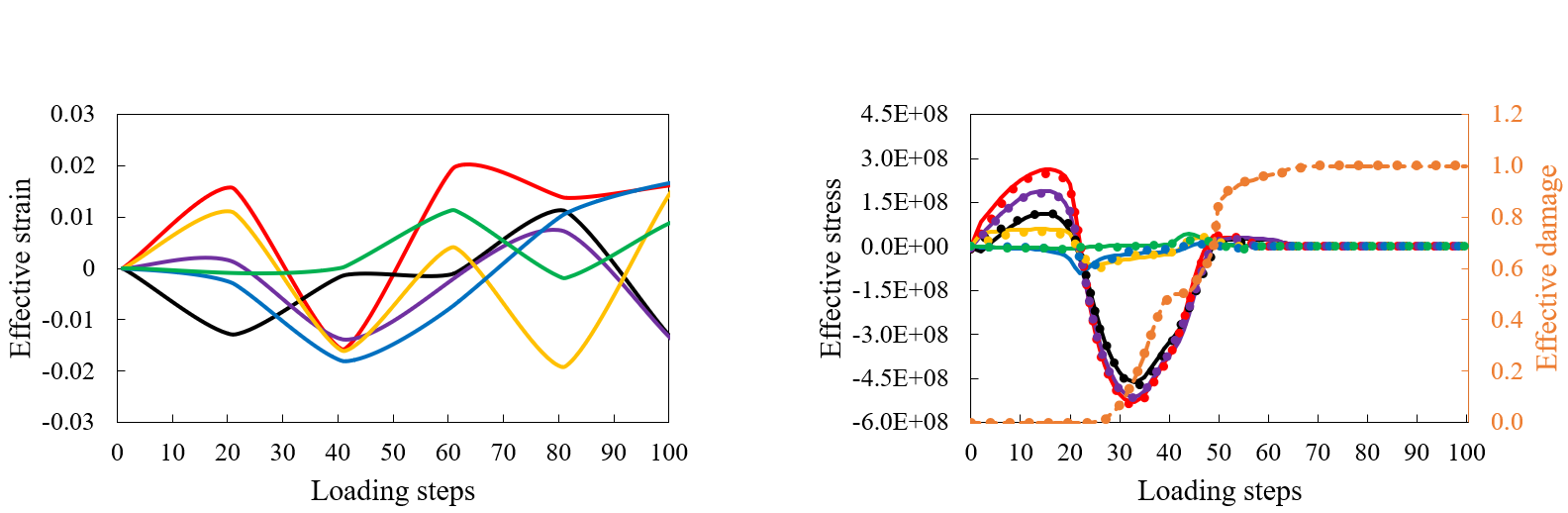}
        \caption{Random strains and effective responses of test sample 2}
    \end{subfigure}
    \begin{subfigure}[b]{0.85\textwidth}
        \centering
        \includegraphics[width=\textwidth]{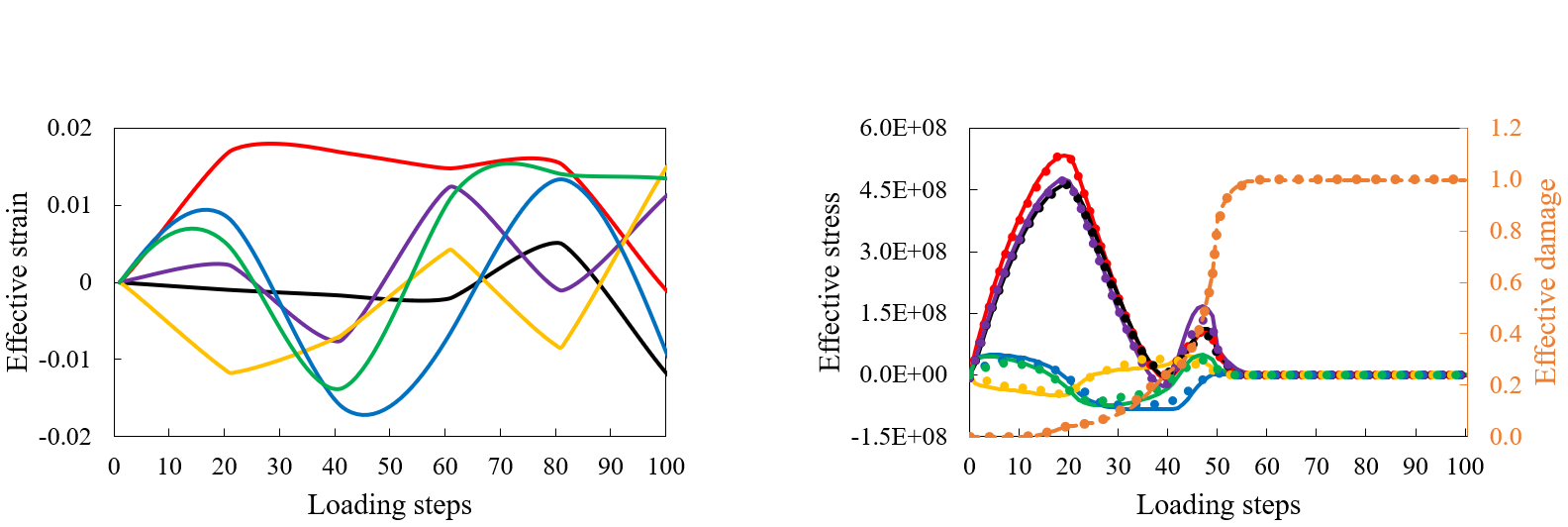}
        \caption{Random strains and effective responses of test sample 3}
    \end{subfigure}
    \begin{subfigure}[b]{0.85\textwidth}
        \centering
        \includegraphics[width=\textwidth]{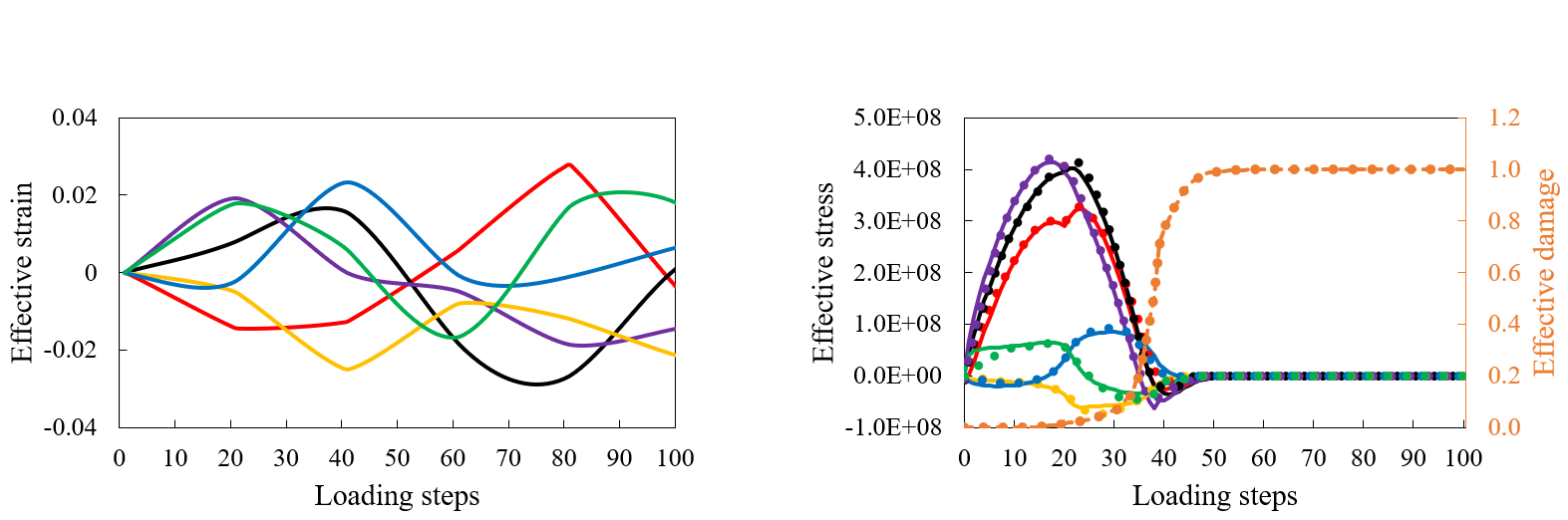}
        \caption{Random strains and effective responses of test sample 4}
    \end{subfigure}
    \caption{\textbf{RVE deformation and responses:} We demonstrate four test samples with different strain paths where each path contains six strain components evolving in the $100$ sequential loading steps (the first column), and their associated effective stress and damage are compared between surrogate predictions and the ground truth (the second column). 
    } \label{fig:RNN_RVE_StrainPaths}
\end{figure}

To visualize the predicted effective responses by our physics-constrained surrogate model, we randomly select four strain paths from the testset, and compare the predictions on their responses by our RNN against the ground truth as shown in Figure \ref{fig:RNN_RVE_StrainPaths}. It is evident that for the four different strain paths with very complex loading histories, our RNN model is able to provide close estimations of both the effective stress and damage variables to the ground truth. In particular, we observe that as the damage variable increases to $1.0$ amid material deformation, the magnitudes of the effective stress are correspondingly reduced, which indicates a significant loss in the RVE's load-carrying capacity. 

\subsubsection{Impacts of teacher forcing} \label{subsubsec:exp_teacherforcing}
As discussed in Section \ref{subsec:RNN_constraints}, teacher forcing, which augments ground truth or predictions from previous steps to the input at the current step, may provide us with more accurate predictions. In order to quantify the impacts of the teacher forcing, we compare the total MSE of the predicted stress and damage variables over the testing dataset between our physics-constrained RNN model with and without teacher forcing technique in Table \ref{tab:MSE_teacherforcing}. 

We implement two teacher forcing models here: the first model with the number of look back step (NLB) of one, and the second model with the NLB of five. From Table \ref{tab:MSE_teacherforcing}, we observe that compared to the model without teacher forcing, the two teacher forcing models improves prediction accuracy by reducing the total MSE by $21.4\%$ (NLB=1) and $24.2\%$ (NLB=5), respectively. Comparing the individual MSEs, we find that the teacher forcing reduces the prediction error of effective damage while not for the stress. Therefore, in the single scale RVE simulations, the teacher forcing improves our RNN's overall prediction accuracy.

\begin{table} 
    \centering
    \begin{tabular}{c c c c} 
        \hline
        & No teacher forcing & Teacher forcing (NLB=1) & Teacher forcing (NLB=5) \\ 
        \hline
        Total MSE & $2.52$ & $1.98$ & $1.91$ \\ 
        \hline
        $\mathrm{MSE_S}$ & $4.14$ & $4.62$ & $4.54$ \\ 
        \hline
        $\mathrm{MSE_D}$ & $17.30$ & $9.67$ & $9.24$ \\ 
        \hline
        \caption{\textbf{Impacts of the teacher forcing on the prediction accuracy:} Comparison of the total MSE and individual MSEs ($10^{-5}$) of the predicted effective stress and damage variables over the testing set between our physics constrained RNN model (trained by $29,800$ sequences) without teacher forcing and the same model with teacher forcing using one or five look back steps.} 
        \label{tab:MSE_teacherforcing}
    \end{tabular}
\end{table}

\subsection{Surrogate for multiscale damage analysis}\label{subsec:exp_Lbracket}

After we demonstrate that our RNN can accurately predict microstructural effective responses under various deformation paths in Section \ref{subsec:exp_RVE}. We can now use the RNN as a faithful surrogate to replace the computationally expensive microstructural analysis in multiscale simulations. 

\begin{figure}
    \centering
    \begin{subfigure}[b]{0.5\textwidth}
        \centering
        \includegraphics[width=\textwidth]{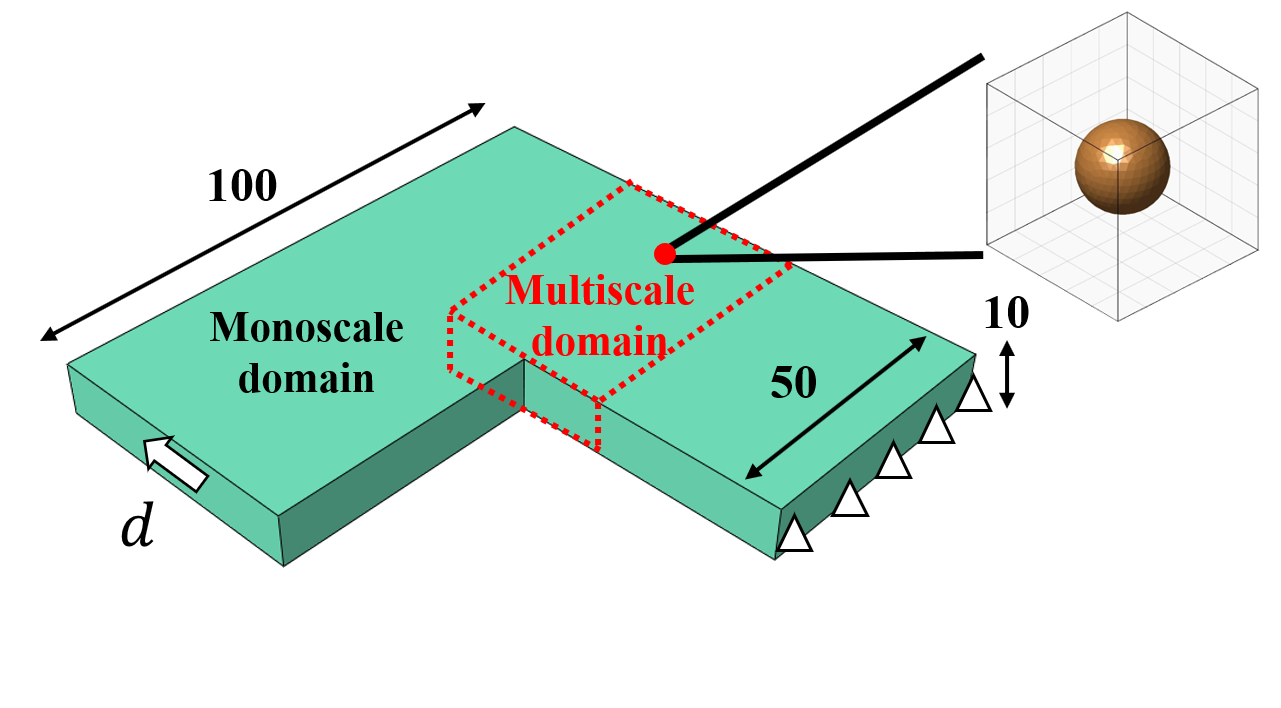}
        \caption{Geometry, dimensions (unit:mm) and boundary conditions}
    \end{subfigure}
    \begin{subfigure}[b]{0.45\textwidth}
        \centering
        \includegraphics[width=\textwidth]{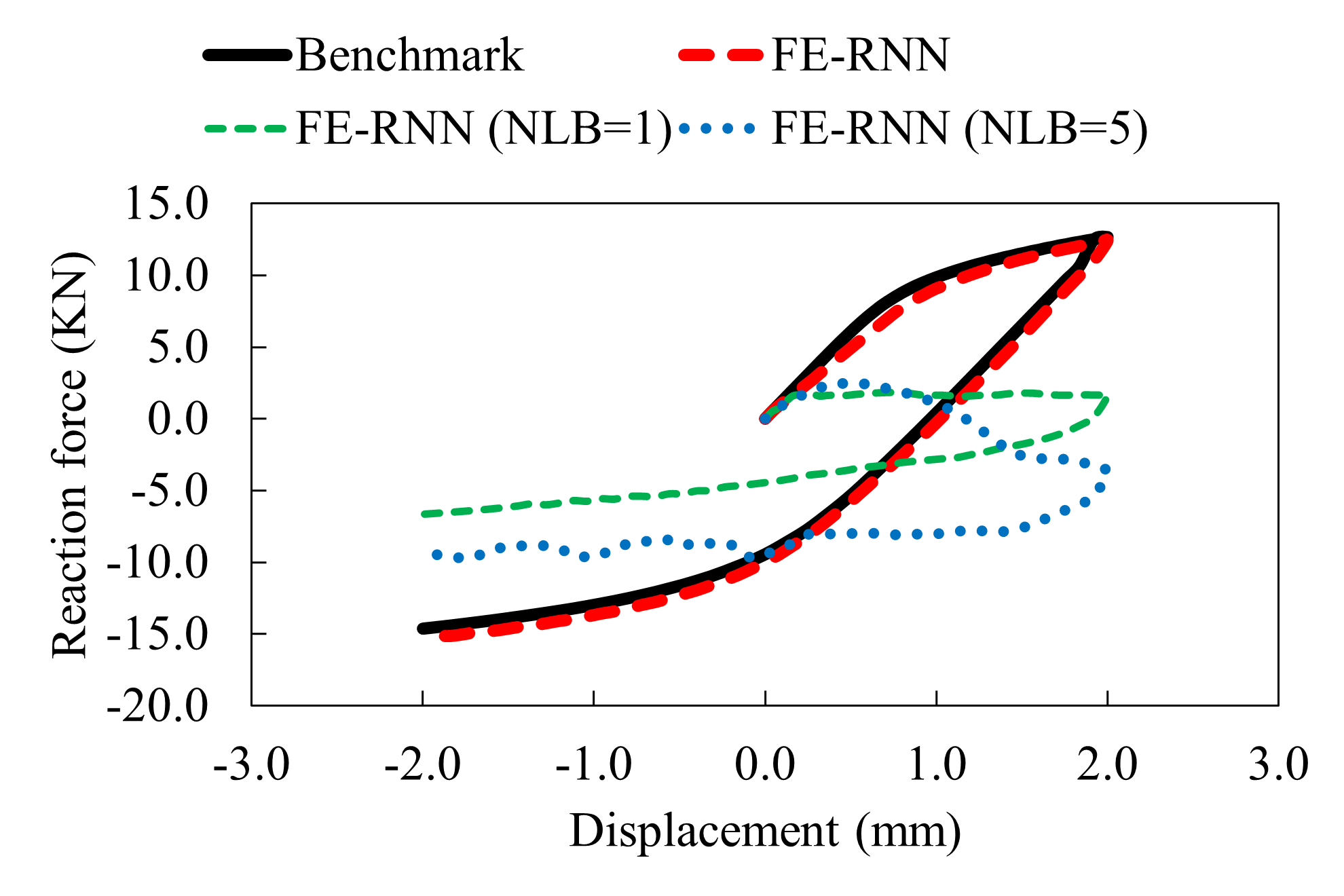}
        \caption{Reaction force and displacement}
    \end{subfigure}
    \caption{\textbf{Multiscale model of the L-shape bracket:} (\textbf{a}) Every macroscale IP in the multiscale domain is associated with a microscale porous RVE; and (\textbf{b}) Comparison of the load-displacement curves of the elasto-plastic hardening behaviors between benchmark and the proposed FE-RNN models with and without teacher forcing.}
    \label{fig:Lbracket_model}
\end{figure}

Our first multiscale simulation is performed on a 3D L-shape bracket as shown in Figure \ref{fig:Lbracket_model}(a). The bracket is subject to a Dirichlet boundary condition on the left side while its right surface is fully fixed. We assume the bracket contains a multiscale domain around the sharp corner where we expect strain concentrations to occur. Specifically, we assume each IP of the multiscale domain to associate with a porous RVE as illustrated in Figure \ref{fig:RVE_geom_mesh}. To save computational costs, we assume there is a mono-scale domain outside the multiscale domain. We assume the IPs of the mono-scale domain are not associated to any RVEs. For computational analysis, we mesh the bracket by $5,300$ tetrahedral elements of reduced integration. The multiscale domain contains $360$ elements, and the each of them are associated with an RVE that is decomposed by $1,200$ clusters.   

We first demonstrate the accuracy of the multiscale model for elasto-plastic hardening behaviors under complex cyclic loading histories. To this end, we let the bracket subject to a loading-unloading-reloading condition by setting the Dirichlet boundary condition as $d = 0 \rightarrow 2 \rightarrow 0 \rightarrow -2$ mm. We compare the resulting reaction force and displacement curves between our proposed FE-RNN approach and the benchmark FE-ROM method in Figure \ref{fig:Lbracket_model}(b).

We note that there are three data-driven models whose solutions are present in Figure \ref{fig:Lbracket_model}(b): our physics-constrained surrogate model without teacher forcing (FE-RNN), and the surrogates with teacher forcing using one look back step (NLB=1) and five look back steps (NLB=5). We find that while the FE-RNN model provides very close solutions to the benchmark, the solutions of the models with teacher forcing are not trustworthy. The underlying reason is that, in nonlinear multiscale simulations, the iterative error of macro responses depends on the surrogate accuracy on all IPs within the multiscale region from the previous and current loading steps. More specifically, small discrepancy between benchmark and the surrogate at each local IP accumulates to the global error of the macro responses. Additionally, when teacher forcing feeds such error from previous loading steps back to the RNN inputs at the current step, the global error continues to grow along iterations and eventually results in solution divergence. We emphasize that this multiscale simulation is different from the scenario in Section \ref{subsubsec:exp_teacherforcing} where we use RNN to surrogate the effective responses of a single scale RVE which shows marginally small errors. It is for this reason, we adopt our FE-RNN model without teacher forcing for all multiscale simulations in the following experiments.

\begin{figure} 
    \centering
    \includegraphics[width = 0.8\textwidth]{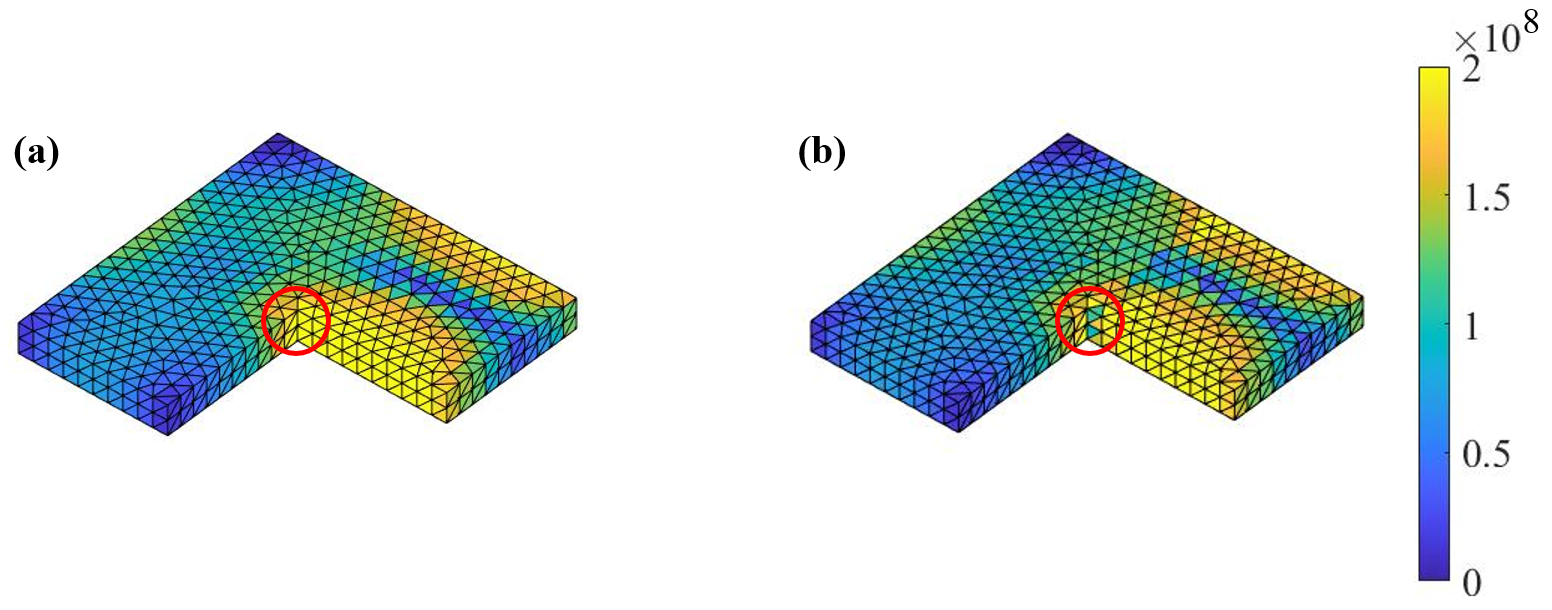}
    \caption{\textbf{Comparison of Von-Mises stress distributions in hardening simulation:} (\textbf{a}) The distribution of Von-Mises stresses (unit: Pa) by benchmark; and (\textbf{d}) Von-Mises stresses by the proposed FE-RNN model.
    } \label{fig:Lbracket_hardening_stress}
\end{figure}

We compare the Von-Mises stress distributions between the benchmark and our FE-RNN model by setting the boundary condition as $d = -2$ mm in the Figure \ref{fig:Lbracket_hardening_stress}. We observe a generally good agreement between the two models despite minor local discrepancy at the sharp corner as highlighted in the figure. A plausible reason for the discrepancy is that RNN's prediction accuracy decreases for extreme values with insufficient training data points or poor extrapolation ability. 

\begin{figure}
    \centering
    \begin{subfigure}[b]{0.35\textwidth}
        \centering
        \includegraphics[width=\textwidth]{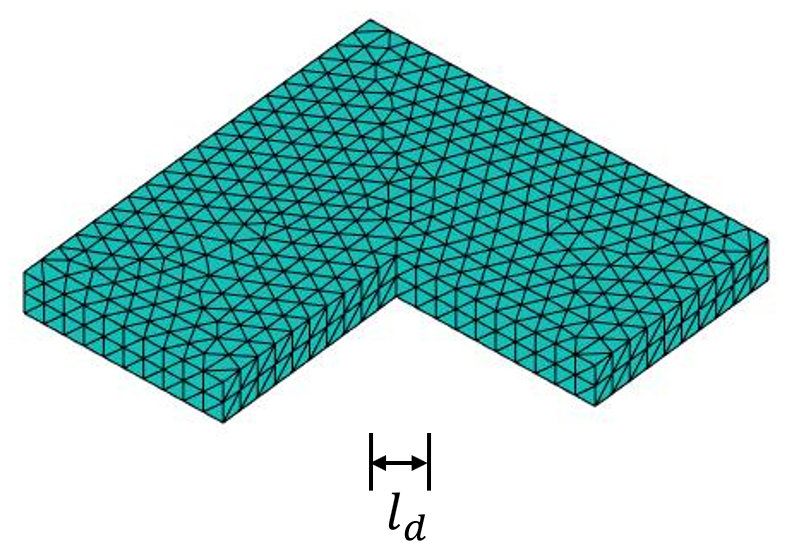}
        \caption{Macroscale mesh and strain localization bandwidth (\textit{$l_d$}) }
    \end{subfigure}
    \begin{subfigure}[b]{0.45\textwidth}
        \centering
        \includegraphics[width=\textwidth]{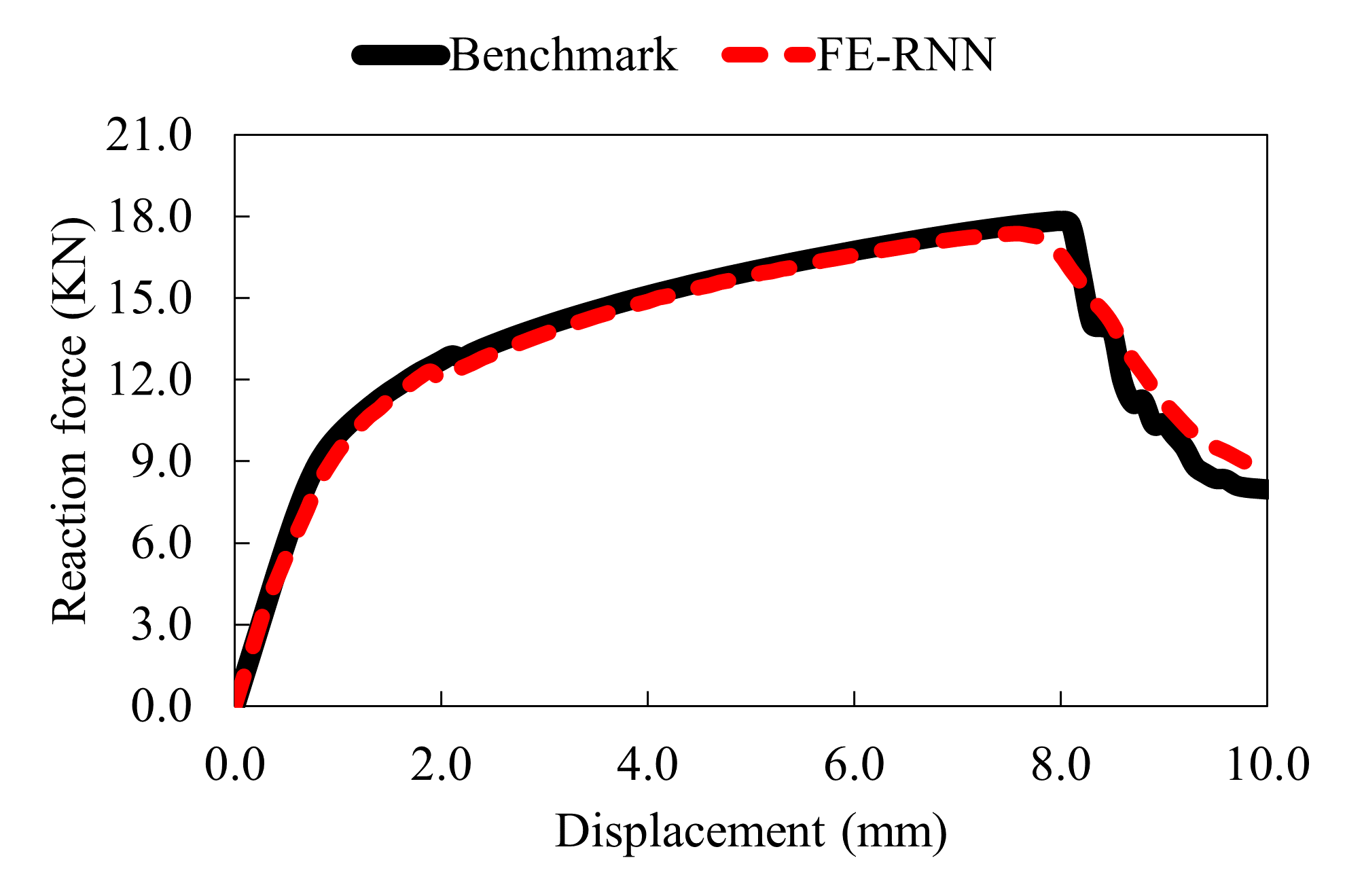}
        \caption{Reaction force and displacement}
    \end{subfigure}
    \caption{\textbf{Damage simulations of the L-shape bracket:} (\textbf{a}) The macroscale discretization and strain localization bandwidth applied in the damage function; and (\textbf{b}) Comparison of the softening load-displacement curves between benchmark and the proposed FE-RNN model.}
    \label{fig:Lbracket_softening}
\end{figure}

Our second multiscale experiment is to simulate the elasto-plastic hardening and softening on the same L-shape bracket where its Dirichlet boundary condition is set as $d = 10$ mm. To prevent the occurrence of the non-physical single-layer fracture bands as discussed in Section \ref{subsec:strainSoftening}, we apply a non-local damage function (see Equation \ref{eqn:nonlocal_damageFunc}) with the strain localization bandwidth of $l_d = 15$ mm. We illustrate its length comparison to the mesh size of the bracket in the Figure \ref{fig:Lbracket_softening}(a). The force-displacement curves are compared in \ref{fig:Lbracket_softening}(b) where the general trends of the two methods match well especially for the hardening section. Minor discrepancy manifests in the softening regime where the data-driven model tends to break earlier which underestimates the component's load-carrying capacity by about $2.5\%$. The underlying reason is that softening behaviors dramatically increase the complexity of the material's governing equations, as it increases the difficulty for our RNN to match with the benchmark model. 

\begin{figure} 
    \centering
    \includegraphics[width = 0.8\textwidth]{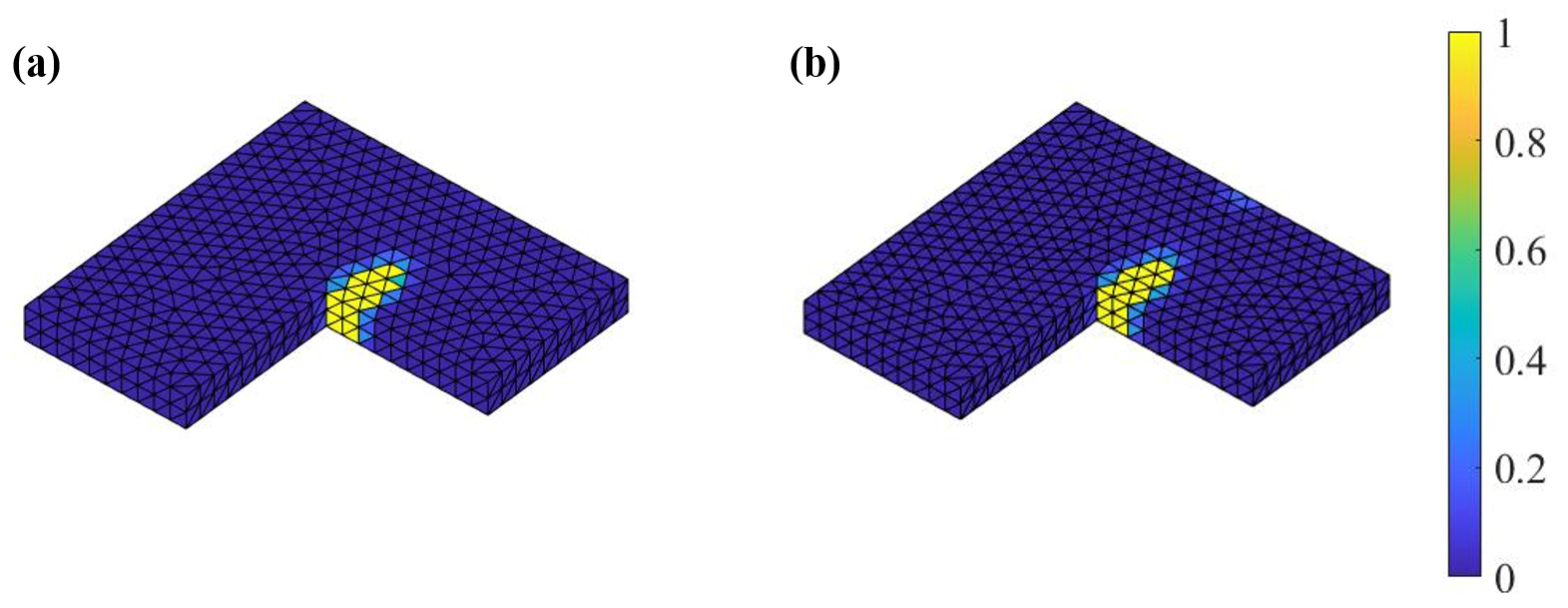}
    \caption{\textbf{Comparison of damage patterns:} (\textbf{a}) The distribution of benchmark damage variables; and (\textbf{d}) Damage variables via our FE-RNN model where color yellow indicates a full material fracture while blue represents an intact state. 
    } \label{fig:Lbracket_softening_damage}
\end{figure}

\begin{figure} 
    \centering
    \includegraphics[width = 0.8\textwidth]{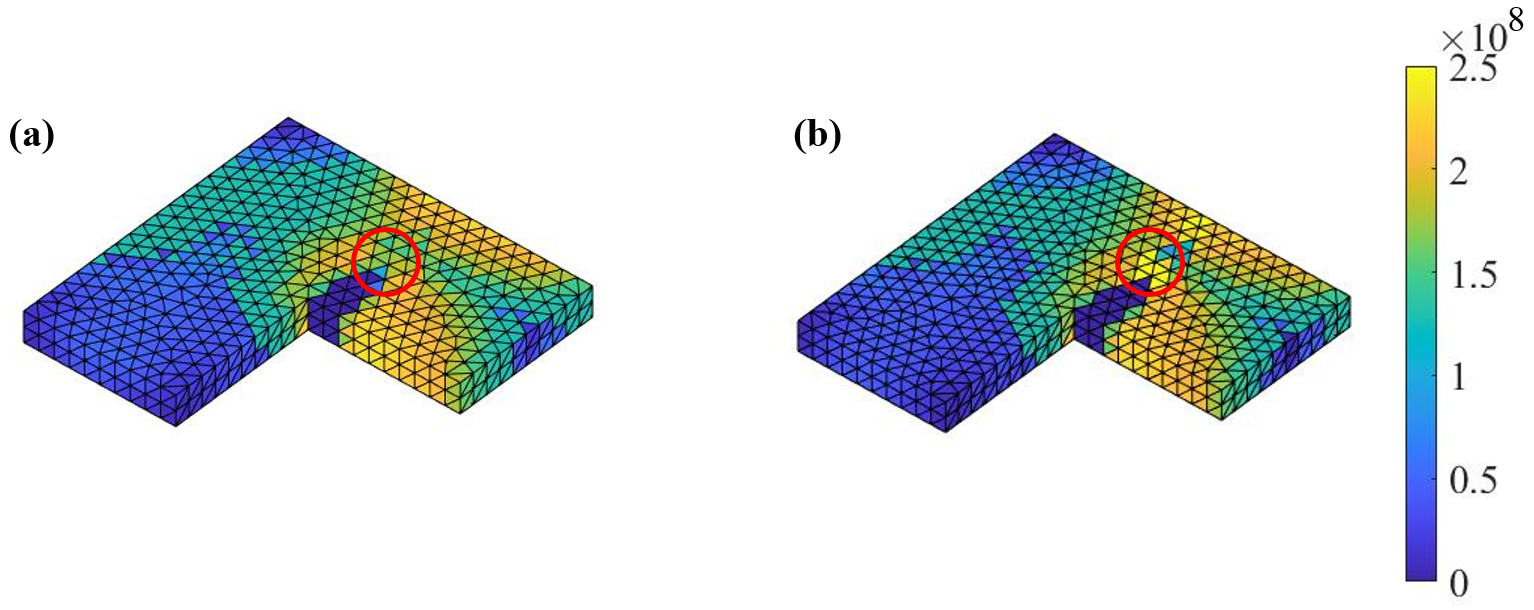}
    \caption{\textbf{Comparison of Von-Mises stress distributions in softening simulation:} (\textbf{a}) The distribution of benchmark Von-Mises stresses (unit: Pa); and (\textbf{d}) Von-Mises stresses by our FE-RNN model.
    } \label{fig:Lbracket_softening_stress}
\end{figure}

We further compare the distributions of damage variables and Von-Mises stresses when the boundary condition is set as $d = 10$ mm in Figures \ref{fig:Lbracket_softening_damage} and Figure \ref{fig:Lbracket_softening_stress}, respectively. We see both field variables' distributions have good agreements between the two approaches. In Figure \ref{fig:Lbracket_softening_damage}, we observe fracture bands initiate from the sharp corner and stretch towards the right surface. We can also clearly see the effects of imposing non-local damage functions in avoiding non-physical single-layer fracture bands. As for the stress distributions in Figure \ref{fig:Lbracket_softening_stress}, both approaches show that the local stress values are significantly reduced within fracture bands that indicates a loss of load-carrying capacity in the fractured elements. We note minor discrepancy of local stresses at the front tip of the fracture bands between the two methods: while the benchmark indicates relatively low stresses at the highlighted region, our FE-RNN model suggests stress concentrations which triggers more damage if the component is further deformed. We can use such stress concentrations to understand the reason why our data-driven model predicts an earlier damage occurrence than the benchmark in Figure \ref{fig:Lbracket_softening}(b).

\begin{figure}
    \centering
    \begin{subfigure}[b]{0.45\textwidth}
        \centering
        \includegraphics[width=\textwidth]{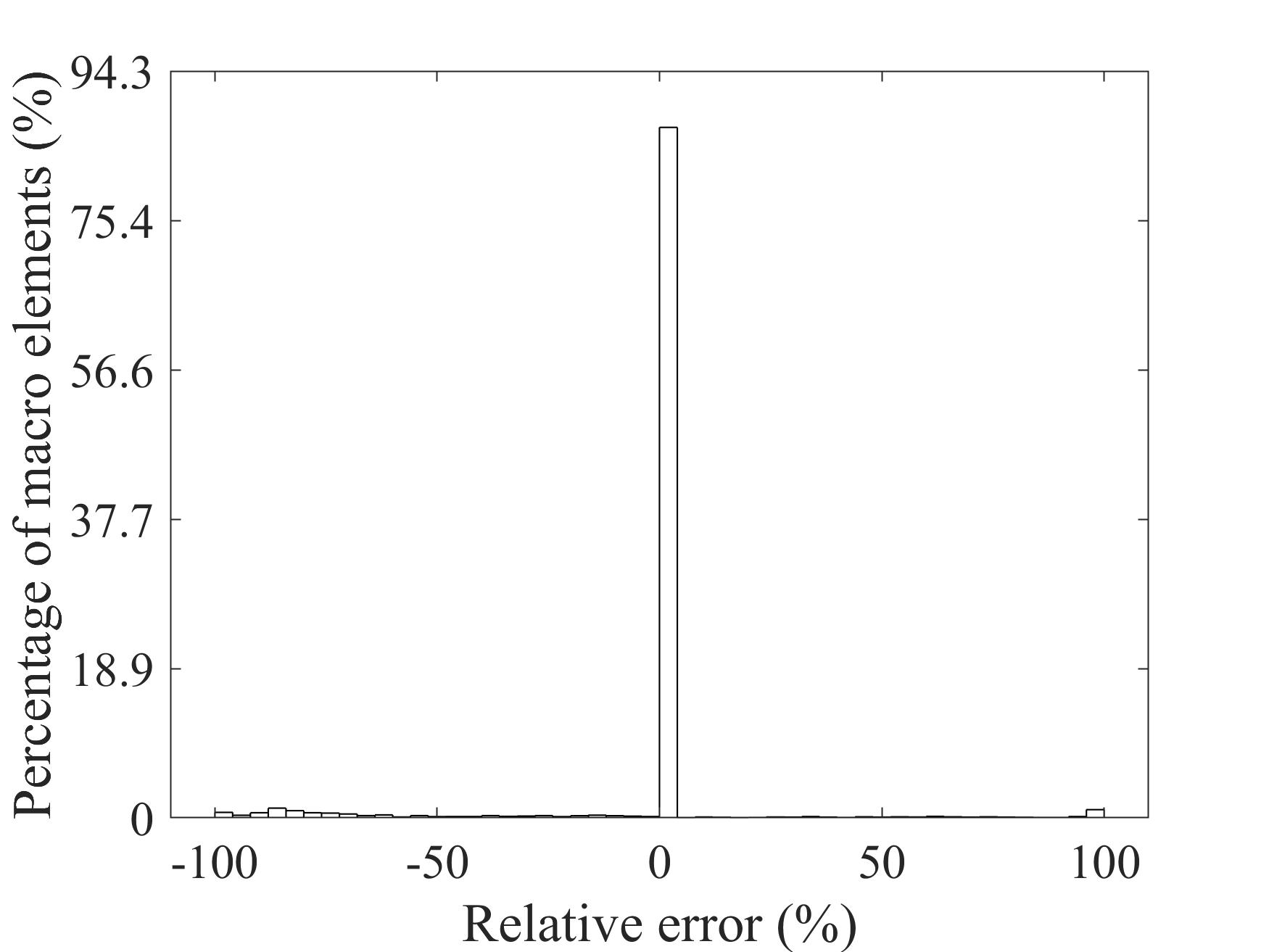}
        \caption{Relative errors of damage variables}
    \end{subfigure}
    \begin{subfigure}[b]{0.45\textwidth}
        \centering
        \includegraphics[width=\textwidth]{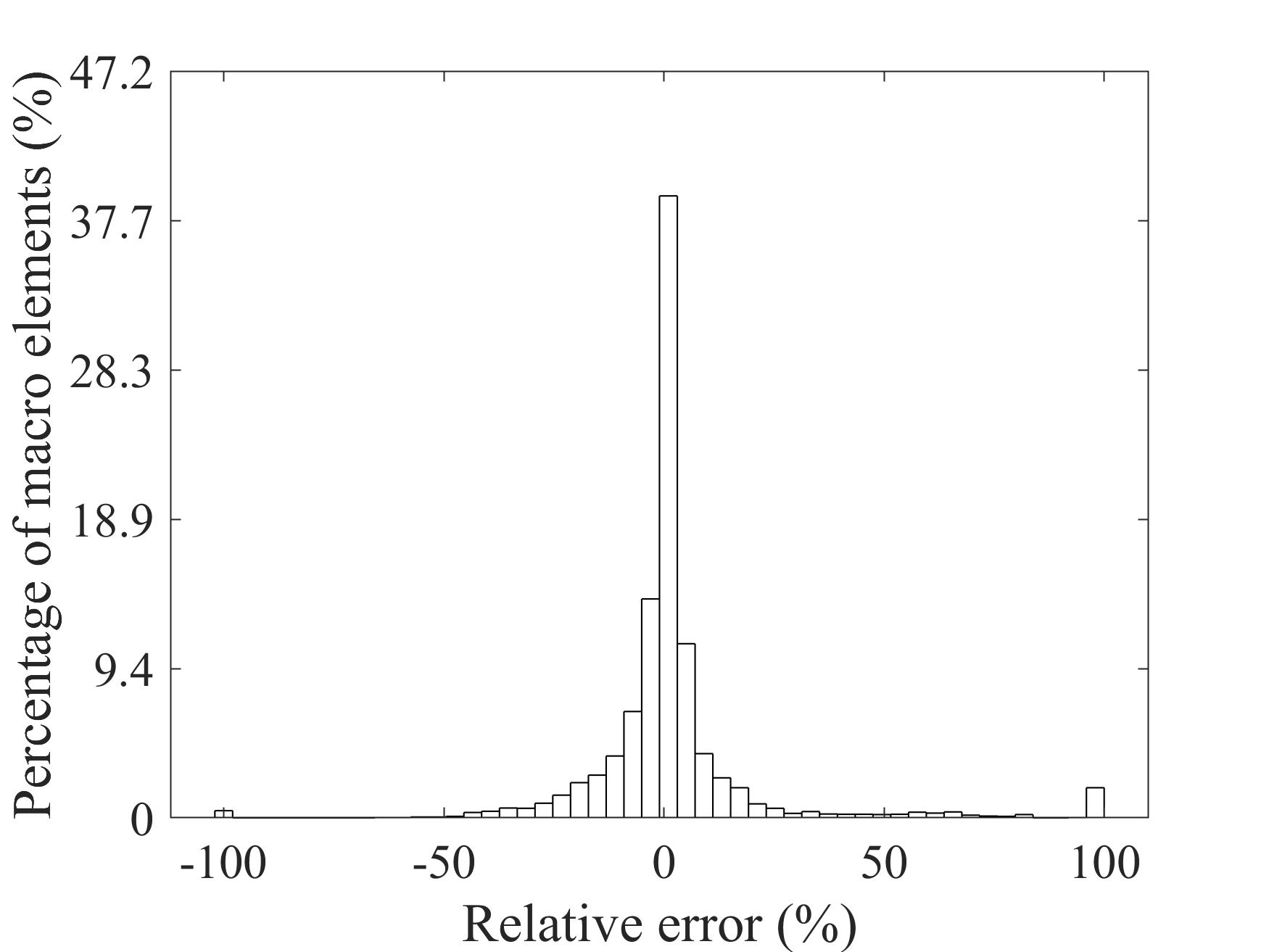}
        \caption{Relative errors of Von-Mises stresses}
    \end{subfigure}
    \caption{\textbf{Histogram of relative errors of field variables:} (\textbf{a}) Relative errors of the values of damage variables between benchmark and our FE-RNN model in Figure \ref{fig:Lbracket_softening_damage}; and (\textbf{b}) Relative errors of the values of Von-Mises stresses in Figure \ref{fig:Lbracket_softening_stress}.}
    \label{fig:Lbracket_StressDamage_errhist}
\end{figure}

The discrepancy between the proposed model and benchmark can be further quantified by the histogram of errors as shown in Figure \ref{fig:Lbracket_StressDamage_errhist}. In terms of damage variables, it is quite clear from Figure \ref{fig:Lbracket_StressDamage_errhist}(a) that the two approaches yield identical solutions in the majority (more than $80\%$) of elements. Based on the distributions of stress errors in Figure \ref{fig:Lbracket_StressDamage_errhist}(b), we can see relatively large errors in about $2\%$ of all elements. It should be noted, however, for most elements, their absolute errors are smaller than about $10\%$ between the two methods, which still suggests a good agreement.  

In order to quantify computational costs, we break down the computational costs of different steps in this multiscale model as shown in Table \ref{tab:Lbracket_time}. Comparing to the mechanistic models (FE-ROM and FE\textsuperscript{2}), our data-driven model (FE-RNN) requires additional costs on database generation and model training. Even though expensive, we only need to perform the two steps once, and after training we can deploy the trained RNN model for any multiscale simulations without extra costs. In terms of the online clock time, our model shows superior efficiency to the mechanics models (FE-ROM and FE\textsuperscript{2}) with about $125 \times$ and $1240 \times$ accelerations, respectively. It is noted we do not directly perform the FE\textsuperscript{2} due to its prohibitively demanding costs, its computational time is estimated by comparing to the ROM on a smaller multiscale simulation whose time comparison is demonstrated in Figure \ref{fig:cube_sols} of Appendix \ref{sec:appendix_A}. We also note that, while we perform the training process on two GPU processors, we carry out both the database generation of the FE-RNN and the multiscale simulations of FE-ROM and FE\textsuperscript{2} by paralleling $60$ CPU cores with $360$ GB RAM on a HPC. Comparatively, our proposed RNN model only needs four CPU cores on a desktop computer for the multiscale computation, providing feasible solutions to the engineers without accessibility to large computational resources.   

\begin{table} 
    \centering
    \begin{tabular}{c c c c} 
        \hline
        & FE-RNN (proposed) & FE-ROM (benchmark) & FE\textsuperscript{2} (estimation) \\ 
        \hline
        Database generation & $239.5 \times 60$ CPU-hour & - & - \\ 
        \hline
        RNN training & $4.3 \times 2$ GPU-hour & - & - \\ 
        \hline
        Multiscale simulation & $0.4 \times 4$ CPU-hour & $49.8 \times 60$ CPU-hour & $494.0 \times 60$ CPU-hour \\ 
        \hline
        \caption{\textbf{Breakdown of computational time of the L-shape bracket model:} Despite considerable costs in database generation and training of the RNN, its efficiency (measured by clock time) in the multiscale simulations is about $125 \times$ and $1240 \times$ higher than ROM and FE\textsuperscript{2}, respectively, where the time estimation of FE\textsuperscript{2} comes from the time comparison in Figure \ref{fig:cube_sols}(b).} 
        \label{tab:Lbracket_time}
    \end{tabular}
\end{table}

\subsection{Mesh convergence study by multiscale damage surrogate}\label{subsec:exp_DoubleNotch}

One of the major challenges of using continuum mechanics to simulate softening behaviors is to prevent fracture bands residing in single element wide layers. One popular solution is to apply non-local functions to constrain damage patterns at different spatial discretization levels. To this end, we apply the proposed RNN model to a new 3D model in this section, and assess its robustness by a mesh convergence study on damage behaviors.  

\begin{figure}
    \centering
    \begin{subfigure}[b]{0.35\textwidth}
        \centering
        \includegraphics[width=\textwidth]{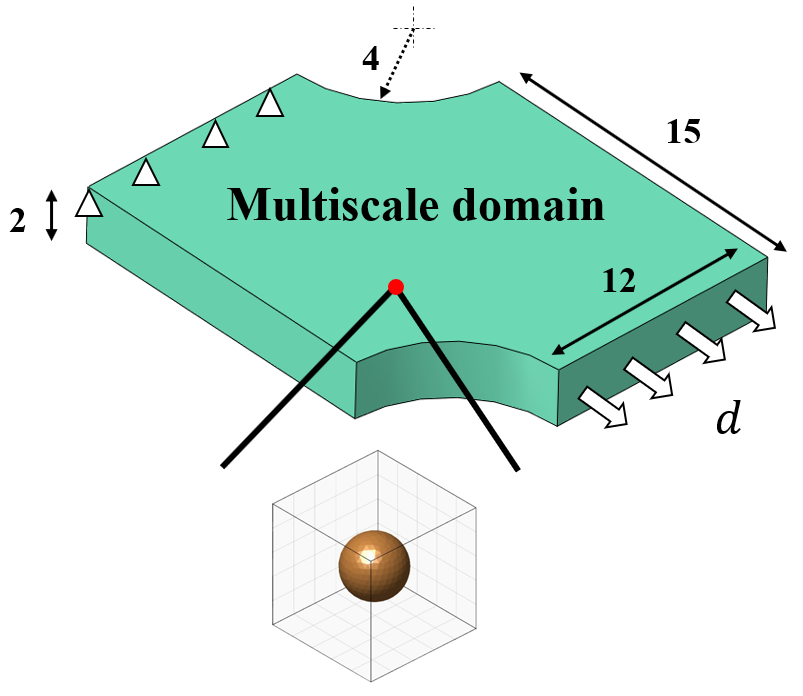}
        \caption{Geometry, dimensions (unit: mm) and boundary conditions}
    \end{subfigure}
    \begin{subfigure}[b]{0.45\textwidth}
        \centering
        \includegraphics[width=\textwidth]{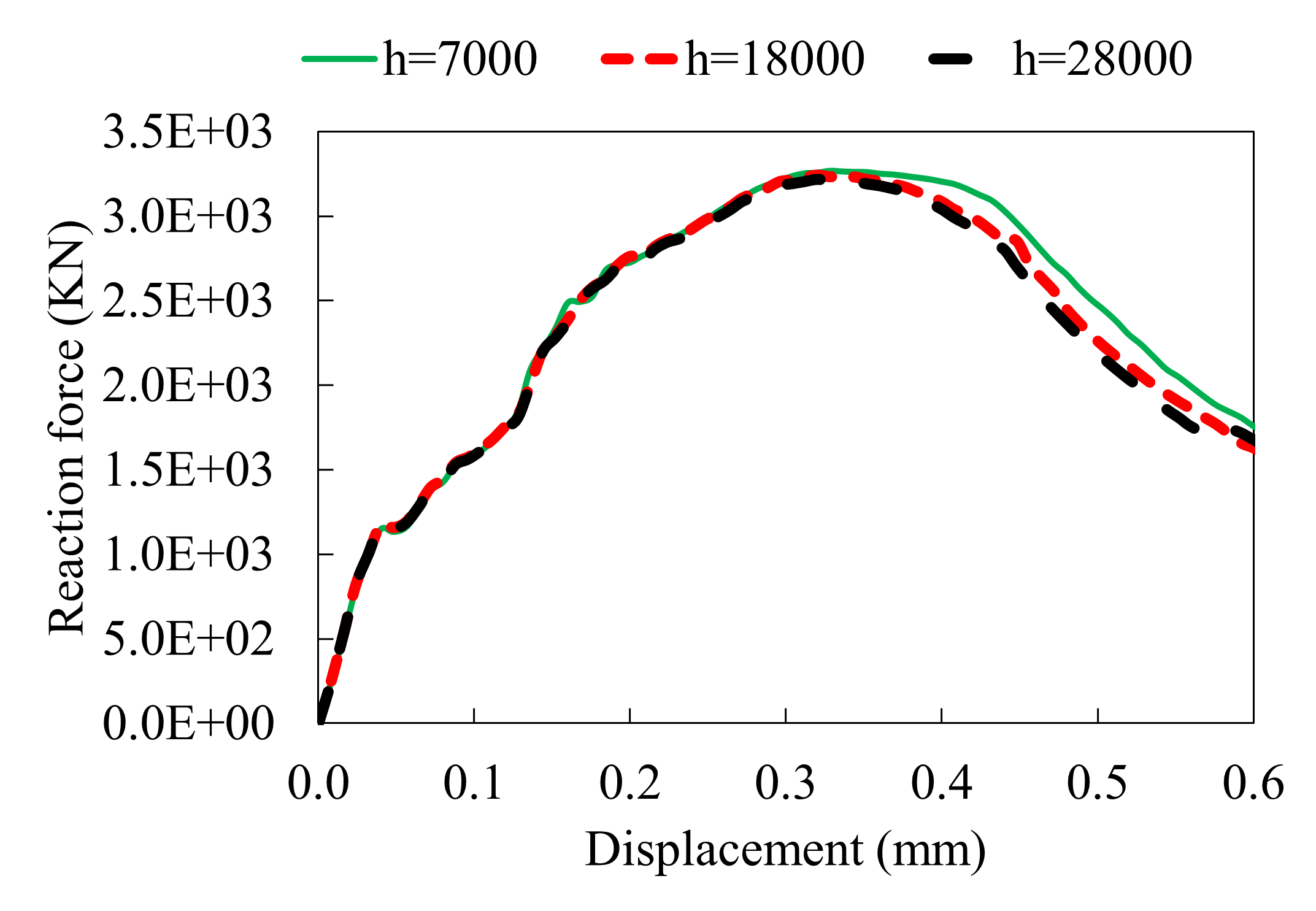}
        \caption{Reaction force and displacement}
    \end{subfigure}
    \caption{\textbf{Multiscale model of double-notched specimen:} (\textbf{a}) Every macroscale integration point is associated with a microscale porous RVE; and (\textbf{b}) Convergence study of the softening load-displacement curves with different macro discretization levels.}
    \label{fig:notch_model}
\end{figure}

The geometry, dimensions and boundary conditions of the double notched specimen is demonstrated in Figure \ref{fig:notch_model}(a). The specimen is fully fixed at the left surface, and its right surface is subject to an extension with a displacement boundary condition of $d = 0.6$ mm. In this experiment, the whole specimen is assumed as the multiscale domain where each IP is associated with a porous RVE. For the mesh convergence study, we discretize the macro specimen with three different mesh sizes: a coarse mesh with $7,000$ elements, a medium mesh with $18,000$ elements, and a fine mesh with $28,000$ elements. 

We demonstrate the reaction force-displacement curves as in Figure \ref{fig:notch_model}(b). From the figure, we notice that the three mesh levels achieve very close elasto-plastic hardening responses, but are slightly different in the softening regime. Specifically, we note that with the mesh level increasing from medium to fine level, the post-failure force-displacement responses tend to converge. 

\begin{figure}
    \centering
    \begin{subfigure}[b]{0.9\textwidth}
        \centering
        \includegraphics[width=\textwidth]{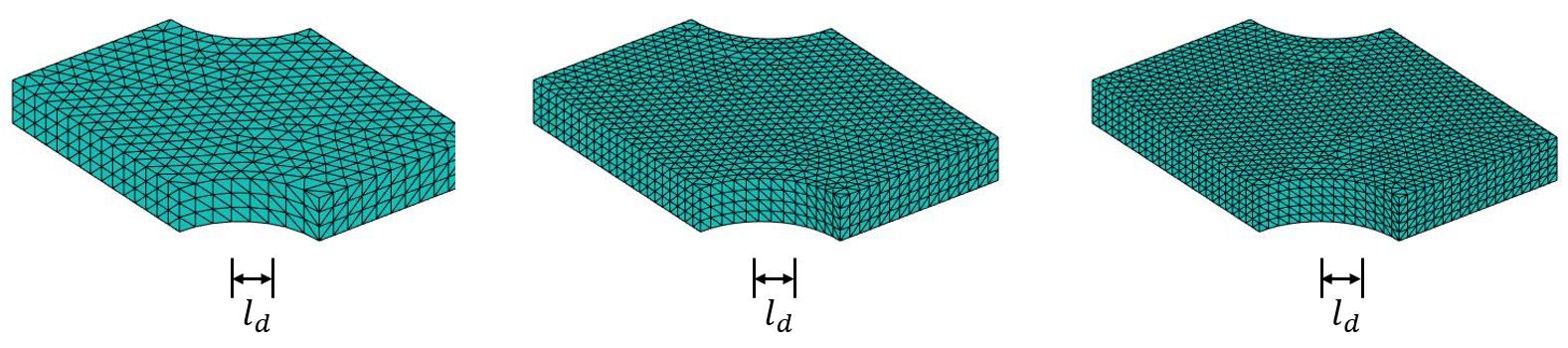}
        \caption{Mesh sizes and strain localization bandwidth (\textit{$l_d$}) }
    \end{subfigure}
    \\[\smallskipamount]
    \begin{subfigure}[b]{0.9\textwidth}
        \centering
        \includegraphics[width=\textwidth]{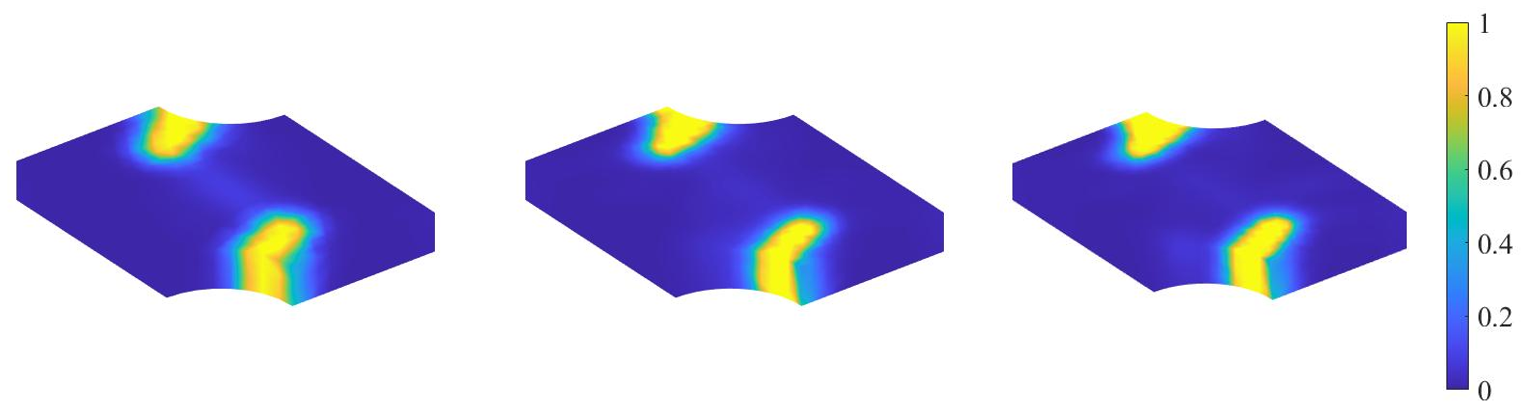}
        \caption{Damage patterns}
    \end{subfigure}
    \\[\smallskipamount]
    \begin{subfigure}[b]{0.9\textwidth}
        \centering
        \includegraphics[width=\textwidth]{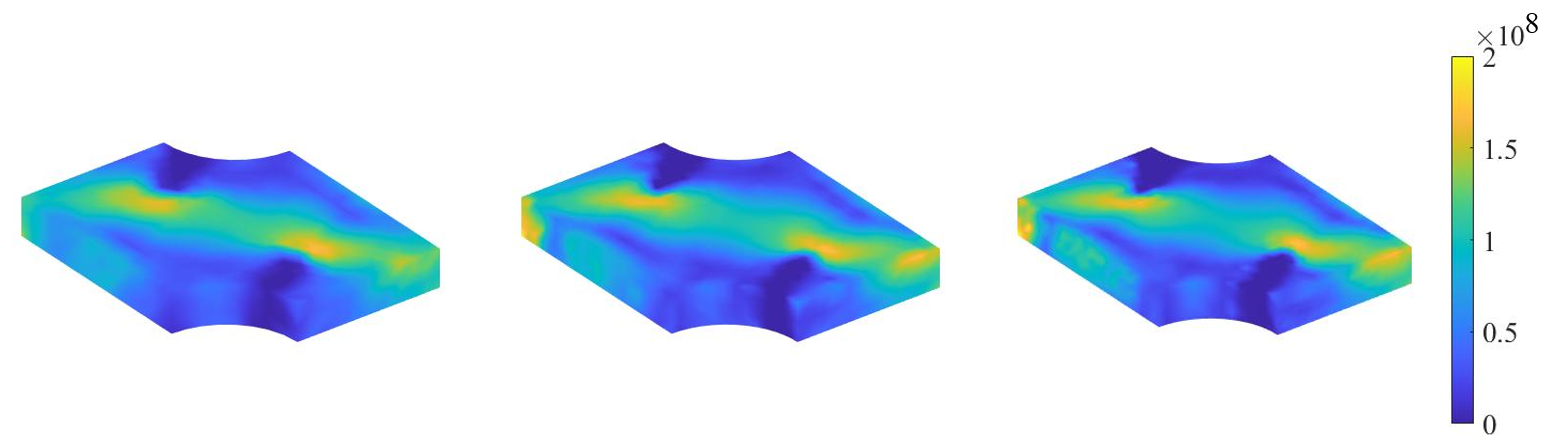}
        \caption{Von-Mises stress distributions}
    \end{subfigure}
    \caption{\textbf{Mesh convergence study of field variables by our FE-RNN:} With the increase of the number of macro-elements in (\textbf{a}), both the damage patterns in (\textbf{b}) and stress distributions in (\textbf{c}) show convergence.}
    \label{fig:notch_sols}
\end{figure}

We can also observe the convergence by inspecting the stress distributions and damage patterns from Figure \ref{fig:notch_sols} with the displacement boundary condition of $d = 0.6$ mm. On the one hand, from Figure \ref{fig:notch_sols}(a) and (b), we can clearly see that at all mesh levels, the damage initiates from the inner circular surfaces and propagates across the specimen as it is further loaded. The influence of imposing non-local function is evident: it not only  successfully avoids non-physical single-element-wide damage layers, but also constrains the fracture bandwidth regardless of the mesh sizes. On the other hand, with finer meshes (using $18,000$ and $28,000$ elements), we fine that stress concentrations consistently appear at both fracture front tips and around sharp corners.

We report the simulation time of this multiscale double notched specimen in Table \ref{tab:NotchModel_time}. We emphasize that due to the superior efficiency, we can apply our trained FE-RNN to any multiscale models with no extra costs of data generation and model training. In specific, our trained FE-RNN mode is so memory lightweight that it can run on a desktop with four CPU cores and 16GB RAM. Base on the time comparison in Table \ref{tab:Lbracket_time}, simulation of the multiscale model with 28,000 elements requires the clock time of 1,304.8 hours (54.4 days) and 12,942.8 hours (539.3 days) by paralleling 60 CPU cores with 360 GB RAM for FE-ROM and FE\textsuperscript{2} methods, respectively.

\begin{table} 
    \centering
    \begin{tabular}{ c c } 
        \hline
        Number of macro-elements & Multiscale simulation time \\
        \hline
        $7,000$ & $3.1 \times 4$ CPU-hour  \\ 
        \hline
        $18,000$ & $7.5 \times 4$ CPU-hour \\ 
        \hline
        $28,000$ & $10.5 \times 4$ CPU-hour \\ 
        \hline
        \caption{\textbf{Computational time of the double notched model:} The multiscale simulations of the double notched specimen are performed by the proposed FE-RNN model on a modest desktop with four CPU cores where the computational costs of one-time data generation and model training are reported in Table \ref{tab:Lbracket_time}.} 
        \label{tab:NotchModel_time}
    \end{tabular}
\end{table}

\section{Conclusions} \label{sec:conclusions}

In this contribution, we propose a physics-constrained deep learning model to surrogate the homogenized nonlinear path-dependent microstructural material behaviors in 3D large-scale multiscale simulations. Our deep learning model builds on the RNN which is trained on a database containing sequences of random microstructural deformation and responses. To reduce the computational costs of the database generation while preserving data generality, we create the sequential data by using GP interpolations in a DoE that is confined by two DoE constraints to reduce the number of sampling sequences. In addition, we adopt mechanistic ROM for the simulation of microstructural effective responses to reduce the computational cost per deformation sequence. 

To facility the training and inference of our surrogate model, we modify RNN architecture by incorporating two physics constraints. The first constraint comes from thermodynamically consistent microstructural energy analysis, and it is reinforced as a soft constraint by adding a penalty term to our RNN's loss function. The second constraint enforces the irreversible nature of damage processes, and we implement it as a hard constraint by directly manipulating the temporal variation of the outputs within RNN architecture. In addition, we incorporate the teacher forcing technique into our RNN model and demonstrate its impacts in both the single scale and multiscale simulations.

We validate the accuracy of our model by comparing with the benchmark of microstructural effective responses that are unseen from the training process. Its accuracy is further verified in multiscale models that are subject to complex and cyclic loading conditions by providing quite close solutions to that of benchmark concurrent solver. We demonstrate that our surrogate model is robust enough to provide reliable multiscale softening solutions that are post-failure convergent and mesh independent. 

Our experiments reveal that while the costs of database generation and model training are considerable, our trained model shows superior efficiency in online computations. For example, our data-driven model is about four orders of magnitude faster than classic FE\textsuperscript{2} approach in terms of CPU hours. Such high efficiency makes our model promising for many computationally intensive tasks that would require large computational resources (multi-core CPUs and GPUs) or need long simulation time.

We can extend our data-driven model to different directions in future work. First, minimization of inference error is critical especially for iterative solvers. While teacher-forcing shows accuracy improvement in single scale simulation, we are interested in how its performance improvement can be translated to the online iterative multiscale computations without available ground truth values. Second, we are also interested in studying the impacts of spatially varying material properties and microstructural morphology on the behaviors of macro components. However, adding such variations would dramatically increase the dimension of sampling space and therein the number of sampling points. In such scenario, to reduce sampling efforts, an adaptive sampling strategy \cite{bengio2015scheduled} for sequence learners needs to be investigated. Last but not the least, our work can be potentially extended for studying uncertainty quantification \cite{bostanabad2018uncertainty} and design optimization for material and structural designs \cite{osanov2016topology,ma2006multidomain,deng2016multi,deng2015multi}.

\noindent \textcolor{navy}{\textbf{Declaration of competing interest}}

The authors declare that they have no known competing financial interests or personal relationships that could have appeared to influence the work reported in this paper.

\noindent \textcolor{navy}{\textbf{Acknowledgments}}

The authors appreciate the supports from ACRC consortium members. The authors also thank Dr. Ling Wu and Dr. Ludovic Noels for helpful discussions and constructive suggestions. Ramin Bostanabad also acknowledges NSF funding (award number OAC-$2103708$).

\appendix
\addcontentsline{toc}{section}{Appendices}
\section*{Appendices}

\section{Deflated clustering analysis} \label{sec:appendix_A}
\setcounter{equation}{0}
\renewcommand{\theequation}{\thesection-\arabic{equation}}
\setcounter{figure}{0}
\renewcommand\thefigure{\thesection.\arabic{figure}}

\begin{figure}[!t]
    \centering
    \includegraphics[width = 0.6\textwidth]{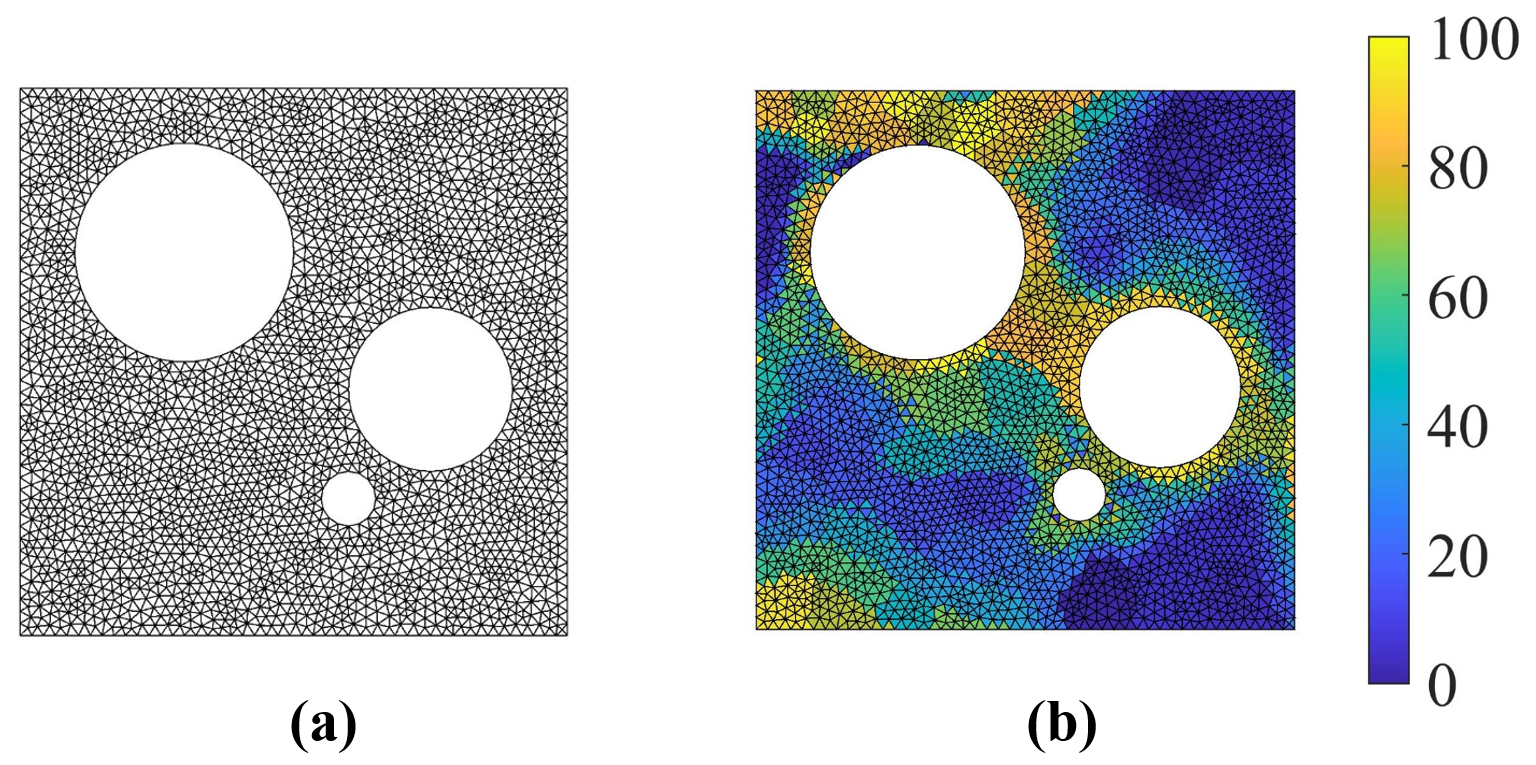}
    \caption{\textbf{Demonstration of clustering in ROM:} The domain of a generic 2D RVE with $5,000$ elements in \textbf{(a)} are decomposed into $100$ clusters in \textbf{(b)} where elements in the same cluster are assigned with the same color.} 
    \label{fig:2DRVE_clutering}
\end{figure}

Simulation of microstructural softening via classic FE\textsuperscript{2} method involves demanding computational costs, which is prohibitive for generating big training data for machine learning models. To accelerate the database generation, we adopt our previously developed mechanistic ROM, i.e., deflated clustering analysis (DCA) \cite{deng2022reduced,deng2022concurrent}. Its high efficiency comes from two facts: (1) the number of unknown variables in the system is dramatically reduced from a large number of finite elements to a few clusters by agglomerating elements via clustering as shown in Figure \ref{fig:2DRVE_clutering}, and (2) the algebraic equations of the reduced system contains much fewer close-to-zero eigenvalues that results in better convergence comparing to the classic FE system.     
  
Our DCA utilizes k-means clustering, i.e., an unsupervised machine learning technique for data interpretation and grouping, to agglomerate neighboring elements into a set of interactive irregular-shape clusters. The clustering begins with feeding the coordinates of element centroids into a feature space where randomly scattered cluster seeds serve as initial cluster means. Clusters accepts or rejects elements by iteratively minimizing the within-cluster variance until all elements are assigned to a cluster. The clustering procedure can be mathematically stated as a minimization problem as: 

\begin{equation}
    \textbf{C} =  \min\limits_{\textbf{C}}\sum\limits_{I = 1}^k \sum\limits_{n \in C^I} \lVert \pmb{\varphi}_n - \Bar{\pmb{\varphi}}_I \rVert^2
\end{equation}

\noindent where \textbf{C} represents the k clusters with $\textbf{C} = \{C^1, C^2, \dots, C^k\}$. $\pmb{\varphi}_n$ and $\Bar{\pmb{\varphi}}_I$ indicate the coordinates of the centroid of the $n^{th}$ element and the mean of the coordinates of the $I^{th}$ cluster, respectively. A clustering example is illustrated in Figure \ref{fig:2DRVE_clutering} where the discrete domain of a 2D generic RVE with $5,000$ elements are decomposed into $100$ clusters.

We construct clustering-based reduced mesh via Delaunay triangularization by connecting cluster centroids where the topological relations between clusters are preserved from the original FE mesh. By assuming the motions of cluster centroids are directly related to clustering nodes, we can compute the nodal displacements via polynomial augmented radian point interpolation \cite{liu2009meshfree} as:

\begin{equation}
    \textbf{u}_c = \textbf{R}\textbf{a} + \textbf{Z}\textbf{b}
\end{equation}

\noindent where $\textbf{u}_c$ represent the displacements of cluster centroids. $\textbf{a}$ is the coefficient vector of the radial basis function matrix $\textbf{R}$, and $\textbf{b}$ is the coefficient vector of the polynomial basis matrix $\textbf{Z}$. Meanwhile, the radial coefficient and the polynomial basis need to satisfy the following equation for every node per cluster and every polynomial basis function to ensure solution uniqueness \cite{liu2009meshfree} as:

\begin{equation}
    \textbf{Z}\textbf{a} = \textbf{0}
\end{equation}

The displacements of cluster centroids are augmented with rotational degrees of freedom to represent the six rigid body motions in a 3D deflation space \cite{jonsthovel2009preconditioned}, including three translations and three rotations. Upon the completion of non-linear analysis on the reduced mesh, the displacement solutions can be projected back to the original FE mesh by:

\begin{equation}
    \textbf{u}_{i}^{j} = \textbf{W}_{i}^{j} \pmb{\lambda}_{j}
\end{equation}

\noindent where $\textbf{u}_{i}^{j}$ represents the displacement vector at the $i^{th}$ node in the $j^{th}$ cluster. In addition, $\pmb{\lambda}_{j}$ is the rigid body motion of the centroid of the $j^{th}$ cluster, while the $\textbf{W}_{i}^{j}$ indicates the deflation matrix for the $i^{th}$ node in the $j^{th}$ cluster as:

\begin{equation}
    \pmb{\lambda}_{j} = [u_{jx}, u_{jy}, u_{jz}, \theta_{jx}, \theta_{jy}, \theta_{jz}]^{T};
    \quad
    \textbf{W}_{i}^{j} = \begin{bmatrix} 1 & 0 & 0 & 0 & z_{i}^{j} & -y_{i}^{j} 
    \\ 0 & 1 & 0 & -z_{i}^{j} & 0 & x_{i}^{j}
    \\ 0 & 0 & 1 & y_{i}^{j} & -x_{i}^{j} & 0 \end{bmatrix}
\end{equation}

\noindent where $u_{jx}$ and $\theta_{jx}$ are the displacement and rotation of the $j^{th}$ cluster along x axis, and the ($x_{i}^{j}$, $y_{i}^{j}$, $z_{i}^{j}$) are the relative 3D coordinates of the $i^{th}$ node with respect to the centroid of the $j^{th}$ cluster. By assuming all elements in the same cluster share identical stress and strain fields, microstructural effective responses can be reproduced in a highly efficient manner such that the unknown variables are dramatically decreased from FE system that accounts for distinct field variables per element to the reduced system with much fewer distinct solutions per cluster.

\begin{figure}
    \centering
    \begin{subfigure}[b]{0.6\textwidth}
        \centering
        \includegraphics[width=\textwidth]{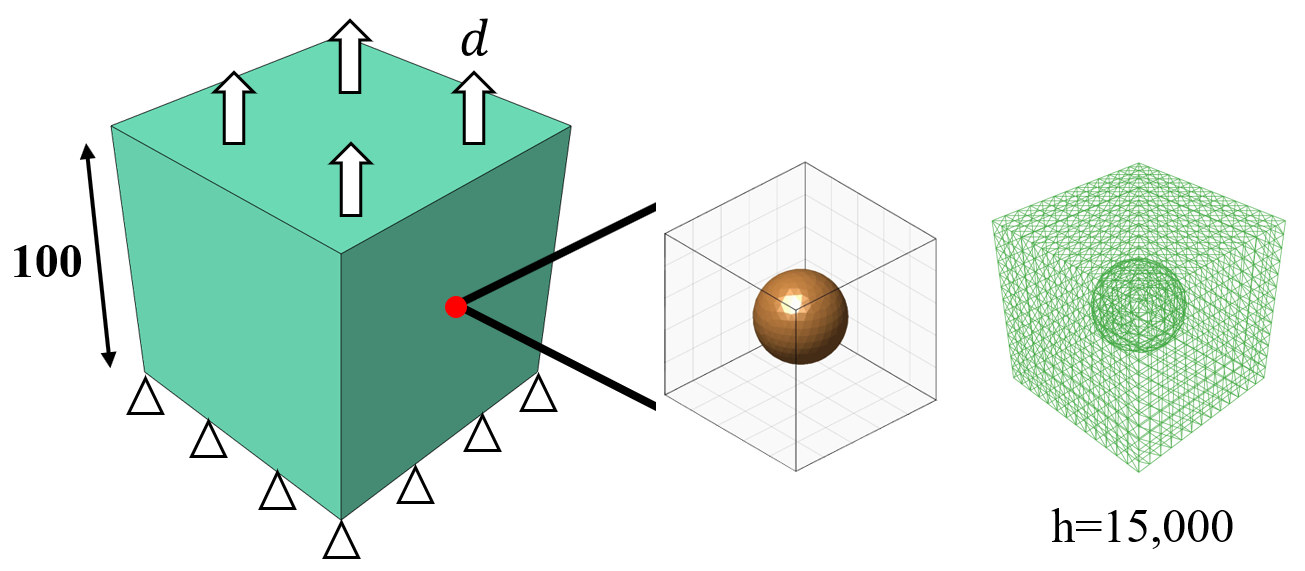}
        \caption{Geometry, dimensions (unit: mm), and boundary conditions of the macro-model; Geometry and the finite element mesh of the porous RVE}
    \end{subfigure}
    \\[\smallskipamount]
    \begin{subfigure}[b]{0.4\textwidth}
        \centering
        \includegraphics[width=\textwidth]{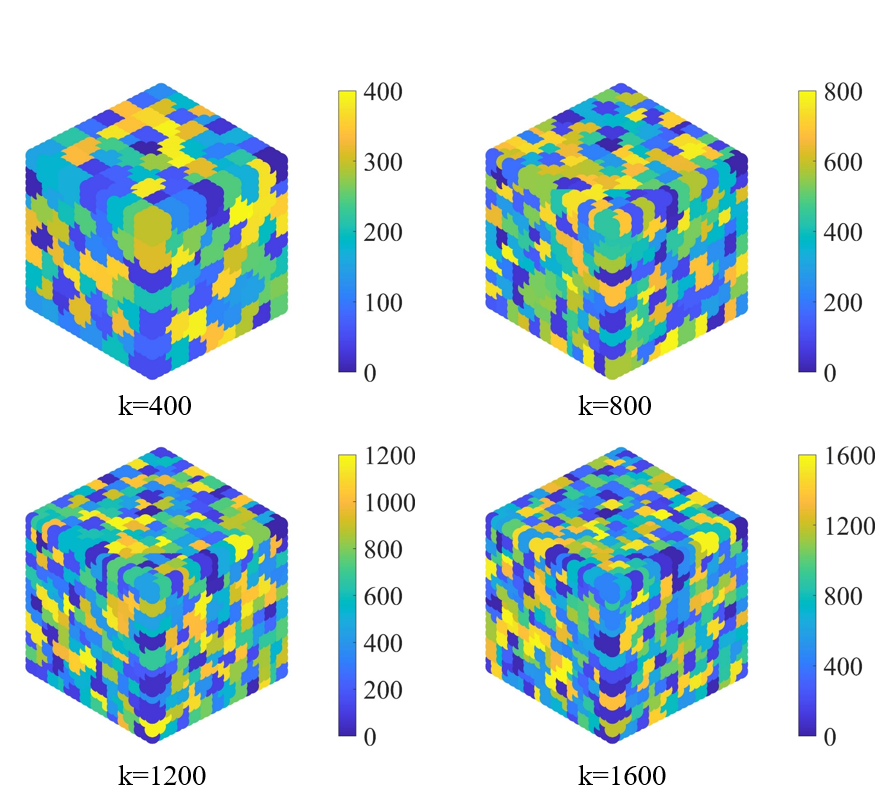}
        \caption{RVE clustering}
    \end{subfigure}
    \caption{\textbf{Multiscale cube model:} (\textbf{a}) Every integration point of the macro-cube model is associated with a porous RVE; and (\textbf{b}) The RVE domain is discretized by different numbers of clusters.}
    \label{fig:cube_model}
\end{figure}

To demonstrate the efficacy of DCA-based ROM, we compare its simulation results on a 3D multiscale cube against the classic FE\textsuperscript{2} method in Figure \ref{fig:cube_model}. The macro-cube is fully constrained at its bottom surface, and it is subject to an upward extension on the top surface with $d = 7$ mm. The cube is meshed with 12 tetrahedral elements of reduced-integration (one IP at the center of each tetrahedron). We assume each macro-IP is associated with the same porous RVE containing one spherical pore in the middle as shown in Figure \ref{fig:cube_model}(a). To illustrate the effects of clustering on the RVE's effective softening behaviors, we adopt four clustering levels on the same RVE mesh ($15,000$ elements) with the number of clusters ($k$) as $400$, $800$, $1,200$ and $1,600$ as in the Figure \ref{fig:cube_model}(b).

\begin{figure}
    \centering
    \begin{subfigure}[b]{0.45\textwidth}
        \centering
        \includegraphics[width=\textwidth]{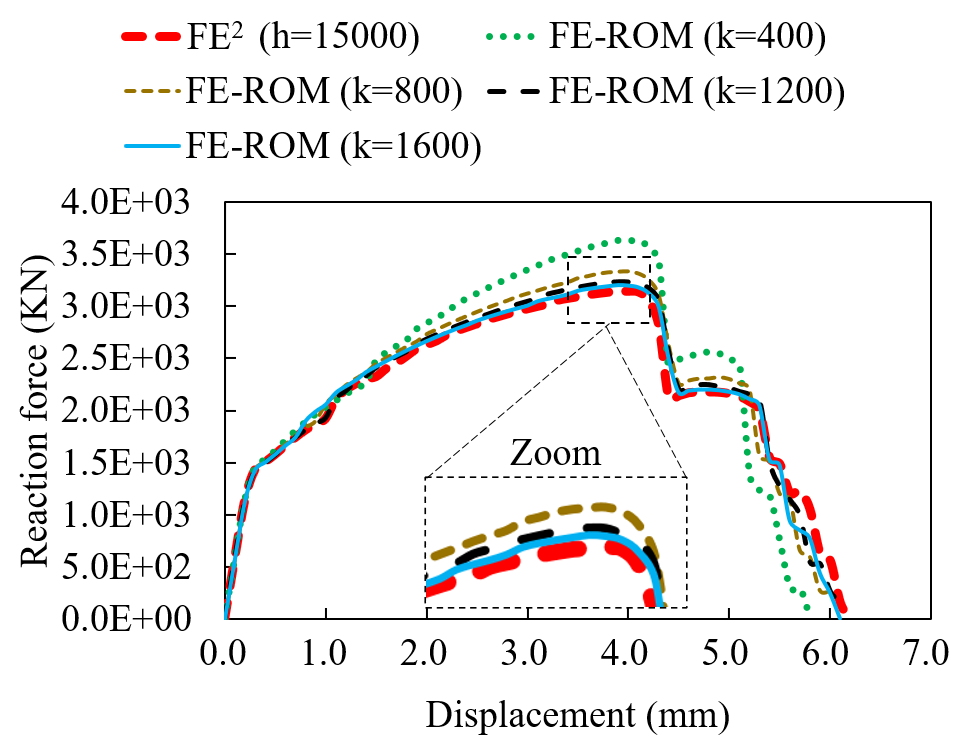}
        \caption{Reaction force and displacement}
    \end{subfigure}
    \hfill
    \begin{subfigure}[b]{0.45\textwidth}
        \centering
        \includegraphics[width=\textwidth]{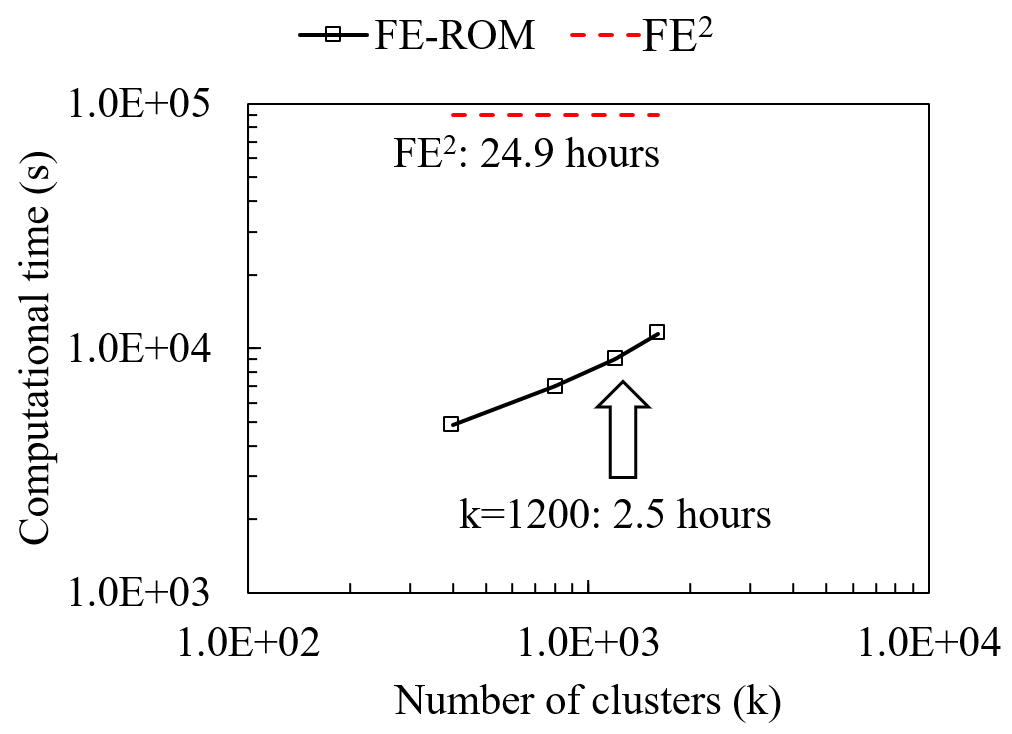}
        \caption{Computational time}
    \end{subfigure}
    \caption{\textbf{Results of the multiscale cube model:} (\textbf{a}) Comparison of the softening load-displacement curves between FE\textsuperscript{2} and FE-ROM with different clusters; and (\textbf{b}) Comparison of computational time.}
    \label{fig:cube_sols}
\end{figure}

We compare the reaction force-displacement curves from FE\textsuperscript{2} and FE-ROM in Figure \ref{fig:cube_sols}(a). By considering the FE\textsuperscript{2} solutions as the benchmark, we observe that: (1) the FE-ROM solutions of $k = 400$ generally overestimates the component's strength which is due to insufficient clustering in the RVE that results in artificially strong material responses as discussed in \cite{liu2016self,deng2022reduced}; and (2) With more clusters, the FE-ROM responses, especially the post-failure behaviors, become more and more closer to the benchmark. In specific, we observe that when the numbers of clusters increase to $1,200$ and $1,600$, FE-ROMs achieve sufficiently accurate results compared to FE\textsuperscript{2} benchmark.

We further quantify the computational costs of the different solvers in Figure \ref{fig:cube_sols}(b). While all experiments are performed on a HPC by paralleling $60$ CPU cores with $360$ GB RAM, the clock time of FE\textsuperscript{2} is the longest, accounting for $24.9$ hours. The clock time of the ROM with $1,200$ and $1,600$ clusters is about $2.5$ and $3.2$ hours, resulting in the acceleration factors of $9.9$ and $7.8$, respectively. Considering the fact that the ROM of $k = 1,200$ is about $28\%$ faster than its counterpart of $k = 1,600$ while achieving similar accuracy, we adopt $k = 1,200$ as the clustering of choice for the generation of the RVE softening database and for the FE-ROM multiscale simulations in Section \ref{sec:experiments}.

\section{Gated recurrent unit} \label{sec:appendix_B}
\setcounter{equation}{0}
\renewcommand{\theequation}{\thesection-\arabic{equation}}
\setcounter{figure}{0}
\renewcommand\thefigure{\thesection.\arabic{figure}}

In order to avoid the vanishing and exploding gradient issues of traditional RNN in processing long sequential data, more advanced memory cells are developed for sequential learners, including the long short term memory (LSTM) and the gated recurrent unit (GRU). Specifically, GRU is a variant of the LSTM, showing comparable performance to LSTM while exhibiting higher efficiency due to its compacted internal structures and fewer math operations. It is for this reason we choose GRU as the memory cell in our proposed RNN architecture as in Figure \ref{fig:GRU_graph}.

\begin{figure}
    \centering
    \includegraphics[width = 0.8\textwidth]{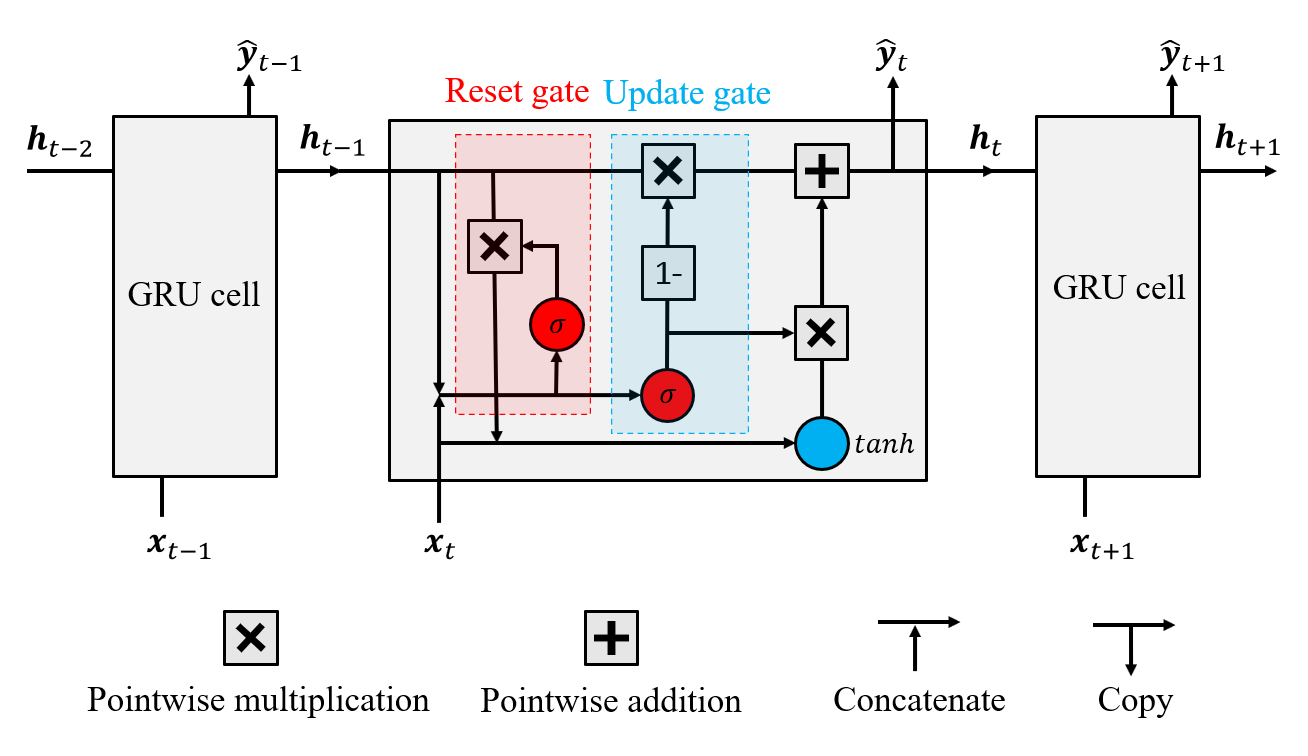}
    \caption{\textbf{Architectures of GRU layer and cells:} The internal structure and mathematical operations are demonstrated in the GRU cell at time step $t$.
    } \label{fig:GRUCell}
\end{figure}

To demonstrate the working mechanism, we illustrate a GRU layer of three interconnected cells in Figure \ref{fig:GRUCell}. In a GRU layer, a typical cell at an arbitrary time step $t$ generates predictions $\hat{\textbf{y}}_t$ and internal memory-like hidden variables $\textbf{h}_t$ after reading in the current inputs $\textbf{x}_t$ and the hidden variables $\textbf{h}_{t-1}$ from the previous cell. Comparing to the RNN cell in Figure \ref{fig:RNN_compGraph}(b), the GRU cell has a more sophisticated gate structure to regulate its internal information flow. Specifically, a GRU cell includes two gate operations: a reset gate and an update gate. 

The reset gate $\textbf{r}_t$ begins to read the current input state $\textbf{x}_t$ and the previous hidden state $\textbf{h}_{t-1}$. It determines a candidate hidden state $\hat{\textbf{h}}_t$ by filtering out less important information passing from the previous cell. Its operations include:
\begin{subequations}
\begin{align}
\mathbf{r}_t &=\sigma\left(\mathbf{W}_{h r} \mathbf{h}_{t-1}+\mathbf{W}_{x r} \mathbf{x}_t+\mathbf{b}_r\right) \label{eq:gru1}\\
\tilde{\mathbf{h}}_t &=\operatorname{tanh}\left(\mathbf{r}_t \odot \mathbf{W}_{h \tilde{h}} \mathbf{h}_{t-1}+\mathbf{W}_{x \tilde{h}} \mathbf{x}_t+\mathbf{b}_{\tilde{h}}\right) \label{equ:GRU_reset}
\end{align}
\end{subequations}
\noindent where $\sigma$ is the sigmoid activation function that returns a value in the range of $[0,1]$, $tanh$ is the hyperbolic tangent function, and $\odot$ represents the element-wise product operation. $\mathbf{W}_{hr}$, $\mathbf{W}_{xr}$, $\mathbf{W}_{h \tilde{h}}$, $\mathbf{W}_{x \tilde{h}}$ are the weighting matrices associated with the hidden state, the input state, the hidden-to-candidate hidden state and the input-to-candidate hidden state, respectively. $\mathbf{b}_r$ and $\mathbf{b}_{\tilde{h}}$ are the biases applied to the sigmoid function in the reset gate and the hyperbolic tangent function, respectively. 

In a similar manner, the update gate operates on $\textbf{x}_t$ and $\textbf{h}_{t-1}$ but using different weights and biases terms. More precisely, the update gate linearly interpolates the previous hidden state $\textbf{h}_{t-1}$ and the candidate hidden state $\tilde{\mathbf{h}}_t$ to update the memory-like hidden state $\textbf{h}_t$ passing onto the next time step:
\begin{subequations}
\begin{align}
\mathbf{u}_t &=\sigma\left(\mathbf{W}_{hu} \mathbf{h}_{t-1}+\mathbf{W}_{x u} \mathbf{x}_t+\mathbf{b}_u\right) \label{eq:gru2} \\
\mathbf{h}_t &=\mathbf{u}_t \odot \mathbf{h}_{t-1}+\left(1-\mathbf{u}_t\right) \odot \tilde{\mathbf{h}}_t+\mathbf{b}_h 
\label{equ:GRU_update}
\end{align}
\end{subequations}
\noindent where $\mathbf{W}_{hu}$ and $\mathbf{W}_{xu}$ are the weights applied onto the hidden state and input state in the update gate. $\mathbf{b}_u$ and $\mathbf{b}_h$ are the two biases associated to the sigmoid function and the generation of current hidden state. In the end, the cell output at the current time step $\hat{\mathbf{y}}_t$ is linearly transformed from the hidden state as:
\begin{equation}
\hat{\mathbf{y}}_t =\mathbf{W}_{h y} \mathbf{h}_t+\mathbf{b}_y  
\end{equation}
\noindent where $\mathbf{W}_{hy}$ and $\mathbf{b}_y$ are the weights and biases associated to the current output state $\hat{\mathbf{y}}_t$. We note that all the weights and biases of the GRU networks are iteratively updated by BPTT during training.

\newgeometry{left=1in,right=1.05in, top=1in, bottom=1.05in}
{\setstretch{1.5}
\printbibliography
}



\end{document}